\documentclass[10pt,twoside,english]{book}

\usepackage{thesis}
\usepackage{latexsym}
\usepackage{graphicx}
\usepackage{times}
\usepackage{amsmath}
\usepackage{amssymb}
\usepackage{amsthm}
\usepackage{float}
\usepackage[format=hang,
	font=small,
	labelfont=bf,up,
	textfont=up
]{caption}
\usepackage{subfig}
\usepackage{multirow}
\usepackage[table]{xcolor}
\usepackage[final]{changes}
\usepackage[square,
	comma,
	numbers,
	sort&compress
]{natbib}
\usepackage{tocvsec2}
\usepackage[bookmarks=true,
	bookmarksopen=true,
	bookmarksnumbered=true,
	plainpages=false,
	colorlinks=true,
	urlcolor=black,
	citecolor=black,
	linkcolor=black,
	pdfstartview=Fit,
	pdftitle={PhD thesis of Matthias Degroote},
	pdfauthor={Matthias Degroote},
	pdfsubject={Faddeev Random Phase Approximation for Molecular Systems},
	pdfkeywords={},
	pdfpagelabels=true
]{hyperref}
\usepackage[chapter]{algorithm}
\usepackage{algpseudocode}
\usepackage{tweaklist}
\usepackage{longtable}
\usepackage[UKenglish,
	dutch
]{babel}
\usepackage{hyphenat}
\usepackage{pdfpages}
\usepackage[papersize={16cm,24cm},
	margin=2cm,
	includehead,
	includefoot]
{geometry}
\usepackage{booktabs}
\usepackage{subfig}
\usepackage{rotating}
\usepackage{rotfloat}

\newcommand{\Equ}[1]{Eq.~(\ref{equ:#1})}

\newcommand{\Fig}[1]{Figure~\ref{fig:#1}}
\newcommand{\Tab}[1]{Table~\ref{tab:#1}}
\newcommand{\Cha}[1]{Chapter~\ref{cha:#1}}
\newcommand{\Hoo}[1]{Hoofdstuk~\ref{cha:#1}}
\newcommand{\Sec}[1]{Section~\ref{sec:#1}}
\newcommand{\App}[1]{Appendix~\ref{app:#1}}

\newcommand{\tablerowcolor}[1]{\ifodd #1 \rowcolors{#1}{gray!25}{gray!10}\else\rowcolors{#1}{gray!10}{gray!25}\fi}
\newcolumntype{o}{>{\columncolor{gray!10}}c}
\newcolumntype{e}{>{\columncolor{gray!25}}c}
\newcolumntype{w}{>{\columncolor{white}}c}
\newcommand{\clearnewpage}{\clearpage{\pagestyle{empty}\cleardoublepage}}

\newcommand{\addtocchapter}[1]{\phantomsection\addcontentsline{toc}{chapter}{#1}}

\renewcommand{\enumhook}{\setlength{\topsep}{0pt}\setlength{\itemsep}{-0.8ex}}
\renewcommand{\itemhook}{\setlength{\topsep}{0pt}\setlength{\itemsep}{-0.8ex}}
\renewcommand{\descripthook}{}

\renewcommand{\familydefault}{cmss}

\fontencoding{\encodingdefault}
\fontfamily{\familydefault}
\fontseries{\seriesdefault}
\fontshape{\shapedefault}
\selectfont

\newfont{\sfq}{cmssq8 scaled 1200}

\renewcommand{\bibname}{Bibliography}
\renewcommand{\contentsname}{Contents}
\renewcommand{\listfigurename}{List of Figures}
\renewcommand{\listtablename}{List of Tables}

\newcommand{\samenvattingname}{Samenvatting}
\newcommand{\prefacename}{Preface}
\newcommand{\papername}{List of Published Papers}

\renewcommand{\emph}[1]{\textit{#1}}

\begin{document}

\selectlanguage{UKenglish}
\graphicspath{
   {generalities/}
   {summaries/dutch/}
   {summaries/english/}
   {chapters/introduction/}
   {chapters/methods/}
   {chapters/frpa/}
   {chapters/results/}
   {chapters/twoblackbox/}
   {chapters/oneblackbox/}
   {chapters/multiple/}
   {chapters/applications/}
   {chapters/conclusion/}
   {appendices/pictures/}
   {appendices/tango/}
   {appendices/timedependent/}
   {appendices/changes/}
}

\pagestyle{empty}
\pagenumbering{roman}

\begin{titlepage}
  \setlength{\unitlength}{1mm}
	\begin{sffamily}
		\begin{figure}[!h]
			\vspace{6.0cm}
		\end{figure}
		\begin{large}
		  \begin{flushleft}
			  Faddeev random phase benadering toegepast op moleculen																						\\
			\end{flushleft}
		\end{large}
		\begin{large}
		  \begin{flushleft}
			  Faddeev random phase approximation applied to molecules																																\\
			\end{flushleft}
		\end{large}
		\begin{large}
		  \begin{flushleft}
    		Matthias Degroote
			\end{flushleft}
		\end{large}
		\vspace{4.0cm}
		\begin{normalsize}
		  \begin{flushleft}
    		Promotor: Prof. dr. Dimitri Van Neck																																									\\
    		Proefschrift ingediend tot het behalen van de graad van																																	\\
    		Doctor in de Wetenschappen: Natuurkunde
			\end{flushleft}
		\end{normalsize}
		\begin{normalsize}
		  \begin{flushleft}
    		Vakgroep Fysica \& Sterrenkunde																																	\\
    		Voorzitter: Prof. dr. Dirk Ryckbosch																																									\\
    		Faculteit Wetenschappen																																												\\
    		Academiejaar 2011-2012
			\end{flushleft}
		\end{normalsize}
	\end{sffamily}
	\vspace{-2.5cm}
	\begin{figure}[h]
		\begin{flushright}
		  \includegraphics{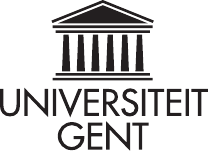}
		\end{flushright}
	\end{figure}
\end{titlepage}

Supervisor:																																																											\\
\vspace{0.2cm}																																																									\\
\begin{tabular}{l}
  Prof. dr. Dimitri Van Neck																																																	\\
\end{tabular}																																																										\\
\vspace{1.0cm}																																																									\\
Research facility:																																																							\\
\vspace{0.2cm}																																																									\\
\begin{tabular}{l}
	Center for Molecular Modelling																																							\\
	Technologiepark 903																																																		\\
	B-9052 Zwijnaarde																																																									\\
	Belgium																																																												\\
\end{tabular}																																																										\\
\vspace{1.0cm}																																																									\\
Members of the examination committee:																																																	\\
\vspace{0.2cm}																																																									\\
\begin{tabular}{l@{\qquad}l@{\qquad}l}
   Prof.~dr.~Carlo Barbieri							&	University of Surrey																												\\
	Prof.~dr.~Freddy Callens (chairman)                  & Ghent University  \\
	Prof.~dr.~Wim Dickhoff							&	Washington University at St. Louis																																			\\
   Prof.~dr.~Jan Ryckebusch                  & Ghent University  \\
   Prof.~dr.~Peter Schuck							& Institut de Physique Nucl\'eaire d'Orsay 																																	\\
                                             & Laboratoire de Physique et Mod\'elisation   \\
                                             & des Milieux Condens\'es \\
   dr.~Kris Van Houcke                       & Ghent University   \\
   Prof.~dr.~Dimitri Van Neck (supervisor)						&	Ghent University																											\\
   Prof.~dr.~ir.~Veronique Van Speybroeck    & Ghent University  \\
   Prof.~dr.~Michel Waroquier                & Ghent University  \\
\end{tabular}																																																										\\
\vspace{4.5cm}																																																										\\
The research presented in this dissertation was supported by a Ph. D. fellowship of the Research Foundation - Flanders (FWO).
												\clearnewpage

\pagestyle{plain}

\addtocchapter{\prefacename}
\chapter*{\prefacename}

Het is moeilijk iedereen te bedanken. Er zijn heel veel mensen die, over de vier jaar die mijn doctoraat geduurd heeft, een invloed gehad hebben en die op een of andere manier een bijdrage geleverd hebben aan het succesvol afronden van dit werk.

Op de eerste plaats wil ik de mensen van de vakgroep bedanken. Zowel op het INW als op het CMM. De koffiepauzes, sociale activiteiten, voetbalwoensdagen, avondlijke afspraken, kerstfeestjes, barbecues en zoveel meer hebben ertoe bijgedragen dat ik mij thuis gevoeld heb. Veel van de collega's zijn dan ook uitgegroeid tot echte vrienden. Natuurlijk dank ik ook mijn promotor voor de samenwerking en de ondersteuning en het hoofd van de onderzoeksgroep voor de mogelijkheden die een plaats als het CMM biedt.

Onderweg heb ik de vriendschap van veel mensen mogen ontvangen. Vrienden uit Ninove, uit Gent, van de voetbal, in de cit\'e en overal waar ik kwam. Mensen die onlosmakelijk met deze vier intensieve jaren verbonden zijn en van wie ik hoop dat ik ze verder op mijn pad mag blijven tegenkomen.

De belangrijksten waren echter mijn familie. Mijn ouders en grootouders, broer en zus hebben mij aangemoedigd en ondersteund wanneer het nodig was. En dan was daar opeens Clara, die zonder het te beseffen de eindsprint verlicht heeft met haar aanwezigheid.

Ik kan jullie niet genoeg bedanken,

\begin{flushright}
	Gent, Juni 2012\\
	Matthias Degroote
\end{flushright}
														\clearnewpage
\addtocchapter{\samenvattingname}
\selectlanguage{dutch}
\chapter*{\samenvattingname}

De komst van ultrasnelle computers heeft een enorme vooruitgang betekend voor de simulatie van kwantumveeldeeltjesproblemen. Ze zijn een essentieel onderdeel geworden van de studie van complexe systemen gaande van biologische toepassingen tot de materiaalwetenschappen. Deze systemen vereisen allemaal de numerieke oplossing van de elektronische Schr\"odinger-vergelijking zodanig dat de totale energie, de elektronendichtheid of andere eigenschappen van het systeem voor handen zijn. Wanneer de Schr\"odinger-vergelijking in zijn exacte vorm wordt opgelost, botst men al snel op de exponenti\"ele toename van de hoeveelheid informatie die vereist is. Dit maakt dat een exacte oplossing onmogelijk is en er benaderingen dienen gemaakt te worden.

Het doel van de kwantumchemie is de bepaling van de elektronische structuur tot op chemische nauwkeurigheid. Dit betekent: vergelijkbaar met experimentele waarden en met voorspellende kracht. Elke ontwikkelde methode moet onderworpen worden aan zorgvuldige screening van deze eis door middel van testsystemen.

Hartree-Fock is een theorie die soms voldoende betrouwbare resultaten oplevert. Maar toch is het met deze techniek onmogelijk om correcte voorspellingen te bekomen voor chemische processen waarin elektroncorrelatie belangrijk wordt. De zoektocht naar een techniek die zowel betrouwbaar is als computationeel goed schaalt, gaat dus onvermoeibaar verder.

Het ontwikkelen van nieuwe kwantummechanische methoden gebeurt deels met het oog op het gebruik in industrieel onderzoek door niet-specialisten. Daardoor is het belangrijk om ervoor te zorgen dat de methode zonder veel parameters of instellingen kan toegepast worden. \textit{Ab initio} methoden hebben als voordeel dat er geen empirische data vereist is (en dus ook geen fundamenteel begrip) om berekingen te doen. Deze methoden gebruiken enkel fundamentele waarden verbonden aan het systeem. Dit maakt ze ideale kandidaten voor dagelijks gebruik in kwantumchemische berekeningen. Het is echter niet eenvoudig een kwantummechanische techniek te vinden die over een wijd spectrum van systemen kan toegepast worden.

Greense functies zijn de aangewezen grootheden voor de directe bepaling van de elektronische structuur van atomen en molecules. In plaats van de volledige elektronische golffunctie te berekenen, volstaat het om de eigenschappen van enkele elektronen in hun omgeving te beschouwen. De overeenkomstige fysische grootheden zijn de Greense functies of propagatoren. Er zijn verscheidene voordelen aan het gebruik van Greense functies bij de beschrijving van de fysica van veeldeeltjessystemen. Ten eerste vormen ze een directe berekeningsmethode voor fysische grootheden. Eens de Greense functie gekend is, is het mogelijk de verwachtingswaarde van elke \'e\'endeeltjesoperator en de grondtoestandsenergie te bepalen zonder verdere berekeningen. Ook de ionisatie-energie\"en en spectrale sterkte is binnen handbereik. Ten tweede laten ze een eenvoudige fysische interpretatie toe. De kwantummechanische Greense functie bepaalt de evolutie van toestanden in de tijd en ruimte, net als de klassieke tegenhanger. De diagrammatische representatie in Feynman-diagrammen biedt een gemakkelijke en universele mogelijkheid om theorie\"en te vergelijken en te verduidelijken. Op deze manier kan men ook nieuwe methoden ontwikkelen.

In dit werk introduceren we een Greense-functietechniek voor de berekening van grondtoestandsenergie\"en en ionisiatie-energie\"en in kwantumveeldeeltjessystemen. De elementaire bouwstenen van deze theorie zijn de RPA (random phase approximatie) excitaties. Deze worden gebruikt om een benadering voor de zelfenergie op te bouwen via een stelsel Faddeev-vergelijkingen. De resulterende FRPA (Faddeev RPA) is een potentieel wijd toepasbare Greense functie techniek. Het gebruik van de RPA heeft als gevolg dat de techniek bruikbaar is voor uitgestrekte systemen zoals het uniform elektronengas en nucleaire materie, maar ook voor eindige systemen zoals atomen en moleculen. Wanneer de FRPA in zijn volledig zelfconsistente vorm wordt uitgevoerd, voldoet deze aan de Baym-Kadanoff voorwaarden en zal er bijgevolg aan belangrijke behoudswetten voldaan zijn.

Eerst worden de theoretische concepten ge\"introduceerd die nodig zijn om een benadering in te voeren voor de zelfenergie. De Dyson-vergelijking wordt geformuleerd in functie van de zespunts vertexfunctie. Dit maakt de koppeling tussen \'e\'endeeltjestoestanden en 2p1h (two-particle-one-hole) en 2h1p (two-hole-one-particle) toestanden natuurlijk. Daarna worden de polarisatiepropagator en de tweedeeltjes Greense functie ingevoerd. De toepassing van de RPA voor deze objecten vormt een benadering voor de ph en pp excitaties.

In \Hoo{frpa} wordt de FRPA techniek afgeleid. Het resultaat is een methode die pp (particle-particle) en ph (particle-hole) interacties van het RPA niveau vermengt op gelijke voet. De procedure die daarvoor zorgt, is ontwikkeld in een ruimte die naast de \'e\'endeeltjestoestanden ook de 2p1h en 2h1p toestanden omvat. De Faddeev-vergelijkingen zorgen ervoor dat de pp en ph excitaties op een correcte manier gekoppeld worden tot 2p1h en 2h1p toestanden. De eigenwaarden en transitie-amplitudes uit het RPA probleem worden gebruikt om de Faddeev-matrices op te bouwen. Het is absoluut noodzakelijk dat de Faddeev-techniek gebruikt wordt om excitaties te kunnen koppelen van het RPA type. Het is op dit vlak dat de gebruikte techniek verschilt van de gangbare ADC(3) (algebraic diagrammatic construction tot op derde orde); een techniek die interacties bevat van het TDA (Tamm-Dancoff approximation) type. Niettegenstaande zijn ze beide toch erg gelijkend. De ADC(3)-vergelijkingen kunnen afgeleid worden uit de FRPA-vergelijkingen door het weglaten van de achterwaards propagerende amplitudes van het RPA-probleem. We noemen ADC(3) vanaf nu dan ook FTDA (Faddeev TDA). Deze equivalentie was al duidelijk uit diagrammatische overwegingen, maar wij hebben dit nu ook op het vlak van de matrixvergelijkingen aangetoond. Doordat de FRPA de TDA transitie-amplitudes en eigenwaarden vervangt door hun RPA tegenhangers, hoort deze beter de correlaties te beschrijven in uitgestrekte systemen. Op het eerste gezicht lijkt het alsof de FRPA de matrixdimensies met een factor drie doet toenemen. Wanneer men echter de aard van de oplossingen bekijkt, dan kan men besluiten dat twee derden van deze oplossingen onfysisch zijn en kunnen weggelaten worden. De fysische en onfysische oplossingen kunnen gemakkelijk onderscheiden worden. Een matrixprojectie herleid de matrixdimensies tot die van een 2p1h en 2h1p diagonalisatie.

\Hoo{results} bevat de resultaten die berekend zijn met de FRPA. Als eerste toepassing is de elektronische structuur van een reeks atomen berekend. Deze berekeningen tonen aan dat zelfs voor erg kleine systemen, waar de RPA het Pauli-principe hard schendt, de berekende waarden van goede kwaliteit zijn. Algemeen kunnen we zeggen dat FRPA en FTDA zeer gelijklopende resultaten geven voor de lichtste atomen. Naarmate het atoomgetal toeneemt, worden de additionele diagrammen in de RPA van groter belang en vormen de FRPA-waarden een verbetering ten opzichte van FTDA. Beide methoden volgen de trend van CCSD (coupled cluster met singles en doubles) in de limiet van oneindig grote basis set en de afwijking ten opzichte van het experiment ligt binnen de afgeschatte extrapolatiefout. De bijna ontaarding in het Be atoom vormt een groter probleem. De fouten, die bij elke methode voorkomen, zijn waarschijnlijk vooral te wijten aan tekortkomingen van de basis set. Voor de ionisatie-energie\"en gelden dezelfde conclusies: de kleinere twee-elektron atomen vertonen geen verschillen tussen FRPA en FTDA, terwijl voor alle andere atomen de FRPA het resultaat in de richting van het experiment duwt. Ook hier vormt het Be atoom een problematisch geval.

Na deze berekeningen voor atomen is het logisch om over te gaan tot het bepalen van de eigenschappen van molecules. Bij molecules vormt de internucleaire afstand een extra parameter. Rond de evenwichtsgeometrie vinden we dat de resultaten gelijklopend zijn met die uit de atomaire berekeningen. Voor kleinere molecules vertonen FRPA en FTDA weinig verschillen, terwijl voor grotere molecules de energie\"en verder uit elkaar liggen. Het aflopen van verschillende internucleaire afstanden toont in elk geval aan dat de zelfconsistentie voor het Hartree-Fock diagram noodzakelijk is. Zonder deze iteratieve procedure wordt er voor sommige molecules zelfs geen minimum gevonden in de potenti\"ele-energiecurve. De ionisatie-energie\"en zijn ongeveer even goed in FRPA als in FTDA. Er is net als bij atomen ook een problematisch geval bij het bepalen van de ionisatie-energie\"en: de derde ionisatie-energie van de N$_2$ molecule wijkt sterk af van het experiment. In dit hoofdstuk werd ook een poging ondernomen om de basis set limiet te bereiken, maar dit ligt nog buiten de computationele mogelijkheden.

Moleculaire berekeningen in de dissociatielimiet leggen het zwakke punt van de FRPA bloot. Doordat de FRPA voor de opbouw van de matrices zeer sterk afhangt van de RPA, kan de FRPA ook enkel toegepast worden daar waar de RPA een zinvol resultaat produceert. Het is bekend dat de ph RPA een triplet instabiliteit vertoont wanneer de internucleaire afstand van de diatomische moleculen vergroot wordt. Bij de kleinere molecules vinden we niet dat de kwaliteit van de resultaten verslechtert in de buurt van de instabiliteit. Voor de grotere moleculen is dit echter niet geldig, zoals de berekeningen voor de HF molecule aantonen. Bij molecules met een hogere correlatie-energie wijkt het FRPA-resultaat sterk af van de verwachte resultaten in de buurt van de instabiliteit. Dit betekent dat de FRPA vooral een methode is die rond de evenwichtsgeometrie geldig is.

De Hubbard-molecule is een testsysteem dat hetzelfde gedrag vertoont als een molecule in de dissociatielimiet. Dit model laat toe om de problematiek van dissociatie te bekijken in de eenvoudigst mogelijke vorm. De exacte resultaten voor dit model vormen een leidraad voor het aanbrengen van verbeteringen in de huidige theorie. Door fragmentatie toe te laten in de \'e\'endeeltjespropagator is het mogelijk om voorbij de instabiliteit van de RPA toch berekeningen te doen met de FRPA.
																	\clearnewpage
\selectlanguage{UKenglish}

\maxtocdepth{subsection}
{\setlength{\parskip}{0pt}
   \addtocchapter{\contentsname}
   \tableofcontents																				\clearnewpage
      \addtocchapter{\listfigurename}
   \listoffigures																					\clearnewpage
      \addtocchapter{\listtablename}
   \listoftables																						\clearnewpage
}

\pagestyle{headings}
\pagenumbering{arabic}
\setcounter{page}{1}

\chapter{Introduction}
\label{cha:introduction}

The development of modern computers over the last 50 years has enabled the simulation of the quantum many-body problem\cite{Dickhoff2005,Fetter1971,Ring1980}. They have become an essential tool for studying complex systems such as those arising in biology and material science. The underlying technology is the computational solution of the electronic Schr\"odinger equation, i.e. to calculate the electronic energy, electron density and other properties. In its exact form, this Schr\"odinger equation becomes exponentially harder to solve with increasing number of electrons. Hence, a brute force solution is out of the question and an approximation given the positions of the nuclei and the number of electrons has to be made.

A large number of chemical properties can be calculated, among which the electronic structure and spectroscopic constants of molecules. The ambition of the quantum chemist has always been to predict these properties with chemical accuracy, meaning that the results obtained are able to compete with experimental data. Every method that is developed needs to be submitted to careful testing and benchmarking on test systems to prove its value.

Hartree-Fock theory produces results that are reasonably accurate, but is incapable of yielding correct results in chemical events where correlation is important. Thus, finding a robust treatment of electron correlation that exhibits a tractable scaling in computational effort with the size of the system, is the key problem for present-day model developers. 

The aim of a newly developed method is always the widespread use by non-specialists in research and industry. With this in mind, the black-box character of a method greatly determines its applicability in the wider field. Ab initio methods constitute a class of simulation techniques that do not rely on empirical input (and thus deeper understanding) from the user. These methods use only fundamental knowledge of the system of interest. This means that they are the perfect candidates for every-day use in quantum chemical calculations. Nevertheless, it is not easy to find an ab initio method that can be used for any system from the smallest toy models to extended systems such as solids.

\section{Motivation}
\label{sec:motivation}

Green's function methods\cite{Dickhoff2005,Fetter1971,Linderberg2005} represent a very useful set of tools for direct calculation of electronic structure properties of molecules\cite{Danovich2011,VonNiessen1984,Tsui1974,Kutzelnigg2009}. They rely on the idea that for the description of various properties of the molecule, it is not necessary to know the exact information of every electron. It suffices to describe one or two particles within their environment. The corresponding physical quantities are the one- and two-particle Green's functions. 

The advantages of using Green's functions to describe physical phenomena in many-body systems are numerous. First of all, the most important physical quantities can be calculated directly. When the Green's function is known, deriving these variables requires no further calculations. Secondly, they have a simple physical interpretation. Just like the classical analogon, the quantum mechanical Green's function can be seen as the object that propagates the particles of interest in time and space. Another advantage is the easy and systematic way of expressing Green's functions in terms of Feynman diagrams. These graphical representations can be used as a tool for the interpretation of certain approximations and are a powerful communication tool about different theories. They also present a succesful construction path for new approximation schemes.

The Green's function formalism provides a good alternative to widely spread many-body techniques such as quantum Monte Carlo (QMC), configuration interaction (CI) and coupled cluster (CC) theory. 

The remainder of this introduction presents a short overview of the historical evolution of methods as well as the relevant literature that has led to the development of the approach considered in the present work. 
\subsection{Extended systems}
\subsubsection{Electron gas}

\begin{figure}
\begin{center}
\includegraphics[scale=0.5,clip=true]{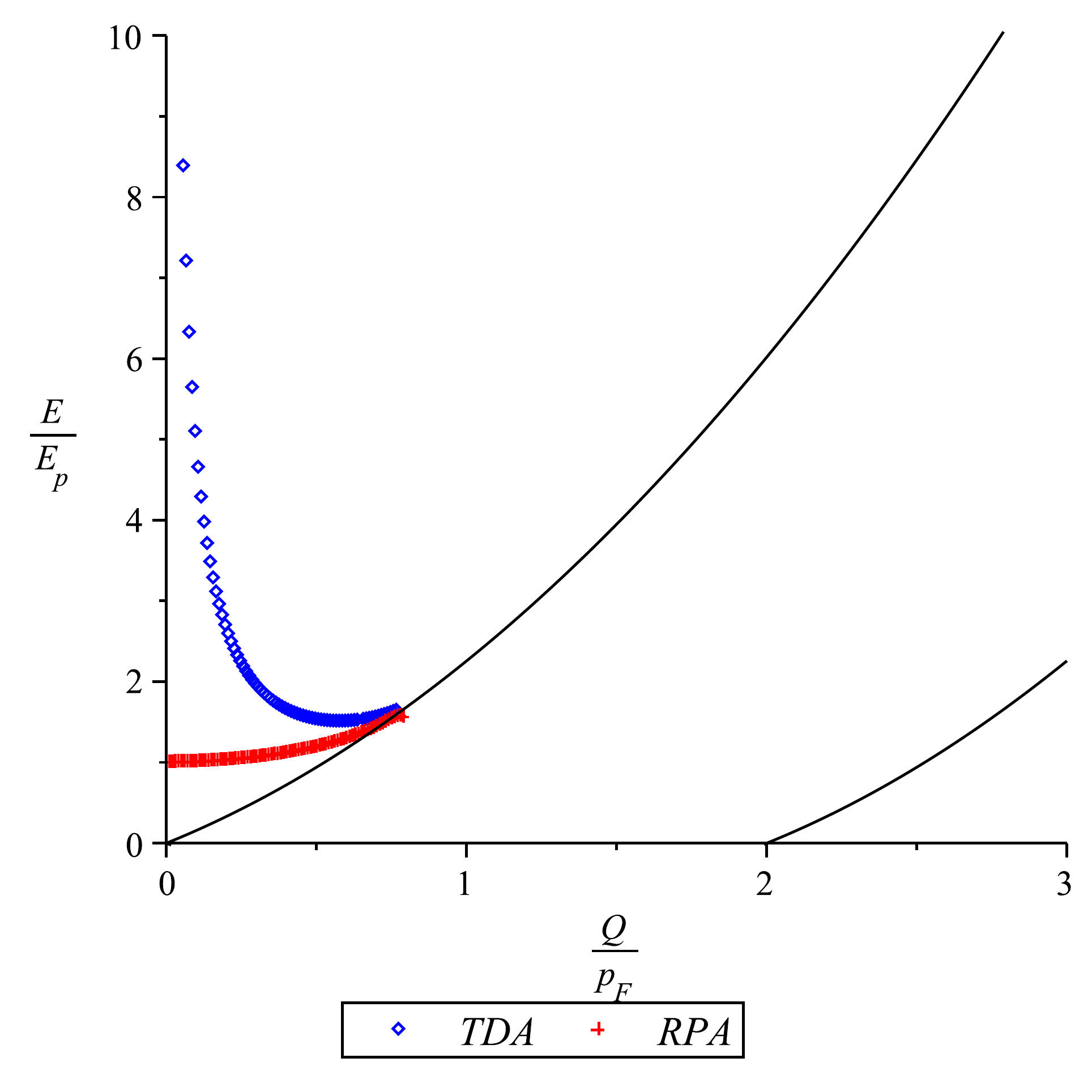}
\caption[Plasmon pole in the TDA and RPA]{
   \label{fig:introduction:TDA}
   This figure shows the low momentum behavior of the polarization propagator for the uniform electron gas for the Wigner-Seitz radius\cite{Dickhoff2005} $r_s=2$. This value indicates that the electron gas is in its metallic regime. The discrete pole energy versus $\frac{Q}{p_F}$ in the TDA is given by the blue diamonds, while the red crosses represent values obtained with the RPA. The true plasmon frequency should be at $\frac{E}{E_p}=1$ as obtained with the RPA. In the area between the two black lines, the polarization propagator has continuous poles. The derivation is given in \App{electrongas}.
}
\end{center}
\end{figure}

A wide array of test systems has been developed to study the behavior of the Coulomb interaction in atoms, molecules and solids. The uniform electron gas consists of a uniform negative charge distribution that is governed by a Coulombic force against a backdrop of positive charge. There has been much interest pioneered by Pines\cite{Pines1971} in describing the response of this system to an external perturbation. The object that holds information about this response is the polarization propagator. Apart from a continuum spectrum of particle-hole (ph) poles, this polarization propagator exhibits a discrete pole in the low momentum region corresponding to the plasmon excitation. When describing the polarization propagator in the Tamm-Dancoff approximation, the pole energy diverges due to the $\frac{1}{Q^2}$ behavior of the Coulombic interaction. This is illustrated in \Fig{introduction:TDA}. The lowest level of theory that describes the plasmon pole correctly is the random phase approximation (RPA)\cite{Bohm1951,Bohm1952,Bohm1953}. The RPA polarization screens the Coulomb interaction in such a way that the classical plasmon energy
\begin{eqnarray}
E_p = \sqrt{\frac{4\pi\rho e^2\hbar^2}{m}}
\end{eqnarray}
is correctly reproduced. Here $\rho$ is the uniform density of the electron gas. This means that for a method to be applicable in this low momentum limit, correlations at least on the RPA level or beyond need to be included. 

The GW approximation (GWA)\cite{Hedin1965,Arya1998,Onida2002} applied in the electron gas theory is one of the methods that benefits from the succesful description of the screening of the interaction in RPA. This method can be seen as a version of Hartree-Fock where the bare interaction is replaced by a dynamically screened interaction.
The success for the electron gas raises the question whether this method can also be applied to smaller systems. These systems are characterized by a stronger electronic relaxation effect and weaker screening than extended systems. Research has been conducted in continuous\cite{Rostgaard2010} and discrete\cite{SanHuang2011} versions of molecular codes. These show an improvement of the ionization energies compared to Hartree-Fock and Density Functional Theory (DFT).
Still, there remain several problems with this procedure.
\begin{enumerate}
\item The inclusion of the anti-symmetric Coulomb matrix elements (Generalized GW (GGW)) seems to deteroriate the results, while this is essential both for extended and finite systems. 
\item The coupling of the single-particle states with two-particle-one-hole and two-hole-one-particle states involves only two propagating electrons. This inherently violates the Pauli exchange with the third electron. 
\item In the GWA, only partial diagonalization in the particle-hole channel is performed. In doing so, only the particle-hole interaction is screened, while in the low density limit this leads to the failure of the GWA. In this limit it is important to take into account the particle-hole and particle-particle (pp) channels on the same footing. 
\item The best results are obtained by doing calculation on the G$_0$W$_0$ level. This means that increasing self-consistency actually makes the results worse, instead of systematically increasing the accuracy.
\end{enumerate}

The electron gas in 2D and 3D has also been extensively studied using quantum Monte Carlo methods. Provided that the sign problem is under control, these accurate methods provide exact results or at least exact constraints on the properties of the electron gas that can be used as guidance for the development of other approximations or as reference values. More information on the continuous formulation of QMC can be found in Ref.~\cite{Ceperley2004} and references therein.

The electronic behavior of solids bears great resemblance to the electron gas problem. Consequently, the same methods are applied to the calculation of the selfenergy and spectral function of solids. The GWA describes the long-range behavior of the system in an edequate manner, but lacks the correct short-range interaction. A natural extension of the GW formalism therefore includes these interactions by using the GW Green's function in T-matrix theory, as such describing multiple scattering between two holes or particles\cite{Springer1998}. This improves the description of the satellite structure of the spectral function, allowing for the occurence of a satellite a few electronvolts below the main peak, which is also found in experiments. This success gives a hint that the simultaneous inclusion of particle-hole and particle-particle scattering is mandatory.

\subsubsection{Nuclear Matter}
Nuclear matter (i.e. a homogeneous gas of nucleons) is a system for which the particle-particle channel has been studied extensively. The interaction between nucleons is very repulsive at short distances and the probability of finding two nucleons close together is consequently small. As a result, a mean-field description starting from the bare nucleon-nucleon potential is not meaningful. However, considering the interaction between particles at a more sophisticated level, one creates correlations that lead to an effective interaction. The dressing of the interaction can be attained within the so-called ladder approximation\cite{Galitskii1958}. In this approximation only graphs with one hole line are retained and an infinite sum of all ladders is performed. This approximation corresponds in fact to the low density limit of nuclear matter.

\subsection{Finite systems}
\subsubsection{Molecular physics}
For closed shell atoms, the spectrum obtained in photoionization processes is fairly simple. There is a main peak for every orbital that carries the larger part of the spectral strength, while the rest of the strength is placed in a number of smaller satellite lines. For these systems, a single-pole approximation describes the ionization physics respectably well. However, in some cases this quasi-particle picture breaks down. The breakdown mostly occurs when the occupied levels are energetically close to the vacant levels. Whereas for atoms this breakdown of the quasi-particle picture is rather exceptional, it is frequently seen in molecules\cite{Cederbaum1977}. This behavior is due to the small gap between occupied and unoccupied orbitals in some molecules and the large number of occupied orbitals in the valence region. As a result, there is mixing of higher excitations with the single-particle states, which implies the breakdown of the quasi-particle picture.

A successful approach to go beyond the quasi-particle picture is by including two-particle-one-hole (2p1h) and two-hole-one-particle (2h1p) configurations into the eigenstates of the anion or kation\cite{Schirmer1978}. This approximation is called the two-particle-one-hole Tamm-Dancoff approximation (2p1h-TDA) and can be seen as a configuration interaction expansion including the 2p1h and 2h1p states into excited states or as a Green's function method with a selfenergy approximation that exhibits the right analytical properties of the exact selfenergy. This method leads to qualitatively correct ionization properties\cite{Cederbaum1980}, but takes the ground-state correlations only partly into account. As a result, outer valence ionization energies are systematically too low. Such behavior is cured by refining to an approximation that is exact up to third-order of perturbation theory. The extended 2p1h-TDA\cite{Walter1981} can be shown to coincide with the third order algebraic diagrammatic construction (ADC(3))\cite{Schirmer1983} technique. From the diagrammatical derivation of this technique, it is clear that, in the selfenergy of ADC(3), interactions between two lines are included up to the Tamm-Dancoff (TDA) level. This method yields very satisfactory results for a wide range of small molecular systems\cite{VonNiessen1984}, but the discussion on the plasmon pole shows that the restriction to TDA excitations will possibly lead to disastrous results when applied to extended systems. 

To deal with the computational cost of an all-electron calculation, the electron ionization and attachment are sometimes treated separately. This leads to the so-called non-Dyson ADC (nD-ADC) approximations\cite{Schirmer1998}. In this technique the ionization and affinity parts are analytically decoupled from the beginning, prohibiting the logarithmic divergence of the static part of the selfenergy as shown in Ref.\cite{Deleuze1995}. The separation of addition and removal parts of the Green's function calculation remedies the violation of particle number that causes this problem. To make the method applicable to large bio-molecules, a more drastic simplification was introduced\cite{Ortiz1998}. Here, the terms that are unimportant for the ionization of closed-shell molecules are discarded. The resulting nondiagonal, renormalized second-order (NR2) theory presents a significantly better scaling with particle number than ADC(3).

\subsubsection{Nuclei and atoms}
Based on the successes of the techniques using RPA-like interactions and the indications that coupling to higher excited states have to be included in the theory, there have been efforts to unify these qualities in one and the same technique. This unification proved to be an extremely challenging problem to solve. Extensions of the second-order diagram of the selfenergy\cite{Rijsdijk1992} stay limited to either the particle-hole or particle-particle channel. Seeing that both these channels are important in some cases\cite{Arita1971}, the inclusion of both channels on the same footing was investigated\cite{Schuck1973}. This resulted in a matrix eigenvalue equation that is very similar to the simple RPA equations. These equations were formulated in terms of the solutions of the ph and pp RPA. On close inspection, however, there are some severe problems with this approach\cite{Rijsdijk1996}. It was shown that the technique used to derive the 2p1h RPA leads to the inclusion of unphysical solutions. Nevertheless, the results of the calculations were encouraging and particularly the proton spectral strength at low missing energies for $^{48}$Ca nuclei was closer to the experimental value than the 2p1h-TDA result. This was thanks to the extra ground-state correlations that induce an extra energy-dependent diagram in the selfenergy that is absent in 2p1h-TDA. The extra diagram results in a stronger coupling of the phonons for states where both the ph and pp channels are important.

In a recent work\cite{Barbieri2001} a formalism has been introduced that includes the ph and pp channel at the level of the RPA. The Faddeev random phase approximation (FRPA) starts from an irreducible 2p1h/2h1p propagator for the selfenergy, so the Pauli principle is fully accounted for. The breakthrough that enables the coupling of ph and pp channels on an equal footing is the use of a Faddeev technique\cite{Faddeev1960} to resum the diagrams to all orders. The Faddeev route is familiar when solving the three-body problem.  This results in an improved description of the long range correlations that need to be included in the single particle propagator. In this approach, the spurious solutions also appear\cite{Adhikari1979}, but they are easily separated from the physical solutions\cite{Navratil1999} for the three-body problem. In a many-body calculation, the separation is less simple but can still be performed\cite{Barbieri2001}. Another advantage is that this method is easily extended with different interactions. If one wants to go beyond the standard RPA level, this can easily be done by replacing the corresponding intermediary excitations\cite{Barbieri2003}. The FRPA approach has been successfully applied to nuclei\cite{Barbieri2002} and atoms\cite{Barbieri2007}. Fragmentation effects were also investigated for nuclei. The results obtained with this dressed FRPA are in good agreement with the experiments and for some effects it is advisable to use the dressed propagators. Overall, the FRPA is a method that is applicable both for finite and extended electronic systems because of the RPA screening of the interaction.

In this work we will study the applicability of the FRPA for a set of small molecules \cite{Degroote2011,Degroote2010}. In view of the good results obtained in ADC(3) and the superiority of the RPA over the TDA, we expect that this method will result in a better description of the excitation energies and ground-state energies. We will also address the accuracy of the method when the basis set is enlarged\cite{Barbieri2012}. To do this we will extrapolate the results for a series of atoms to the basis set limit. At every time we will compare our results to the experiment where available, and to the current golden standard in molecular and atomic calculations, the CC with singles, doubles and perturbative triples (CCSD(T)).

As an alternative reference method one could also look at the QMC results. It is computationally possible to obtain ground and exited state information about small nuclei up to $^{12}$C\cite{Wiringa2009} and even in atomic and molecular physics\cite{Ceperley1996} QMC methods are becoming more and more established.

\section{Outline}
\label{sec:outline}

In \Cha{methods} we will present all the theoretical tools that are needed to construct the FRPA. The definition of the single-particle Green's function . The Dyson equation will be given in the form that requires the irreducible 2p1h/2h1p propagator. After this the definition of the polarization propagator and two-particle Green's function will be given. These propagators will be used in a suitable approximation called the RPA.

\Cha{frpa} will deal with the derivation of the FRPA mechanism. For this the six-point vertex function has to be reduced from a six-time object to a two-time object. The reduction requires that the propagators are two-time quantities and that the class of diagrams that is resummed is restricted. The Faddeev procedure increases the matrix dimensions by a factor of three. However, by elimination of the spurious solutions, the matrix dimension brought back to the original dimension through a simple projection. The ADC(3) is encompassed by the FRPA as a limit case which can easily be derived.

The results obtained with the previously derived method will be given in \Cha{results}. We have applied the FRPA to a series of closed-shell atoms, a set of simple diatomic molecules and a schematic model. Through this schematic model we try to give an understanding of the problems that occur in the dissociation limit with the FRPA.

In \Cha{conclusion} we will present the conclusions of this work, together with an outlook on possibilities for future research.
							\clearnewpage
\chapter{Methods}
\label{cha:methods}
As the amount of information that needs to be stored for the exact description of the wave function of an $N$-body system explodes exponentially with the number of particles, there has always been a search for other physical quantities that represent the physically relevant part of the wave function. Similar to the electronic density and the reduced density matrix\cite{Verstichel2012}, the single-particle Green's function is one of these quantities that describe the physics of an $N$-body system in a more compact manner. Green's functions form a natural way of describing the evolution of an $N$-body system. Just like the density matrix, they hold information about the ground state and allow for the calculation of the expectation value of any one-body operator, as well as the ground-state energy when interactions are limited to two-body interactions. At the same time, the poles and amplitudes of the Green's function offer access to the $N-1$- and $N+1$-particle systems. For finite systems, all data contained by the single-particle Green's functions can be related to experiments. The addition process can be probed by elastic scattering experiments and the removal process can be witnessed experimentally in (e,2e) experiments.

\section{Description of the system}

The time-independent Schr\"odinger equation can be written as an eigenvalue equation
\begin{equation}
H|\Psi_0^N\rangle=E_0|\Psi^N_0\rangle\label{equ:methods:schroedinger},
\end{equation}
where $H$ is the Hamiltonian of the system and the energy $E_0$ is the eigenvalue of this operator for the ground state $|\Psi_0^N\rangle$ of the $N$-particle system. The nuclear positions change on a much slower timescale than the electrons. One can assume that the fields generated by the ions act as an external potential for the electrons. After separating the nuclear and electronic wave functions using this Born-Oppenheimer approximation, the electronic Hamiltonian can be written as
\begin{equation}
H = \sum_{\alpha\beta} \langle\alpha|T|\beta\rangle a_{\alpha}^{\dagger} a_{\beta} + \frac{1}{4} \sum_{\alpha\beta\gamma\delta}\langle\alpha\beta|V|\gamma\delta\rangle a_{\alpha}^{\dagger}a_{\beta}^{\dagger}a_{\delta}a_{\gamma},
\end{equation}
where $a_{\alpha}^{\dagger}$ creates a state with quantum numbers $\alpha$ and $a_{\alpha}$ destroys such a state. The quantum numbers for diatomic molecules will typically consist of an index labeling the Hartree-Fock orbitals, a spin projection and the projection of the angular momentum on the molecular axis. The matrix elements $\langle\alpha|T|\beta\rangle$ are the matrix elements of the kinetic energy operator and the external potential, while the matrix elements $\langle\alpha\beta|V|\gamma\delta\rangle$ are the expectation values of the anti-symmetrized two-body Coulomb interaction. Unlike in nuclear problems, this Hamiltonian fully describes the non-relativistic system. There are no three-body forces. Since in electronic systems the interaction is completely know, the two-body matrix elements can be written in terms of exact 2-electron integrals
\begin{eqnarray}
\langle\alpha\beta|V|\gamma\delta\rangle &=& \int \mathrm{d}\mathbf{x_1} \int \mathrm{d}\mathbf{x_2} \phi^{\dagger}_{\alpha}\left(\mathbf{x_1}\right)\phi^{\dagger}_{\beta}\left(\mathbf{x_2}\right) \frac{1}{| \mathbf{r_1} - \mathbf{r_2} |} \phi_{\gamma}\left(\mathbf{x_1}\right)\phi_{\delta}\left(\mathbf{x_2}\right) \nonumber\\
&&- \int \mathrm{d}\mathbf{x_1} \int \mathrm{d}\mathbf{x_2} \phi^{\dagger}_{\alpha}\left(\mathbf{x_1}\right)\phi^{\dagger}_{\beta}\left(\mathbf{x_2}\right) \frac{1}{| \mathbf{r_1} - \mathbf{r_2} |} \phi_{\delta}\left(\mathbf{x_1}\right)\phi_{\gamma}\left(\mathbf{x_2}\right).\nonumber\\
&&\label{equ:methods:2int}
\end{eqnarray}
The integration over the variables $\mathbf{x_1}$ and $\mathbf{x_2}$ also implies the summation over the discrete degrees of freedom.

\section{Single particle Green's function}
\label{sec:methods:sp}

To define the single-particle Green's function, we have to switch from the Schr\"odinger picture of quantum mechanics to the Heisenberg picture (\App{pictures}).
In this picture the creation and annihilation operators are time dependent
\begin{eqnarray}
a_{\alpha_H}^{\dagger}(t) &=& \exp\left(\frac{i}{\hbar}H t\right) a_{\alpha}^{\dagger} \exp\left(-\frac{i}{\hbar}H t\right)\label{equ:methods:adagger}\\
a_{\alpha_H}(t) &=& \exp\left(\frac{i}{\hbar}H t\right) a_{\alpha} \exp\left(-\frac{i}{\hbar}H t\right)\label{equ:methods:a}.
\end{eqnarray}
The single-particle Green's function is then defined as
\begin{equation}
G_{\alpha\beta}\left(t,t'\right) = - \frac{i}{\hbar} \langle\Psi_0^N|\mathcal{T}[a_{\alpha_H}(t)a_{\beta_H}^{\dagger}(t')]|\Psi_0^N\rangle\label{equ:methods:greensfunctiontime},
\end{equation}
where $|\Psi_0^N\rangle$ is the exact ground-state wave function from \Equ{methods:schroedinger}. The operator $\mathcal{T}$ is the time-ordering operator, which introduces a minus sign when $t'>t$
\begin{equation}
\mathcal{T}[a_{\alpha_H}(t)a_{\beta_H}^{\dagger}(t')] = \theta(t-t')a_{\alpha_H}(t)a_{\beta_H}^{\dagger}(t') - \theta(t'-t)a_{\beta_H}^{\dagger}(t')a_{\alpha_H}(t).
\end{equation}
The Hamiltonian operators appearing in the exponent of \Equ{methods:adagger} and \Equ{methods:a} can now be resolved by introducing a complete set of eigenstates for the $N+1$ and $N-1$ systems. The Green's function has the form
\begin{eqnarray}
G_{\alpha\beta}\left(t,t'\right) &=& -\frac{i}{\hbar}\left\{\theta(t-t')\sum_m \exp\left(\frac{i}{\hbar}\left(E_0^N-E_m^{N+1}\right)\left(t-t'\right)\right)\right.\nonumber\\
&& \times \left.\langle \Psi_0^N |a_{\alpha}|\Psi_m^{N+1}\rangle \langle\Psi_m^{N+1}| a^{\dagger}_{\beta}|\Psi_0^N\rangle \right.\nonumber\\
&&\left. -\theta(t'-t)\sum_n \exp\left(\frac{i}{\hbar}\left(E_0^N - E_n^{N-1}\right)\left(t'-t\right)\right)\right.\nonumber\\
&& \left.\times\langle\Psi_0^N|a_{\beta}^{\dagger}|\Psi_n^{N-1}\rangle \langle \Psi_n^{N-1}| a_{\alpha} |\Psi_0^N\rangle \right\}\label{equ:methods:greentime}\\
&=& G_{\alpha\beta}\left(t-t'\right),
\end{eqnarray}
which only depends on the time difference $t-t'$. Note that here we only include discrete eigenstates. Taking into account continuum eigenstates would result in the replacement of the discrete sum with an integral. The complete set of eigenstates for the $N+1$ and $N-1$ systems have eigenenergies analogous to \Equ{methods:schroedinger}:
\begin{eqnarray}
H|\Psi_m^{N+1}\rangle &=& E_m^{N+1}|\Psi_m^{N+1}\rangle\\
H|\Psi_n^{N-1}\rangle &=& E_n^{N-1}|\Psi_n^{N-1}\rangle.
\end{eqnarray}
The first term of \Equ{methods:greentime} is the forward propagating or "particle" part; it describes the evolution of an additional particle on top of the $N$-particle system with quantum numbers $\beta$ created at time $t'$ to a state $\alpha$ which is destroyed at time $t>t'$. The second term is the backward propagating or "hole" part which describes the probability of the removal of a particle in state $\alpha$ from the system at time $t$ and the return to the ground state at time $t'>t$ by adding a particle in state $\beta$. By inserting these complete sets, it becomes clear that the Green's function holds information about the $N+1$- and $N-1$-particle system. This becomes even more obvious when we make the Fourier transform to the energy domain where we recover the Lehmann representation of the Green's function
\begin{eqnarray}
G_{\alpha\beta}\left(E\right) &=& \sum_m \frac{\langle \Psi_0^N | a_{\alpha} | \Psi_m^{N+1}\rangle \langle \Psi_m^{N+1}| a_{\beta}^{\dagger} | \Psi_0^N\rangle}{E-(E_m^{N+1}-E_0^M) + i\eta}\nonumber\\
&&+ \sum_{n} \frac{\langle \Psi_0^N| a^{\dagger}_{\beta} |\Psi_n^{N-1}\rangle \langle \Psi_n^{N-1} | a_{\alpha} | \Psi_0^N\rangle}{E-(E_0^N-E_n^{N-1}) -i\eta}\label{equ:methods:lehmann}\\
&=& G_{\alpha\beta}^>(E) + G_{\alpha\beta}^<(E)\label{equ:methods:forback},
\end{eqnarray}
in which $\eta$ is a small positive number that ensures convergence of the Fourier transform. The forward and backward part are labeled in \Equ{methods:forback}. The quantity that is probed in (e,2e) experiments is the hole spectral function which is related to the imaginary part of the backward propagating part of the Green's function by using the Sokhotski-Plemelj relation
\begin{eqnarray}
S_{\alpha}^<(E) &=& \frac{1}{\pi}\Im G^<_{\alpha\alpha}(E)\\
&=& \sum_n \left|\langle \Psi_n^{N-1}|a_{\alpha}|\Psi_0^N\rangle\right|^2 \delta\left(E-(E_0^N-E_n^{N-1})\right).
\end{eqnarray}
The hole spectral function is a sequence of $\delta$-functions located at the poles of the Green's function, which coincide with the transition energies to the $N-1$-system. At this energy, the spectral function measures the probability of removing a particle with quantum numbers $\alpha$ from the ground state while leaving the residual system in a state of energy $E^N_0 - E$. The deviation from the non-interacting single-particle picture is best captured by the spectroscopic factor of an excited state $n$
\begin{eqnarray}
Z_n &=& \sum_{\alpha} \left|\langle \Psi_n^{N-1}|a_{\alpha}|\Psi_0^N\rangle\right|^2.
\end{eqnarray}
This spectroscopic factor expresses the total probability of removing a particle from the system and arriving at the excited state $n$ of the $N-1$-body problem.

The expectation value of any one-body operator can be expressed in function of the Green's function. This can be seen from the connection between the density matrix and the Green's function. The ground-state expectation value of a single-body operator is obtained as
\begin{eqnarray}
\langle \Psi_0^N| O|\Psi_0^N\rangle &=& \sum_{\alpha\beta} \langle \alpha|O|\beta\rangle \langle \Psi_0^N| a^{\dagger}_{\alpha}a_{\beta} | \Psi_0^N \rangle\\
&=& \sum_{\alpha\beta} \langle \alpha | O | \beta\rangle N_{\alpha\beta},
\end{eqnarray}
where the density matrix was introduced as
\begin{eqnarray}
N_{\alpha\beta} = \langle \Psi_0^N| a^{\dagger}_{\alpha}a_{\beta} | \Psi_0^N \rangle.
\end{eqnarray}
By inserting a complete set, the structure of the numerator of the Lehmann representation of the Green's function appears. The single-particle density matrix is related to the Green's function by
\begin{eqnarray}
N_{\alpha\beta} &=& \sum_n \langle\Psi_0^N| a^{\dagger}_{\alpha} |\Psi_n^{N-1}\rangle\langle\Psi_n^{N-1}|a_{\beta} | \Psi_0^N \rangle\\
&=& \int\frac{\mathrm{d}E}{2\pi i} \exp\left(i\eta E\right) \sum_{n} \frac{\langle \Psi_0^N| a^{\dagger}_{\alpha} |\Psi_n^{N-1}\rangle \langle \Psi_n^{N-1} | a_{\beta} | \Psi_0^N\rangle}{E-(E_0^N-E_n^{N-1}) -i\eta}\\
&=& \int\frac{\mathrm{d}E}{2\pi i} \exp\left(i\eta E\right) G_{\beta\alpha}(E),
\end{eqnarray}
where the factor $\exp\left(i\eta E\right)$ is inserted to select the backward going part of the propagator when integrating over the complex poles.

One would expect that the determination of the ground-state energy is beyond the scope of the single-particle Green's function, as the expectation value of the Hamiltonian involves the evaluation of both one- and two-body operators with respect to the ground-state wave function. The ground-state energy can be retrieved, however, with the Migdal-Galitskii-Koltun sum rule
\begin{eqnarray}
E_0^N &=& \langle \Psi_0^N | H | \Psi_0^N \rangle\\
&=& \frac{1}{2\pi}\int_{-\infty}^{\epsilon_F^-} \mathrm{d}E \sum_{\alpha\beta}\left(\langle \alpha| T | \beta\rangle + E\delta_{\alpha\beta}\right)\Im G^{<}_{\beta\alpha}(E),
\end{eqnarray}
which greatly extends the applicability of the Green's function theory. Here the integral over the energy goes from $-\infty$ to the Fermi-level $\epsilon_F^-$ of the system which is defined as
\begin{eqnarray}
\epsilon_F^{-} &=& E_0^N-E_0^{N-1}\\
\epsilon_F^{+} &=& E_0^{N+1} - E_0^N.
\end{eqnarray}
The Fermi-level also serves as a separation between particle (p) and hole (h) states. The particle states are those with energy larger than or equal to $\epsilon_F^+$, whereas the hole states have energy below or equal to $\epsilon_F^-$.

It is instructive to look at the Green's function for the system without correlations. This is the solution of that part of the Hamiltonian $H_0$ that can be diagonalized in a single-particle basis
\begin{equation}
H = H_0 + H_1\label{equ:methods:hamiltonian}.
\end{equation}
There is a certain degree of freedom for the choice of the operators $H_0$ and $H_1$. One can for instance include the average two-body interaction (mean field) into the one-body operator by defining an auxilary potential
\begin{eqnarray}
H_0 = T + U
\end{eqnarray}
and
\begin{eqnarray}
H_1 = V - U.
\end{eqnarray}
The splitting should be done in a way that the remaining part of the two-body interaction $H_1$ is small in some sense, and perturbation theory can be more easily applied starting from the eigenstates of $H_0$. The operator $H_0$ can be written in its eigenbasis as
\begin{eqnarray}
H_0 = \sum_{\alpha} \epsilon_{\alpha} a_{\alpha}^{\dagger} a_{\alpha},
\end{eqnarray}
where the $\epsilon_{\alpha}$ are the eigenvalues so that
\begin{eqnarray}
H_0 |\alpha\rangle = \epsilon_{\alpha}|\alpha\rangle.
\end{eqnarray}
The N-particle ground state of this Hamiltonian is simply the product state of the $N$ lowest states $\alpha_i$
\begin{eqnarray}
|\Phi_0^N\rangle = \prod_{i=1}^N a^{\dagger}_{\alpha_i}|0\rangle.
\end{eqnarray}
Inserting this ground-state wave function and Hamiltonian into the definition of the Green's function (\Equ{methods:lehmann}), one arrives at the non-interacting Green's function $G^{(0)}$
\begin{eqnarray}
G^{(0)}_{\alpha\beta}(E) &=& \delta_{\alpha\beta}\left(\frac{\theta\left(\epsilon_{\alpha}-\epsilon_F^+\right)}{E-\epsilon_{\alpha}+i\eta}  + \frac{\theta\left(\epsilon_F^- - \epsilon_{\alpha}\right)}{E-\epsilon_{\alpha}-i\eta} \right)\\
&=& G^{(0)>}_{\alpha\beta}(E) + G^{(0)<}_{\alpha\beta}(E).
\end{eqnarray}
$G_{\alpha\beta}^{(0)}$ is diagonal in the eigenbasis of $H_0$ and has a single pole at energy $\epsilon_{\alpha}$ with transition probability equal to $1$. This is also observed in the hole spectral function. Large deviations from this kind of behavior in a Green's function indicate a strong dissimilarity with a non-interacting system and thus important correlation effects.

\section{Polarization propagator and $N\pm 2$ Green's function}
\label{sec:methods:tp}

Similar to the definition of the single-particle Green's function in \Equ{methods:greensfunctiontime}, a propagator for the evolution of two particles can be defined
\begin{eqnarray}
G^{II}_{\alpha\beta,\gamma\delta}\left(t_{\alpha},t_{\beta},t_{\gamma},t_{\delta}\right) &=& -\frac{i}{\hbar} \langle \Psi_0^N | \mathcal{T}\left[a_{\beta_H}(t_{\beta})a_{\alpha_H}(t_{\alpha})a^{\dagger}_{\gamma_H}(t_{\gamma})a^{\dagger}_{\delta_H}(t_{\delta})\right] | \Psi_0^N\rangle.\nonumber\\
&&
\end{eqnarray}
The full $G^{II}$ is not needed for the study of excited states. It suffices to take only the two-time limit of the two-particle propagator, with the indices corresponding to particle-hole (ph) excitations
\begin{eqnarray}
G^{ph}_{\alpha\beta^{-1},\gamma\delta^{-1}}(t-t') &=& \lim_{t_{\beta}\rightarrow t^+}\lim_{t_{\gamma}\rightarrow t'^+} G^{II}_{\alpha\bar{\delta},\bar{\beta}\gamma}(t,t',t_{\beta},t_{\gamma})\\
&=& -\frac{i}{\hbar} \langle \Psi_0^N | \mathcal{T}\left[a^{\dagger}_{\bar{\beta}_H}(t)a_{\alpha_H}(t)a^{\dagger}_{\gamma_H}(t')a_{\bar{\delta}_H}(t')\right] | \Psi_0^N\rangle.\nonumber\\
&&
\end{eqnarray}
The states $\bar{\alpha}$ are related to the states $\alpha$ by time-reversal. For an electron in a spatial orbital $a$ and with spin projection $m_s$, this means
\begin{eqnarray}
|\overline{a,m_s}\rangle &=& (-1)^{\frac{1}{2}+m_s}|a,-m_s\rangle\label{equ:methods:timereversal}.
\end{eqnarray}
The phase factor in this relation is not unambiguously defined, but it is fixed by this definition. Applying the time-reversal relation twice returns the quantum numbers to their normal values but with a minus sign
\begin{eqnarray}
|\overline{\overline{a,m_s}}\rangle = - |a,m_s\rangle.
\end{eqnarray}
By inserting a complete set of states in analogy to \Equ{methods:greentime}, the ph Green's function is written as
\begin{eqnarray}
G^{ph}_{\alpha\beta^{-1},\gamma\delta^{-1}}(t-t') &=& -\frac{i}{\hbar}\langle \Psi_0^N | a^{\dagger}_{\bar{\beta}}a_{\alpha}| \Psi_0^N\rangle \langle \Psi_0^N | a^{\dagger}_{\gamma}a_{\bar{\delta}}| \Psi_0^N\rangle\nonumber\\
&&-\frac{i}{\hbar}\sum_{n\neq0}\theta(t-t')\exp\left(\frac{i}{\hbar}(E_0^N-E_n^N)(t-t')\right)\nonumber\\
&& \times\langle \Psi_0^N | a^{\dagger}_{\bar{\beta}}a_{\alpha}| \Psi_n^N\rangle \langle \Psi_n^N | a^{\dagger}_{\gamma}a_{\bar{\delta}}| \Psi_0^N\rangle\nonumber\\
&&-\frac{i}{\hbar}\sum_{n\neq0}\theta(t'-t)\exp\left(\frac{i}{\hbar}(E_0^N-E_n^N)(t'-t)\right) \nonumber\\
&&\times\langle \Psi_0^N | a^{\dagger}_{\gamma}a_{\bar{\delta}}| \Psi_n^N\rangle \langle \Psi_n^N | a^{\dagger}_{\bar{\beta}}a_{\alpha}| \Psi_0^N\rangle.
\end{eqnarray}
The first term in this propagator still represents a contribution from the ground state. It is customary to study the polarization propagator which includes only terms that are relevant to the study of excited states and that are not already included in the single-particle propagator
\begin{eqnarray}
\Pi_{\alpha\beta^{-1},\gamma\delta^{-1}}(t-t') &=& G^{ph}_{\alpha\beta^{-1},\gamma\delta^{-1}}(t-t') \nonumber\\
&&+ \frac{i}{\hbar}\langle \Psi_0^N | a^{\dagger}_{\bar{\beta}}a_{\alpha}| \Psi_0^N\rangle \langle \Psi_0^N | a^{\dagger}_{\gamma}a_{\bar{\delta}}| \Psi_0^N\rangle.
\end{eqnarray}
The non-interacting polarization propagator is again obtained by replacing the full Hamiltonian $H$ with $H_0$ and the exact ground state $|\Psi_0^N\rangle$ with the non-interacting one $|\Phi_0^N\rangle$. In the eigenbasis of $H_0$ this results in
\begin{eqnarray}
\Pi^{(0)}_{\alpha\beta^{-1},\gamma\delta^{-1}}(t-t') &=& -\frac{i}{\hbar}\theta(t-t')\theta(\epsilon_{\alpha}-\epsilon_F^+)\theta(\epsilon_F^--\epsilon_{\beta})\delta_{\alpha\gamma}\delta_{\beta\delta}\nonumber\\
&& \times \exp\left(\frac{i}{\hbar}(\epsilon_{\alpha}-\epsilon_{\beta})(t-t')\right)\nonumber\\
&& -\frac{i}{\hbar}\theta(t'-t)\theta(\epsilon_F^--\epsilon_{\alpha})\theta(\epsilon_{\beta}-\epsilon_F^+)\delta_{\alpha\gamma}\delta_{\beta\delta}\nonumber\\
&& \times\exp\left(-\frac{i}{\hbar}(\epsilon_{\beta}-\epsilon_{\alpha})(t'-t)\right),
\end{eqnarray}
where time-reversed states have been assumed to have the same energy as normal states so that the time-reversed notation can be suppressed in the right-hand side. The first term describes the simultaneous propagation of a particle $\alpha$ and a hole $\beta$ at time $t'$ to a particle $\gamma$ and a hole $\delta$ at time $t$. It is clear that the content of the non-interacting polarization propagator can be expressed in terms of single-particle propagators of a particle $\alpha$, making the evolution to a particle $\gamma$ and independently the propagation of the hole $\beta$ to a hole $\delta$. This becomes clear from the formula
\begin{eqnarray}
\Pi^{(0)}_{\alpha\beta^{-1},\gamma\delta^{-1}}(t-t') &=& -i\hbar G^{(0)}_{\alpha\gamma}(t-t')G^{(0)}_{\bar{\delta}\bar{\beta}}(t'-t).
\end{eqnarray}
The second term interchanges the time arguments and the roles of the particle and hole states. The product of step functions cancels when the time arguments have opposite sign. Since the polarization propagator is again only a function of a time difference, one can apply a Fourier transformation to get the Lehmann-representation as a function of only one energy
\begin{eqnarray}
\Pi_{\alpha\beta^{-1},\gamma\delta^{-1}}(E) &=& \sum_{n\neq 0} \frac{\langle\Psi_0^N|a_{\bar{\beta}}^{\dagger}a_{\alpha}|\Psi_n^N\rangle\langle\Psi_n^N|a^{\dagger}_{\gamma}a_{\bar{\delta}}|\Psi_0^N\rangle}{E+(E_0^N-E_n^N)+i\eta} \nonumber\\
&&- \sum_{n\neq 0}\frac{\langle\Psi_0^N|a^{\dagger}_{\gamma}a_{\bar{\delta}}|\Psi_n^N\rangle\langle\Psi_n^N|a_{\bar{\beta}}^{\dagger}a_{\alpha}|\Psi_0^N\rangle}{E+(E_n^N-E_0^N)-i\eta}\\
&=&\Pi_{\alpha\beta^{-1},\gamma\delta^{-1}}^>(E) + \Pi_{\alpha\beta^{-1},\gamma\delta^{-1}}^<(E) . 
\end{eqnarray}
The poles of this function, just like with the single-particle propagator, hold information about the excited states of the $N$-particle system, while the numerator is made up of transition amplitudes. The non-interacting polarization propagator in the energy domain is given by
\begin{eqnarray}
\Pi^{(0)}_{\alpha\beta^{-1},\gamma\delta^{-1}}(E) &=& \delta_{\alpha\gamma}\delta_{\beta\delta}\left(\frac{\theta(\epsilon_{\alpha}-\epsilon_F^+)\theta(\epsilon_F^--\epsilon_{\beta})}{E-(\epsilon_{\alpha}-\epsilon_{\beta})+i\eta}\right.\nonumber\\
&& \left.- \frac{\theta(\epsilon_F^--\epsilon_{\alpha})\theta(\epsilon_{\beta}-\epsilon_F^+)}{E+(\epsilon_{\beta}-\epsilon_{\alpha})-i\eta}\right)\\
&=& \Pi^{(0)}_{\alpha\beta^{-1},\alpha\beta^{-1}}(E),
\end{eqnarray}
which consists of a forward and backward propagating term that describe the non-interacting ph- and hp-propagation.

In the same spirit as $G^{ph}$ we can derive the propagator that describes the system with a pair of particles removed or added. Again it is not needed to study the full two-particle Green's function; the object that holds the interesting information is the two-time limit $G^{pp}$
\begin{eqnarray}
G^{pp}_{\alpha\beta,\gamma\delta}(t-t') &=& \lim_{t_{\beta}\rightarrow t^+} \lim_{t_{\delta}\rightarrow t'^-} G^{II}_{\alpha\beta,\gamma\delta}(t,t_{\beta},t',t_{\delta})\\
&=& -\frac{i}{\hbar}\langle\Psi_0^N|\mathcal{T}\left[a_{\beta_H}(t)a_{\alpha_H}(t)a^{\dagger}_{\gamma_H}(t')a^{\dagger}_{\delta_H}(t')\right]|\Psi_0^N\rangle.
\end{eqnarray}
It becomes clear from the Lehmann representation that we are in fact dealing with an object that holds information about the $N\pm2$-particle system
\begin{eqnarray}
G^{pp}_{\alpha\beta,\gamma\delta}(E) &=& \sum_m \frac{\langle\Psi_0^N|a_{\beta}a_{\alpha}|\Psi_m^{N+2}\rangle\langle\Psi_m^{N+2}|a^{\dagger}_{\gamma}a^{\dagger}_{\delta}|\Psi_0^N\rangle}{E-(E_m^{N+2}-E_0^N)+i\eta}\\
&& - \sum_n \frac{\langle\Psi_0^N|a^{\dagger}_{\gamma}a^{\dagger}_{\delta}|\Psi_n^{N-2}\rangle\langle\Psi_n^{N-2}|a_{\beta}a_{\alpha}|\Psi_0^N\rangle}{E-(E_0^N-E_n^{N-2})-i\eta}.
\end{eqnarray}
The non-interacting version of this Green's function is retrieved in the customary manner. It can again be expressed as a product of non-interacting single-particle Green's functions
\begin{eqnarray}
G^{pp(0)}_{\alpha\beta,\gamma\delta}(t-t') = i\hbar\left(G^{(0)}_{\alpha\gamma}(t-t')G^{(0)}_{\beta\delta}(t-t') - G^{(0)}_{\alpha\delta}(t-t')G^{(0)}_{\beta\gamma}(t-t')\right).
\end{eqnarray}
The first term describes the non-interacting propagation of state $\gamma$ to $\alpha$ and $\delta$ to $\beta$ while the second term describes the exchange equivalent of this, where the two initial or final states are exchanged. The propagator in the energy domain shows the possible propagation of either two particle states or two hole states
\begin{eqnarray}
G^{pp(0)}_{\alpha\beta,\gamma\delta}(E)&=& \left(\delta_{\alpha\gamma}\delta_{\beta\delta}-\delta_{\alpha\delta}\delta_{\beta\gamma}\right)\left(\frac{\theta(\epsilon_{\alpha}-\epsilon_F^+)\theta(\epsilon_{\beta}-\epsilon_F^+)}{E-(\epsilon_{\alpha}+\epsilon_{\beta})+i\eta} \right.\nonumber\\
&&\left.- \frac{\theta(\epsilon_F^- - \epsilon_{\alpha})\theta(\epsilon_F^- - \epsilon_{\beta})}{E-(\epsilon_{\alpha}+\epsilon_{\beta})-i\eta}\right).
\end{eqnarray}

\section{Equation of Motion}
Elegant as they are, exact Green's functions are hard to obtain for realistic systems. One has to introduce approximations that describe the correlations in the system to keep the procedure computationally tractable. These approximations have to be proposed in a way that can easily be expanded to higher orders. One such approach is the equation of motion approach for the Green's function. It starts from a Hamiltonian written in the form of \Equ{methods:hamiltonian}. It is possible to write the time derivative of the single-particle propagator as
\begin{eqnarray}
i\hbar\frac{\partial}{\partial t} G_{\alpha\beta}(t-t') &=& \frac{\partial}{\partial t} \langle \Psi_0^N|\mathcal{T}\left[a_{\alpha_H}(t)a^{\dagger}_{\beta_H}(t')\right]|\Psi_0^N\rangle\\
&=& \langle \Psi_0^N|\frac{\partial}{\partial t}\left[\theta(t-t')a_{\alpha_H}(t)a^{\dagger}_{\beta_H}(t')\right.\nonumber\\
&& \left.- \theta(t'-t)a^{\dagger}_{\beta_H}(t')a_{\alpha_H}(t)\right]|\Psi_0^N\rangle\\
&=& \delta(t-t')\langle \Psi_0^N|a_{\alpha_H}(t)a^{\dagger}_{\beta_H}(t')+a^{\dagger}_{\beta_H}(t')a_{\alpha_H}(t)|\Psi_0^N\rangle\nonumber\\
&& + \langle \Psi_0^N|\mathcal{T}\left[\frac{\partial}{\partial t}a_{\alpha_H}(t)a^{\dagger}_{\beta_H}(t')\right]|\Psi_0^N\rangle\\
&=& \delta(t-t')\delta_{\alpha\beta} + \langle \Psi_0^N|\mathcal{T}\left[\frac{\partial}{\partial t}a_{\alpha_H}(t)a^{\dagger}_{\beta_H}(t')\right]|\Psi_0^N\rangle,\nonumber\\
&&\label{equ:methods:eom}
\end{eqnarray}
where the fundamental anti-commutation relation for the fermionic operators was used. It suffices to know the time evolution of the annihilation operator $a_{\alpha_H}(t)$ to determine the equation of motion for the single-particle Green's function. In the Heisenberg picture we find
\begin{eqnarray}
i\hbar \frac{\partial}{\partial t} a_{\alpha_H}(t) &=& \left[a_{\alpha_H}(t),H\right]\\
&=& \epsilon_{\alpha}a_{\alpha_H}(t) - \sum_{\delta} \langle\alpha|U|\delta\rangle a_{\delta_H}(t)\nonumber\\
&&+ \frac{1}{2}\sum_{\delta,\gamma,\zeta}\langle\alpha\delta|V|\gamma\zeta\rangle a^{\dagger}_{\delta_H}(t)a_{\zeta_H}(t)a_{\gamma_H}(t).
\end{eqnarray}
Substituting this in \Equ{methods:eom} we get
\begin{eqnarray}
\left[i\hbar\frac{\partial}{\partial t} -\epsilon_{\alpha}\right]G_{\alpha\beta}(t-t') &=& \delta(t-t')\delta_{\alpha\beta} - \sum_{\delta}\langle\alpha|U|\delta\rangle G_{\delta\beta}(t-t')\nonumber\\
&& +\frac{1}{2} \sum_{\delta,\gamma,\zeta}\langle\alpha\delta|V|\gamma\zeta\rangle G^{II}_{\gamma\zeta,\delta\beta}(t,t,t^+,t')\label{equ:methods:eom2}.
\end{eqnarray}
This equation relates the single-particle propagator to the 2-particle propagator. Following this path, it is possible to express every $N$-particle propagator in function of the $N+1$-particle propagator, building a hierarchy of more complicated objects. Writing the equation of motion for the non-interacting single-particle propagator yields an equation similar to \Equ{methods:eom2}, but without the interaction terms
\begin{eqnarray}
\left[i\hbar\frac{\partial}{\partial t} -\epsilon_{\alpha}\right]G^{(0)}_{\alpha\beta}(t-t') &=& \delta(t-t')\delta_{\alpha\beta}\label{equ:methods:inverseg0}.
\end{eqnarray}
This defines the inverse of the non-interacting propagator as
\begin{eqnarray}
G^{(0)-1}_{\alpha\beta}(t-t') &=& \delta(t-t')\delta_{\alpha\beta}\left[i\hbar\frac{\partial}{\partial t} -\epsilon_{\alpha}\right],
\end{eqnarray}
which is exactly the first factor in \Equ{methods:eom2}. Bringing this factor to the right-hand side of the equation leads to
\begin{eqnarray}
G_{\alpha\beta}(t-t') &=& G^{(0)}_{\alpha\beta}(t-t') - \int \mathrm{d}t_1 \sum_{\gamma,\delta}G^{(0)}_{\alpha\gamma}(t-t_1) \langle\gamma|U|\delta\rangle G_{\delta\beta}(t_1-t')\nonumber\\
&& + \frac{1}{2} \int \mathrm{d}t_1 \sum_{\delta,\gamma,\zeta,\iota} G^{(0)}_{\alpha\iota}(t-t_1) \langle\iota\delta|V|\gamma\zeta\rangle \nonumber\\
&& \times G^{II}_{\gamma\zeta,\delta\beta}(t_1,t_1,t_1^+,t')\label{equ:methods:eom3}.
\end{eqnarray}
One can now split off the reducible parts from the two-particle Green's function as a product of exact Green's functions and replace the interaction with the so-called four-point vertex function $\Gamma^{4-pt}$, which serves as an effective interaction between dressed particles in the medium. It has four time arguments, even though the last term in \Equ{methods:eom3} only has two time arguments. The connection to the external times is also handled by interacting single-particle Green's functions, which results in the following expression for $G^{II}$
\begin{eqnarray}
G^{II}_{\alpha\beta,\gamma\delta}(t,t,t^+,t') &=& i\hbar G_{\alpha\gamma}(t-t^+)G_{\beta\delta}(t-t')-i\hbar G_{\alpha\delta}(t-t')G_{\beta\gamma}(t-t^+)\nonumber\\
&&+(i\hbar)^2 \int \mathrm{d}t_{\epsilon} \int \mathrm{d}t_{\zeta} \int \mathrm{d}t_{\eta} \int \mathrm{d}t_{\theta} \sum_{\epsilon,\zeta,\eta,\theta}G_{\alpha\epsilon}(t-t_{\epsilon})\nonumber\\
&&\times G_{\beta\zeta}(t-t_{\zeta})\langle \epsilon\zeta| \Gamma^{4-pt}(t_{\epsilon},t_{\zeta},t_{\eta},t_{\theta})|\eta\theta\rangle\nonumber\\
&&\times G_{\eta\gamma} (t_{\eta}-t) G_{\theta\delta}(t_{\theta}-t').
\end{eqnarray}
Inserting this equation in \Equ{methods:eom3}, one can define the irreducible selfenergy $\Sigma^{\star}$ 
\begin{eqnarray}
\Sigma^{\star}_{\alpha\beta}(t,t') &=& -\delta(t-t')\langle\alpha|U|\beta\rangle - i\hbar\delta(t-t') \sum_{\gamma,\delta} \langle\alpha\gamma|V|\beta\delta\rangle G_{\delta\gamma}(t-t^+)\nonumber\\
&& +(i\hbar)^2\frac{1}{2} \int \mathrm{d}t_{\zeta} \int \mathrm{d}t_{\eta} \int \mathrm{d}t_{\theta} \sum_{\gamma,\delta,\epsilon,\zeta,\eta,\theta} \langle\alpha\gamma|V|\delta\epsilon\rangle G_{\delta\zeta}(t-t_{\zeta})\nonumber\\
&& \times G_{\epsilon\eta}(t-t_{\eta})\langle\zeta\eta|\Gamma^{4-pt}(t_{\zeta},t_{\eta},t_{\theta},t')|\theta\beta\rangle G_{\theta\gamma}(t_{\theta}-t),\label{equ:methods:selfenergy}
\end{eqnarray}
so that the full Green's function is the solution of the Dyson equation
\begin{eqnarray}
G_{\alpha\beta}(t-t') &=& G^{(0)}_{\alpha\beta}(t-t') + \int\mathrm{d}t_1 \int \mathrm{d}t_2 \sum_{\gamma,\delta} G_{\alpha\gamma}^{(0)}(t-t_1) \Sigma^{\star}_{\gamma\delta}(t_1,t_2)\nonumber\\
&&\times G_{\delta\beta}(t_2-t')\label{equ:methods:dysontime}.
\end{eqnarray}
This is a self-consistency problem. The full Green's function appears in the left- and right-hand side of the equation, as well as in the irreducible selfenergy. The correlated Green's function is equal to the non-interacting Green's function and a part that includes all interactions. Making an approximation to the propagator implies finding an approximate selfenergy.

The irreducible selfenergy $\Sigma^{\star}$ acts as a potential for the Green's function. This can be made clear by going to the energy domain, where the convolution time integrals of \Equ{methods:dysontime} turn into simple products in the energy representation:
\begin{eqnarray}
G_{\alpha\beta}(E) &=& G^{(0)}_{\alpha\beta}(E) + \sum_{\gamma,\delta}G^{(0)}_{\alpha\gamma}(E)\Sigma^{\star}_{\gamma\delta}(E)G_{\delta\beta}(E)\label{equ:methods:dysonenergy}.
\end{eqnarray}
Multiplying by a factor $E-\epsilon_n$, with $\epsilon_n$ a pole of $G$ in the removal part of the Lehmann representation, which naturally does not coincide with a pole of $G^{(0)}$ or $\Sigma^{\star}$ and then taking the limit
\begin{eqnarray}
\lim_{E\rightarrow \epsilon_n}(E-\epsilon_n) \left[G_{\alpha\beta}(E) = G_{\alpha\beta}^{(0)}(E) + \sum_{\gamma,\delta}G_{\alpha\gamma}^{(0)}(E)\Sigma^{\star}_{\gamma\delta}(E)G_{\delta\beta}(E)\right]\label{equ:methods:trick}
\end{eqnarray}
and dividing by a factor $\langle\Psi_0^N|a_{\beta}^{\dagger}|\Psi_n^{N-1}\rangle$, one arrives at an eigenvalue equation for the amplitudes of the Green's function
\begin{eqnarray}
\langle \Psi_n^{N-1}|a_{\alpha}|\Psi_0^N\rangle &=& \sum_{\gamma\delta} G^{(0)}_{\alpha\gamma}(\epsilon_n)\Sigma^{\star}_{\gamma\delta}(\epsilon_n)\langle\Psi_n^{N-1}|a_{\delta}|\Psi_0^n\rangle\label{equ:methods:dysonr}.
\end{eqnarray}
This was derived working in the basis that diagonalizes $H_0$, but \Equ{methods:dysonr} is valid in any basis.

Special attention should go to the normalization of the left-hand side of \Equ{methods:dysonr}. One can prove\cite{Dickhoff2005} that the states are normalized like
\begin{eqnarray}
\sum_{\alpha,\beta} \langle\Psi_n^{N-1}|a_{\alpha}|\Psi_0^N\rangle\left(\delta_{\alpha,\beta}-\frac{\partial\Sigma^{\star}_{\alpha\beta}}{\partial E}(\epsilon_n)\right)\langle\Psi_0^N|a_{\beta}^{\dagger}|\Psi_n^{N-1}\rangle & = & 1.
\end{eqnarray}
The square of the norm of these amplitudes constitutes the normalized spectroscopic factors that are commonly determined in experiments employing (e,2e) or (e,e'p) reactions.

Now one can make the transition to the coordinate representation $\alpha=\mathbf{r}m_s$
\begin{eqnarray}
\langle \Psi_n^{N-1}|a_{\mathbf{r}m_s}|\Psi_0^N\rangle &=& \sum_{m_1,m_2} \int \mathrm{d}\mathbf{r_1}\int \mathrm{d}\mathbf{r_2} G^{(0)}_{\mathbf{r}m_s\mathbf{r_1}}(\epsilon_n)\nonumber\\
&&\times \Sigma^{\star}_{\mathbf{r_1}m_1\mathbf{r_2}m_2}(\epsilon_n)\langle\Psi_n^{N-1}|a_{\mathbf{r_2}m_2}|\Psi_0^n\rangle.
\end{eqnarray}
Using the coordinate representation of \Equ{methods:inverseg0}
\begin{eqnarray}
\sum_{m_1}\int \mathrm{d}\mathbf{r_1} \langle \mathbf{r}m_s|\epsilon_n - H_0 |\mathbf{r_1}m_1\rangle G^{(0)}_{\mathbf{r_1}m_1\mathbf{r'}m_s'}(\epsilon_n) &=& \delta_{m_s m_s'}\delta(\mathbf{r}-\mathbf{r'}),\nonumber\\
&&
\end{eqnarray}
and writing the same equation for the amplitudes $\langle \Psi_n^{N-1}|a_{\mathbf{r}m_s}|\Psi_0^N\rangle$
\begin{eqnarray}
\sum_{m_1}\int \mathrm{d}\mathbf{r_1} \langle \mathbf{r}m_s|\epsilon_n - H_0 |\mathbf{r_1}m_1\rangle \langle\Psi_n^{N-1}|a_{\mathbf{r_1}m_1}|\Psi_0^N\rangle &&\nonumber\\
=  \left[\epsilon_n +\frac{\hbar^2}{2m}\nabla^2-U(\mathbf{r})\right]\langle\Psi_n^{N-1}|a_{\mathbf{r}m_s}|\Psi_0^N\rangle &&
\end{eqnarray}
generates a Schr\"odinger-like eigenvalue equation
\begin{eqnarray}
-\frac{\hbar^2}{2m}\nabla^2 \langle \Psi_n^{N-1}|a_{\mathbf{r}m_s}|\Psi_0^N\rangle + \sum_{m_1} \int \mathrm{d}\mathbf{r_1} \Sigma^{\star '}_{\mathbf{r}m_s\mathbf{r_1}m_1}(\epsilon_n)\langle \Psi_n^{N-1}|a_{\mathbf{r_1}m_1}|\Psi_0^N\rangle &&\nonumber\\
= \epsilon_n \langle \Psi_n^{N-1}|a_{\mathbf{r}m_s}|\Psi_0^N\rangle&&
\end{eqnarray}
where $\Sigma^{\star '}$ represents the irreducible selfenergy without the external potential. From this equation it is clear that the irreducible selfenergy acts as an effective potential for a single particle and takes into account the interactions of the particle with the medium. The same equation can be derived for the forward propagating amplitudes, with the understanding that these states with $\epsilon_n>\epsilon_F^+$ correspond to scattering states. 
\section{Diagrammatic language}
\begin{table}
\begin{center}
\begin{tabular}{c l l}
\toprule
Diagram & Time domain & Energy domain \\
\midrule
\includegraphics[scale=0.5,clip=true]{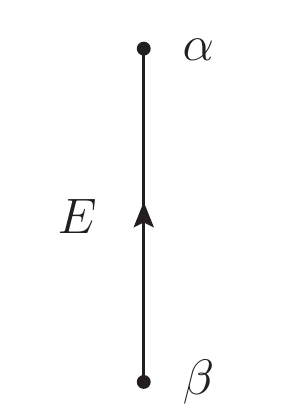} & $G_{\alpha\beta}^{(0)}(t-t')$ & $G_{\alpha\beta}^{(0)}(E)$ \tabularnewline
\includegraphics[scale=0.5,clip=true]{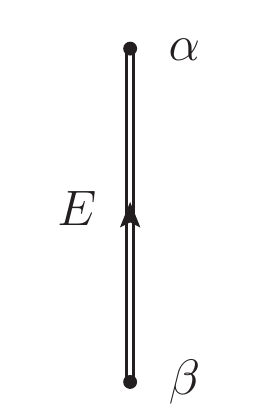} & $G_{\alpha\beta}(t-t')$ & $G_{\alpha\beta}(E)$\\
\includegraphics[scale=0.5,clip=true]{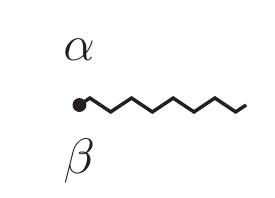} & $\langle\alpha|U|\beta\rangle$ & $\langle\alpha|U|\beta\rangle$ \\
\includegraphics[scale=0.5,clip=true]{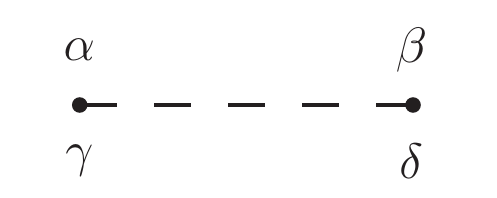} & $\langle\alpha\beta|V|\gamma\delta\rangle$& $\langle\alpha\beta|V|\gamma\delta\rangle$\\
\bottomrule
\end{tabular}
\caption{
\label{tab:methods:time}
Diagrammatic translation of the propagators in the time domain and energy domain
}
\end{center}
\end{table}
One of the great advantages of Green's function techniques is the existence of a graphical language to express complex equations. Every element of the Green's function theory can be represented by a Feynman graph\cite{Mattuck1992,Dickhoff2005}. \Tab{methods:time} presents the propagator and interactions in the time domain that are used in \Sec{methods:sp}. In each of these diagrams, both the forward and backward term are represented in one diagram, i.e. there is no time ordering. The full propagator is presented by two lines, while the non-interacting propagator is represented by a single line. 

The same procedure can be repeated in the energy domain. Every propagator describes the evolution of a state with energy $E$, both forward and backward propagation are captured by the same diagram.

\begin{figure}
\begin{center}
\includegraphics[scale=0.5,clip=true]{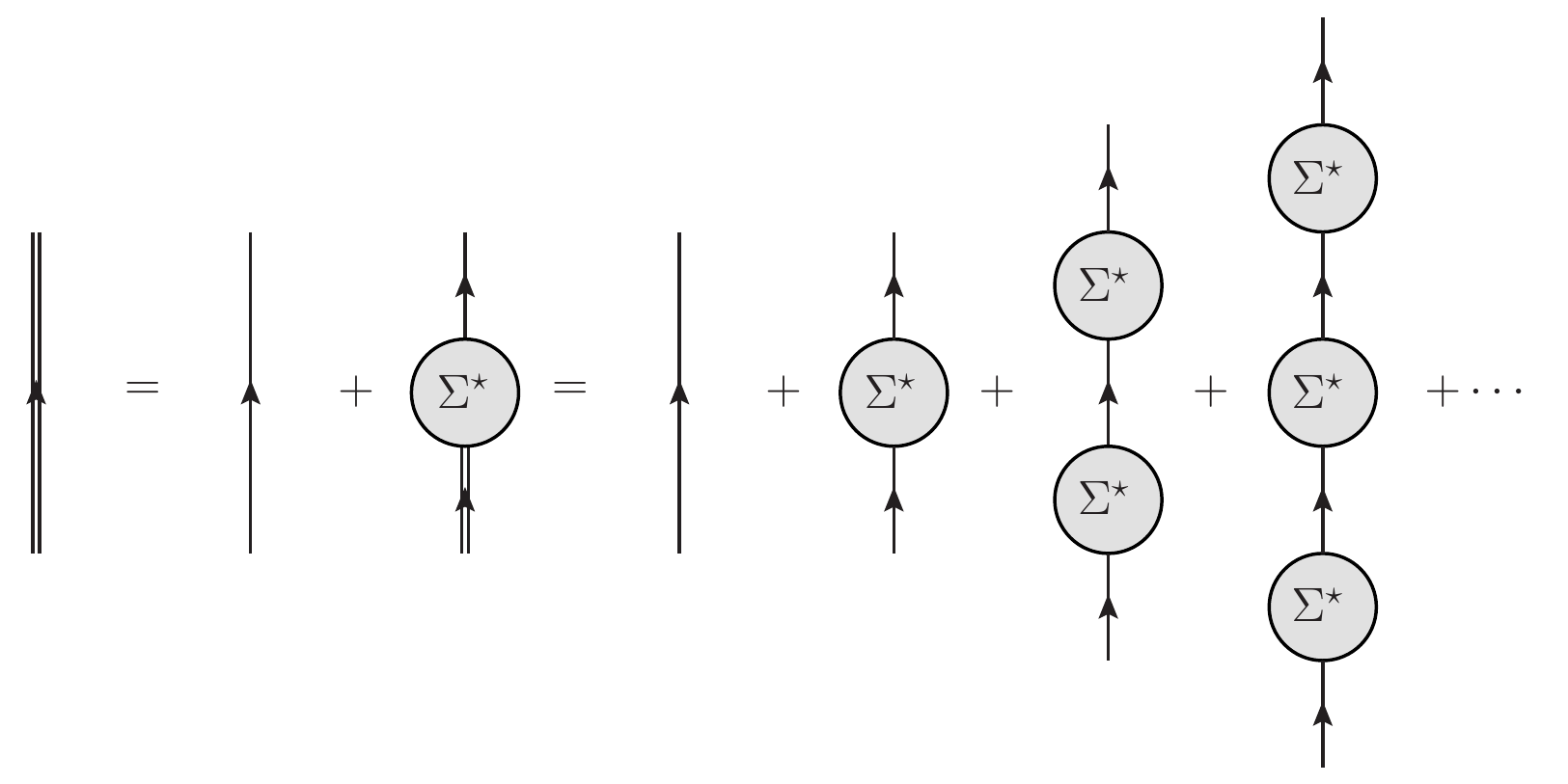}
\caption{
   \label{fig:methods:dyson}
   Diagrammatical representation of the Dyson equation in the energy domain
}
\end{center}
\end{figure}

The Dyson equation \Equ{methods:dysonenergy} is depicted in \Fig{methods:dyson}. From this diagrammatical representation it can easily be seen that the Dyson equation represents an infinite resummation of diagrams: this infinite repetition acts as a renormalization for the interaction. While the Dyson equation is in principle exact, the determination of the complete selfenergy is just as hard as that of the full propagator. In most approximation schemes for the selfenergy, the number of diagrams that is resummed is restricted to a certain class of physically relevant diagrams.

\section{Conserving approximations for the selfenergy}

\begin{figure}
  \centering
  \subfloat[]{\label{fig:methods:selfenergya}\includegraphics[scale=0.5]{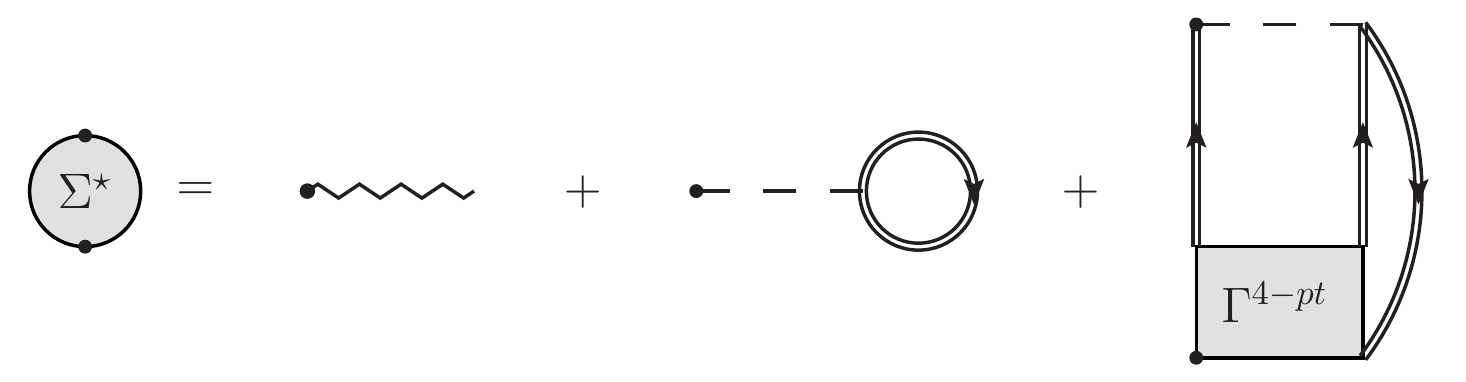}}
  \\
  \subfloat[]{\label{fig:methods:selfenergyb}\includegraphics[scale=0.5]{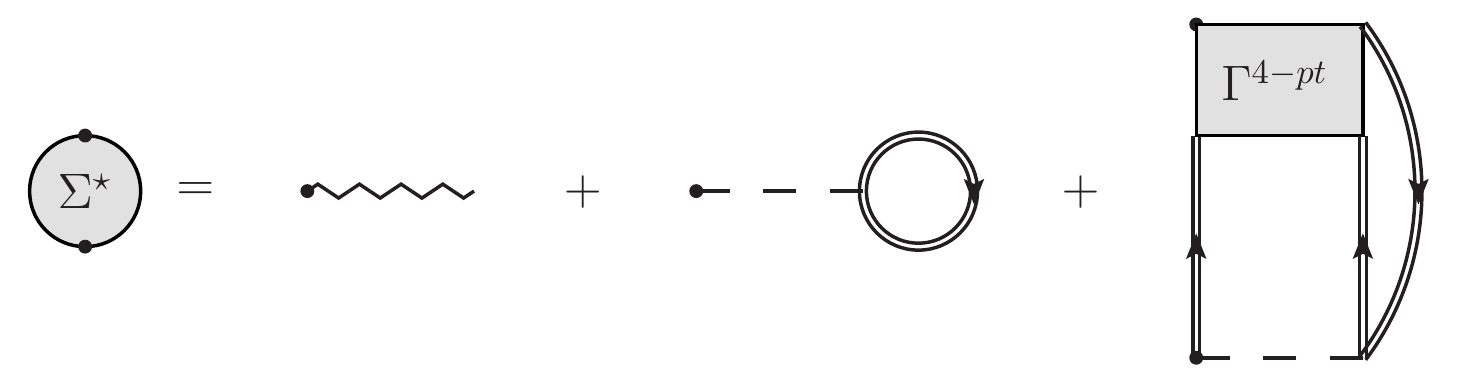}}
  \caption[Selfenergy as a functional of $\Gamma^{4-pt}$]{Two different ways to write the selfenergy $\Sigma^{\star}$ as a functional of the four-point vertex function $\Gamma^{4-pt}$}
  \label{fig:methods:selfenergy}
\end{figure}
\Equ{methods:selfenergy} has a diagrammatical form as displayed in \Fig{methods:selfenergya}. There is, however, another choice possible in the definition of the four-point vertex, displayed in \Fig{methods:selfenergyb}, or as an equation
\begin{eqnarray}
\Sigma^{\star}_{\alpha\beta}(t,t') &=& -\delta(t-t')\langle\alpha|U|\beta\rangle - i\hbar\delta(t-t') \sum_{\gamma\delta} \langle\alpha\gamma|V|\beta\delta\rangle G_{\delta\gamma}(t-t^+)\nonumber\\
&& -(i\hbar)^2\frac{1}{2} \int \mathrm{d}t_{\zeta} \int \mathrm{d}t_{\eta} \int \mathrm{d}t_{\theta} \sum_{\gamma,\delta,\epsilon,\zeta,\eta,\theta} \langle\alpha\zeta|\Gamma^{4-pt}(t,t_{\zeta},t_{\eta},t_{\theta} |\eta\theta\rangle\nonumber\\
&&\times G_{\eta\delta}(t_{\eta}-t')G_{\theta\epsilon}(t_{\theta}-t') \langle\delta\epsilon|V|\beta\gamma\rangle G_{\zeta\gamma}(t-t_{\zeta}).\label{equ:methods:selfenergyb}
\end{eqnarray}
There is no reason why the two different definitions should lead to a different outcome. In fact, it has been shown\cite{Baym1961,Baym1962} that the equivalence of these two diagrams has important physical implications. If these are equivalent and the four-point vertex is symmetric
\begin{eqnarray}
\Gamma^{4-pt}_{\alpha\beta,\gamma\delta} = \Gamma^{4-pt}_{\beta\alpha,\delta\gamma},
\end{eqnarray}
then the particle number, total momentum, total angular momentum and energy are all conserved quantities in this approximaton. This is a very strong statement and approximations that have this property are to be preferred above others.

\begin{figure}
\begin{center}
\includegraphics[scale=0.5,clip=true]{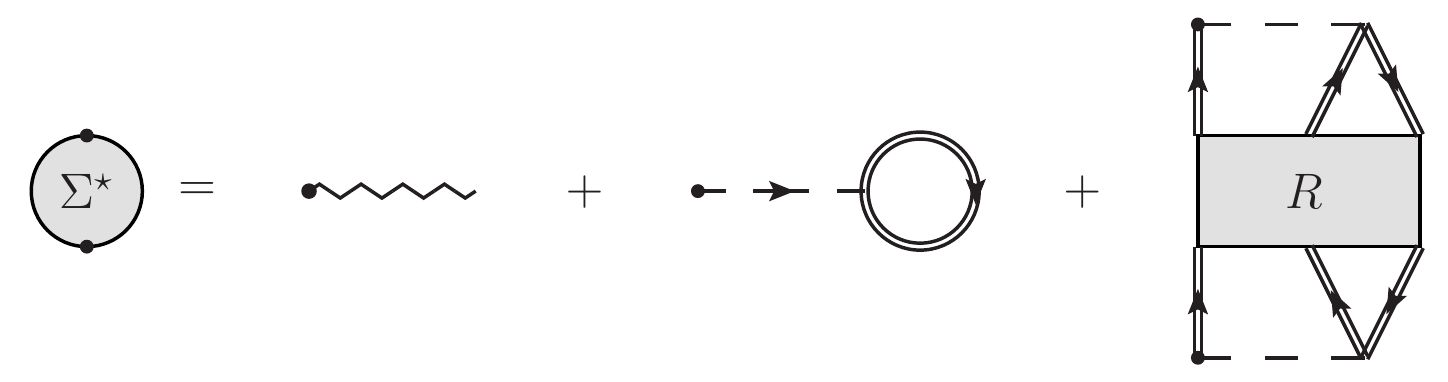}
\caption{
   \label{fig:methods:R}
   Diagrammatical representation of the selfenergy from \Equ{methods:selfR} that fulfills the Baym-Kadanoff requirements
}
\end{center}
\end{figure}

To enforce this symmetry between the two ways to write the selfenergy, it is best to study the equation of motion for $G^{II}_{\alpha\beta,\gamma\delta}(t,t,t^+,t')$, but now with the time derivative with respect to $t'$
\begin{eqnarray}
-i\hbar\frac{\partial}{\partial t'} G^{II}_{\alpha\beta,\gamma\delta}(t,t,t^+,t') &=& \left(\delta_{\alpha\gamma}\delta_{\beta\delta} - \delta_{\alpha\delta}\delta_{\beta\gamma}\right)\delta(t-t')\nonumber\\
&&+ \langle\Psi_0^N|\mathcal{T}\left[a^{\dagger}_{\gamma_H}(t)a_{\beta_H}(t)a_{\alpha_H}(t)\frac{\partial}{\partial t'}a^{\dagger}_{\delta_H}(t')\right]|\Psi_0^N\rangle,\nonumber\\
&&
\end{eqnarray}
using the equation of motion for the creation operator $a_{\delta_H}(t')$, we arrive at
\begin{eqnarray}
\left(-i\hbar\frac{\partial}{\partial t'} -\epsilon_{\delta}\right)G^{II}_{\alpha\beta,\gamma\delta}(t,t,t^+,t') &=& \left(\delta_{\alpha\gamma}\delta_{\beta\delta} - \delta_{\alpha\delta}\delta_{\beta\gamma}\right)\delta(t-t')\nonumber\\
&& - \sum_{\epsilon}G^{II}_{\alpha\beta,\gamma\epsilon}(t,t,t^+,t')\langle\epsilon|U|\delta\rangle\nonumber\\
&&- \frac{1}{2}\frac{i}{\hbar}\sum_{\epsilon\zeta\eta}G^{2p1h}_{\alpha\beta\gamma,\epsilon\zeta\eta}(t-t')\langle\epsilon\zeta|V|\delta\eta\rangle\nonumber\\
&&\label{equ:methods:2p1h},
\end{eqnarray}
which is the definition of a next Green's function in the hierarchy $G^{2p1h}$
\begin{eqnarray}
G^{2p1h}_{\alpha\beta\gamma,\delta\epsilon\zeta}(t-t') &=& -\frac{i}{\hbar}\langle\Psi_0^N|\mathcal{T}\left[a_{\gamma_H}^{\dagger}(t)a_{\beta_H}(t)a_{\alpha_H}(t)a_{\delta_H}^{\dagger}(t')a_{\epsilon_H}^{\dagger}(t')a_{\zeta_H}(t')\right]|\Psi_0^N\rangle.\nonumber\\
&&
\end{eqnarray}
This Green's function describes the simultaneous motion of two particles and one hole in the medium. Just like the single-particle Green's function, it depends on the difference of two times only. The first factor in the right-hand side of \Equ{methods:2p1h} represents a non-interacting propagator $G^{(0)}$. The right-hand side can be rewritten in terms of the second definition of the selfenergy \Equ{methods:selfenergyb}. Using the Dyson equation \Equ{methods:dysontime} this results in the following expression for $G^{II}_{\alpha\beta,\gamma\delta}(t,t,t^+,t')$
\begin{eqnarray}
G^{II}_{\alpha\beta,\gamma\delta}(t,t,t^+,t') &=& \frac{1}{2}\sum_{\epsilon,\zeta,\eta\theta} \int \mathrm{d}t_1 G^{2p1h}_{\alpha\beta\gamma,\epsilon\zeta\eta}(t-t') \langle\epsilon\zeta|V|\theta\eta\rangle G_{\theta\delta}(t_1-t')\nonumber\\
&&-\frac{1}{2}\sum_{\epsilon,\zeta,\eta,\theta,\iota,\kappa}\int \mathrm{d}t_1\int \mathrm{d}t_2 \mathrm{d}t_3 G^{II}_{\alpha\beta,\gamma\epsilon}(t,t,t^+,t_1)\nonumber\\
&& \times G^{-1}_{\epsilon\zeta}(t_1-t_2)G^{II}_{\zeta\eta,\theta\iota}(t_2,t_3^-,t_3,t_3)\langle\eta\theta|V|\kappa\iota\rangle\nonumber\\
&&\times G_{\kappa\delta}(t_3-t')\nonumber\\
&&+\sum_{\epsilon}\left(\delta_{\alpha\gamma}\delta_{\beta\epsilon}-\delta_{\alpha\epsilon}\delta_{\beta\gamma}\right)G_{\epsilon\delta}(t-t').
\end{eqnarray}
Substituting this in \Equ{methods:eom3}, one gets a new definition of the selfenergy in terms of 2p1h excitations\cite{Ethofer1969,Ethofer1969b}, which automatically fulfills the Baym-Kadanoff requirements
\begin{eqnarray}
\Sigma^{\star}_{\alpha\beta}(t-t') &=& -\delta(t-t')\langle\alpha|U|\beta\rangle + \delta(t-t')  \sum_{\gamma\delta} \langle\alpha\gamma|V|\beta\delta\rangle G_{\gamma\delta}(t-t^+)\nonumber\\
&&+\sum_{\gamma,\delta,\epsilon,\zeta\eta\theta} \langle \alpha\gamma|V|\delta\epsilon\rangle R_{\gamma\delta\epsilon,\zeta\eta\theta}(t-t') \langle \zeta\eta|V|\beta\theta\rangle\label{equ:methods:selfR}.
\end{eqnarray}
This is graphically represented in \Fig{methods:R}. The irreducible 2p1h/2h1p propagator $R$ represents the propagation of two-particle-one-hole states and two-hole-one-particle states without intermediary annihilation of a particle or hole state
\begin{eqnarray}
R_{\alpha\beta\gamma,\delta\epsilon\zeta}(t-t') &=& G^{2p1h}_{\alpha\beta\gamma,\delta\epsilon\zeta}(t-t') - \sum_{\eta,\theta} \int \mathrm{d}t_1 \int \mathrm{d}t_2 G^{II}_{\alpha\beta,\gamma,\eta}(t,t,t^+,t_1)\nonumber\\
&& \times G^{-1}_{\eta\theta}(t_1-t_2)G^{II}_{\theta\delta,\epsilon\zeta}(t_2,t'^-,t',t')
\end{eqnarray}

\section{Hartree-Fock}
One can in principle choose any approximation for the irreducible 2p1h/2h1p propagator in \Equ{methods:selfR}, but solving for $G$ becomes too difficult very quickly. The easiest choice is just to set $R$ equal to zero. In that case, only the first two diagrams in \Fig{methods:R} are included in the selfenergy. This approximation is called the Hartree-Fock (HF) approximation. By construction, this method satisfies the Baym-Kadanoff requirements.
\begin{eqnarray}
\Sigma^{HF}_{\alpha\beta}(t-t') &=& -\delta(t-t')\langle\alpha|U|\beta\rangle -i \hbar \delta(t-t')  \sum_{\gamma\delta} \langle\alpha\gamma|V|\beta\delta\rangle G^{HF}_{\delta\gamma}(t-t^+)\nonumber\\
&&
\end{eqnarray}
This equations explain the name self-consistent field equation, the Hartree-Fock quasiparticles move in the field generated by all the other quasiparticles. The selfenergy that is generated in this approximation is an energy-independent one-body operator. The energy independence of the selfenergy implies that this Hartree-Fock prescription is closely related to an independent particle picture. One can construct a basis in which the propagator is diagonal.

The choice of $H_0$ in \Equ{methods:hamiltonian} is fixed only up to an auxilary one-body operator $U$. The choice of $U$ can in principle be completely arbitrary, but it is best to make a choice that already reflects the physics of the many-body system. Since the HF Hamiltonian is a one-body operator only, it can be chosen as $H_0$. The HF propagator can be seen as a suitable candidate for $G^{(0)}$. In the rest of this work, whenever $G^{(0)}$ is used, it is assumed to be the HF Green's function.

\section{Random Phase Approximation}

\begin{figure}
\begin{center}
\includegraphics[scale=0.5,clip=true]{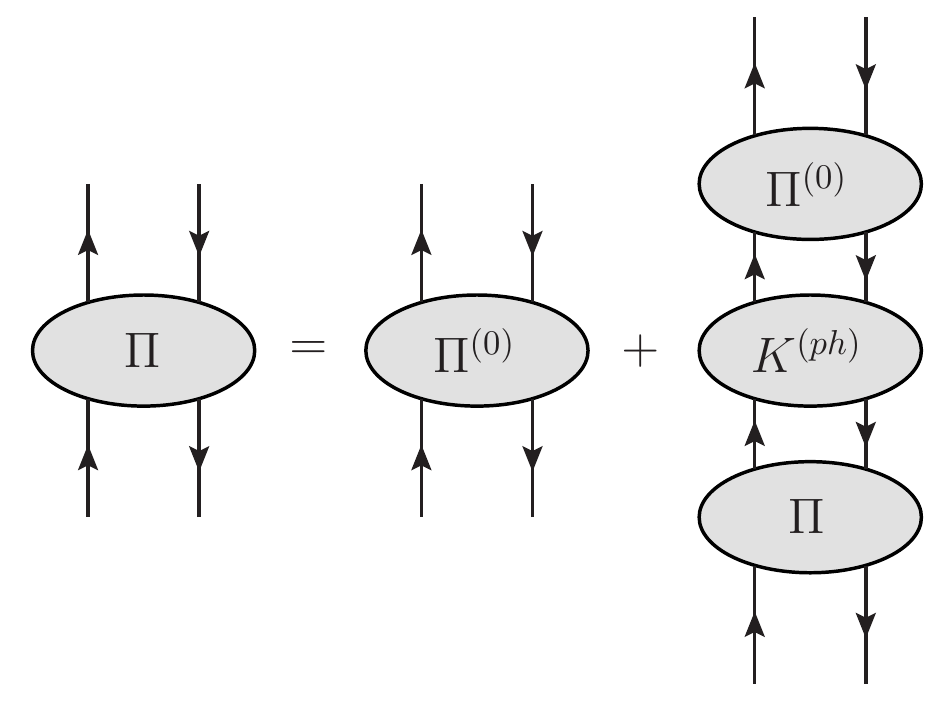}
\caption{
   \label{fig:methods:bethesalpeter}
   Diagrammatical representation of the Bethe-Salpeter equation for the polarization propagator 
}
\end{center}
\end{figure}

Finding a suitable approximation for the polarization propagator $\Pi$ starts with the Bethe-Salpeter equation\cite{Salpeter1951,Nakanishi1969}, which is like a Dyson equation for the polarization propagator. This diagram is shown in \Fig{methods:bethesalpeter}. In principle the interaction $K^{(ph)}$ can be energy-dependent, but in what follows we use a simpler approximation. The logical first step in this approach is to take $K^{(ph)}$ equal to the anti-symmetric two-body interaction $V$. The polarization propagator that is reached in this way is called the Random Phase Approximation (RPA)\cite{Bohm1953} polarization propagator. It represents an infinite resummation of all ring or bubble diagrams. The argument that the diagrams that are not present in this sum have a small contribution due to a randomly changing phase factor and the forthcoming cancellation of terms is the historical reason for the name.

\begin{figure}
\begin{center}
\includegraphics[scale=0.5,clip=true]{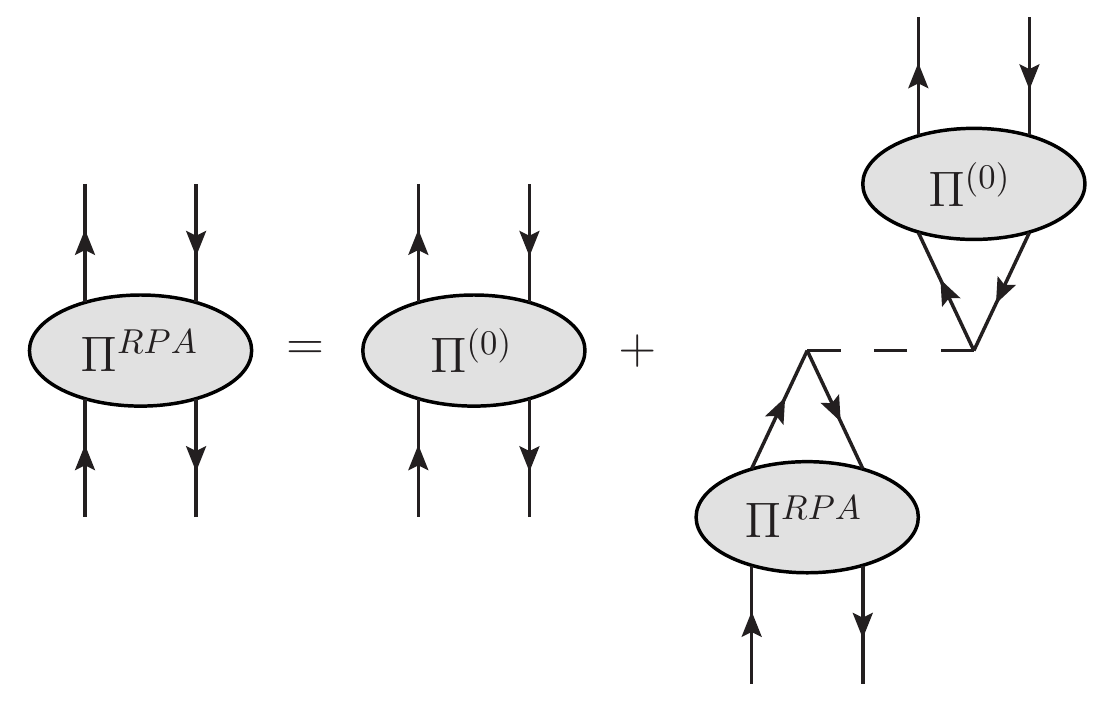}
\caption{
   \label{fig:methods:phRPA}
   Diagrammatical representation of the Bethe-Salpeter equation for the RPA polarization propagator 
}
\end{center}
\end{figure}

The self-consistent equation in the energy domain is given by (see \Fig{methods:phRPA})
\begin{eqnarray}
\Pi^{RPA}_{\alpha\beta^{-1},\gamma\delta^{-1}}(E) &=& \Pi^{(0)}_{\alpha\beta^{-1},\gamma\delta^{-1}}(E)\nonumber\\
&& + \sum_{\epsilon,\zeta} \Pi^{(0)}_{\alpha\beta^{-1},\alpha\beta^{-1}}(E) \langle \alpha\bar{\zeta} |V| \bar{\beta}\epsilon\rangle \Pi^{RPA}_{\epsilon\zeta^{-1},\gamma\delta^{-1}}(E)\label{equ:methods:rpapi}
\end{eqnarray}
and represents both forward and backward propagating rings. In this approximation, it is possible that multiple particle-hole excitations are present at the same time. It is convenient to write the amplitudes for the propagator as
\begin{eqnarray}
\mathcal{X}^{(ph)n}_{\alpha\beta} &=& \langle \Psi^N_n | a^{\dagger}_{\alpha}a_{\bar{\beta}}|\Psi_0^N\rangle^*\\
\mathcal{Y}^{(ph)n}_{\alpha\beta} &=& \langle \Psi^N_n | a^{\dagger}_{\bar{\beta}}a_{\alpha}|\Psi_0^N\rangle^*,
\end{eqnarray}
so that the RPA propagator becomes
\begin{eqnarray}
\Pi^{RPA}_{\alpha\beta^{-1},\gamma\delta^{-1}}(E) &=& \sum_{n\neq 0} \frac{\mathcal{X}^{(ph)n}_{\alpha\beta}(\mathcal{X}^{(ph)n}_{\gamma\delta})^*}{E-\epsilon^{(ph)}_n+i\eta} - \sum_{n\neq 0}\frac{(\mathcal{Y}^{(ph)n}_{\alpha\beta})^*\mathcal{Y}_{\gamma\delta}^{(ph)n}}{E+\epsilon_n^{(ph)}-i\eta},
\end{eqnarray}
where the pole energy is written as
\begin{eqnarray}
\epsilon_n^{(ph)} = E_n^N-E_0^N.
\end{eqnarray}
To calculate the poles and amplitudes one has to resolve to a technique introduced in \Equ{methods:trick}. This results in a non-hermitian eigenvalue equation 
\begin{eqnarray}
\left(\begin{array}{cc}A & B\\ B^* & A^*\end{array}\right)\left(\begin{array}{c} \mathcal{X}^{(ph)n}\\ \mathcal{Y}^{(ph)n}\end{array}\right) &=& \epsilon_n^{(ph)} \left(\begin{array}{cc} 1 & 0 \\ 0 & -1\end{array}\right)\left(\begin{array}{c} \mathcal{X}^{(ph)n}\\ \mathcal{Y}^{(ph)n}\end{array}\right)\label{equ:methods:rpaab},
\end{eqnarray}
with
\begin{eqnarray}
A_{ph,p'h'} &=& \delta_{pp'}\delta_{hh'}\left(\epsilon_p-\epsilon_h\right) + \langle p\bar{h'}|V|\bar{h}p'\rangle\\
B_{ph,p'h'} &=& \langle pp'|V|\bar{h}\bar{h'}\rangle,
\end{eqnarray}
where the indices $p$ and $h$ obviously indicate a particle and a hole state. The solutions are normalized according to
\begin{eqnarray}
\sum_{p,h} (\mathcal{X}^{(ph)n}_{ph})^*\mathcal{X}^{(ph)n'}_{ph} - \sum_{p,h}(\mathcal{Y}^{(ph)n}_{ph})^*\mathcal{Y}^{(ph)n'}_{ph} &=& \delta_{n n'}.
\end{eqnarray}
In the same way there is also the closure relation
\begin{eqnarray}
\sum_{n} \mathcal{X}^{(ph)n}_{ph}(\mathcal{X}^{(ph)n}_{p'h' })^* - \sum_{n}\mathcal{Y}^{(ph)n}_{ph}(\mathcal{Y}^{(ph)n}_{p'h'})^* &=& \delta_{p p '}\delta_{h h'}.
\end{eqnarray}

\begin{figure}
  \centering
  \subfloat[]{\label{fig:methods:tdadiagrams}\includegraphics[scale=0.5]{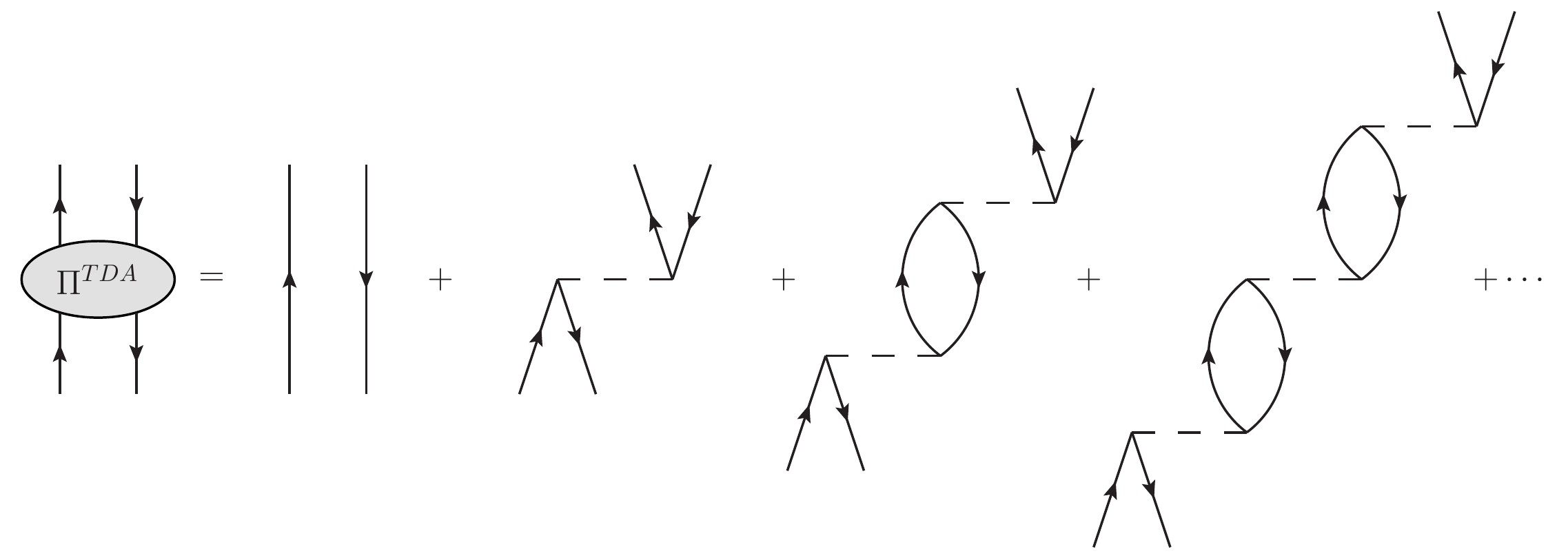}}
  \\
  \subfloat[]{\label{fig:methods:rpadiagrams}\includegraphics[scale=0.5]{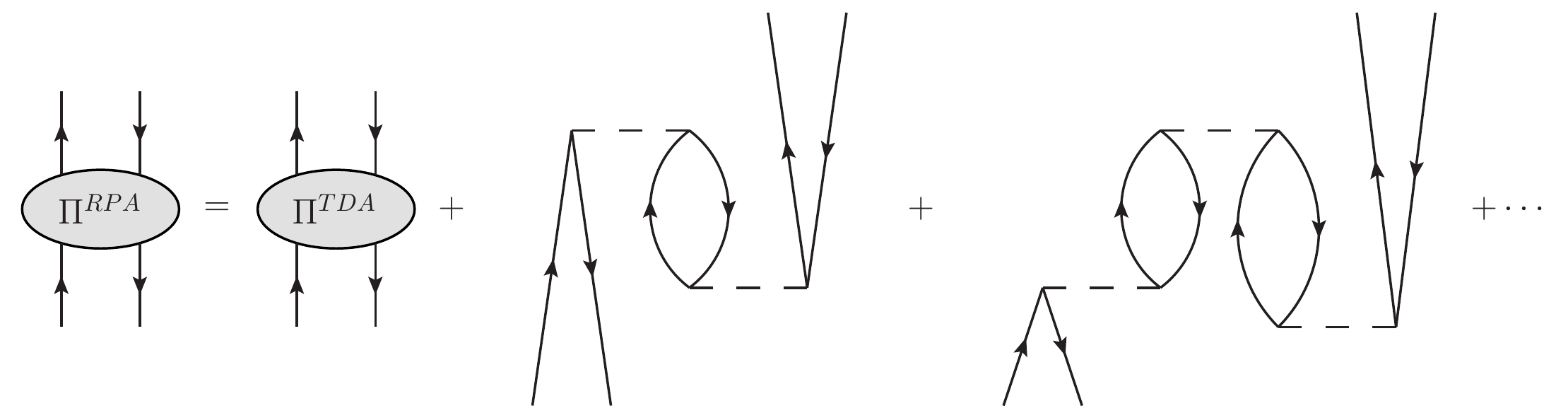}}
  \caption{The difference in polarization propagators in the RPA and TDA}
  \label{fig:methods:rpatda}
\end{figure}

This approach differs from the Tamm Dancoff approximation (TDA) in incorporating correlation also into the ground state, whereas the TDA takes into account correlations only for the excited states. The difference in terms of diagrams is represented by \Fig{methods:rpatda}. The TDA propagator does not include backward going terms, so there is only one excitation present at any time. In the TDA the matrix elements of $B$ are put equal to zero so that the eigenvalue equation reduces to a Hermitian problem. This has the advantage that it eliminates the possibility of complex eigenvalues when the Hartree-Fock approach reaches instability. The result of putting $B$ equal to zero is that there are no backward going amplitudes $\mathcal{Y}^{(ph)}$, which means that these amplitudes can be seen as a measure for the ground-state correlations. Due to the lack of backward going terms, the normalization and closure in the TDA simplify to
\begin{eqnarray}
\sum_{p,h} (\mathcal{X}^{(ph)n}_{ph})^*\mathcal{X}^{(ph)n'}_{ph} &=& \delta_{n n'}
\end{eqnarray}
and
\begin{eqnarray}
\sum_{n} \mathcal{X}^{(ph)n}_{ph}(\mathcal{X}^{(ph)n}_{p'h' })^* &=& \delta_{p p '}\delta_{h h'}\label{equ:methods:closureph}.
\end{eqnarray}

There is another interpretation of the RPA formalism. These equations can be derived by using the equation of motion (EOM) method of Rowe\cite{Rowe1968}. An excitation operator $Q^{(ph)\dagger}$ has to be defined
\begin{eqnarray}
Q^{(ph)n\dagger} &=& \sum_{p,h} \mathcal{X}^{(ph)n}_{ph}a^{\dagger}_{p}a_h - \sum_{p,h} \mathcal{Y}^{(ph)n}_{ph}a^{\dagger}_h a_p\label{equ:methods:qph},
\end{eqnarray}
so that it both creates and destroys a ph pair. Assuming that the hermitian transposed operator destroys the ground state $|\Psi^{(ph)}\rangle$, one can carry out a minimization of the energy with respect to the two amplitudes $\mathcal{X}^{(ph)n}$ and $\mathcal{Y}^{(ph)n}$. This results in the two equations
\begin{eqnarray}
\langle \Psi^{(ph)}|[a^{\dagger}_p a_h, [H,Q^{(ph)n\dagger}]]|\Psi^{(ph)}\rangle &=& \epsilon^{(ph)}_n \langle \Psi^{(ph)}|[a^{\dagger}_p a_h,Q^{(ph)n\dagger}]]|\Psi^{(ph)}\rangle\nonumber\\
&&\\
\langle \Psi^{(ph)}|[a^{\dagger}_h a_p, [H,Q^{(ph)n\dagger}]]|\Psi^{(ph)}\rangle &=& \epsilon^{(ph)}_n \langle \Psi^{(ph)}|[a^{\dagger}_h a_p,Q^{(ph)n\dagger}]]|\Psi^{(ph)}\rangle.\nonumber\\
&&
\end{eqnarray}
When one then again assumes that the ground state is not very different from the HF ground state, the double commutators in the previous equations are much easier to evaluate. The equations thus derived are exactly the same as \Equ{methods:rpaab}. Replacing the RPA ground state with the non-interacting ground state is called the quasi-boson approximation\cite{Brown1961}. The particle hole operators are considered to be perfect boson operators, which is a violation of the Pauli principle, as in fact they are built up out of fermion creation and annihilation operators.

The transition to TDA can also be easily understood from this approach. The same procedure can be applied but with the operator
\begin{eqnarray}
Q^{(ph)n\dagger} &=& \sum_{p,h} \mathcal{X}^{(ph)n}_{ph}a^{\dagger}_{p}a_h,
\end{eqnarray}
which involves only the creation of a ph pair, disconnecting the ph and hp excitations.

\begin{figure}
\begin{center}
\includegraphics[scale=0.5,clip=true]{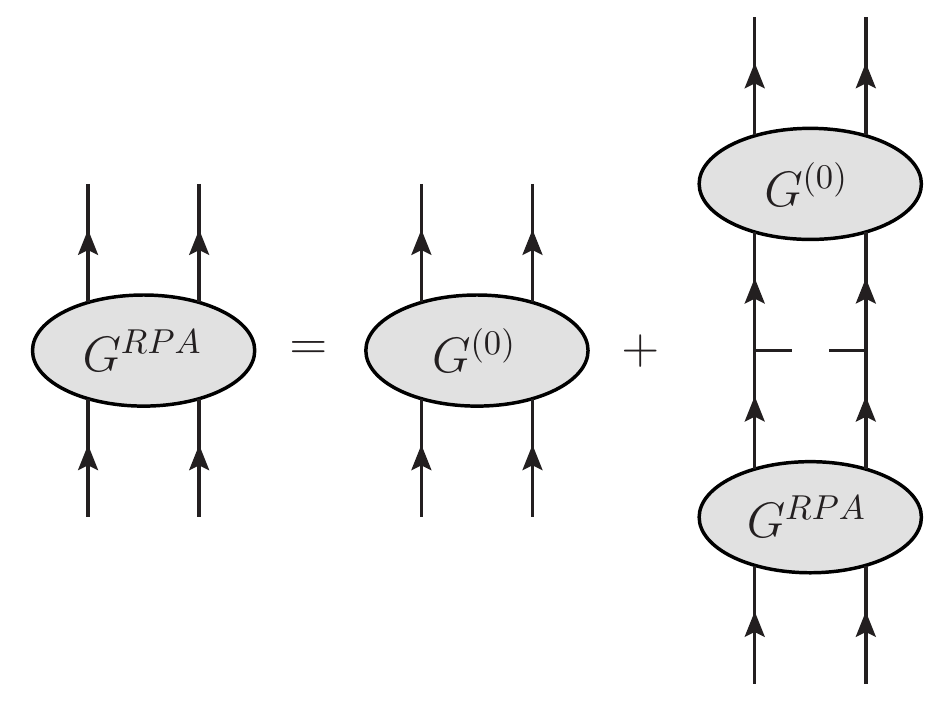}
\caption{
   \label{fig:methods:pprpadiagrams}
   The ladder diagrams included in pp RPA. 
}
\end{center}
\end{figure}

The prescription in \Equ{methods:qph} allows for a straightforward extension of this RPA to pp and hh excitations (see \Fig{methods:pprpadiagrams}). The excitation operator becomes
\begin{eqnarray}
Q^{(pp)n\dagger} &=& \sum_{p_1,p_2} \mathcal{X}^{(pp)n}_{p_1p_2}a^{\dagger}_{p_1}a^{\dagger}_{p_2} - \sum_{h_1,h_2}\mathcal{Y}^{(pp)n}_{h_1h_2}a^{\dagger}_{h_2}a^{\dagger}_{h_1}.
\end{eqnarray}
Following the same procedure as for the ph excitations, the result is a non-hermitian eigenvalue equation
\begin{eqnarray}
\left(\begin{array}{cc}A & B\\ B^{\dagger} & C\end{array}\right)\left(\begin{array}{c} \mathcal{X}^{(pp)m}\\ \mathcal{Y}^{(pp)m}\end{array}\right) &=& \epsilon_m^{(pp)+} \left(\begin{array}{cc} 1 & 0 \\ 0 & -1\end{array}\right)\left(\begin{array}{c} \mathcal{X}^{(pp)m}\\ \mathcal{Y}^{(pp)m}\end{array}\right)
\end{eqnarray}
for the eigenvalues associated with the $N+2$-particle system, labeled by a plus sign and
\begin{eqnarray}
\left(\begin{array}{cc}A & B\\ B^{\dagger} & C\end{array}\right)\left(\begin{array}{c} \mathcal{Y}^{(pp)n}\\ \mathcal{X}^{(pp)n}\end{array}\right) &=& \epsilon_n^{(pp)-} \left(\begin{array}{cc} 1 & 0 \\ 0 & -1\end{array}\right)\left(\begin{array}{c} \mathcal{Y}^{(pp)n}\\ \mathcal{X}^{(pp)n}\end{array}\right)
\end{eqnarray}
for the eigenvalues associated with the $N-2$-particle system, labeled by a minus sign. The matrices involved are the same in both equations
\begin{eqnarray}
A_{p_1p_2,p_1'p_2'} &=& \delta_{p_1p_1'}\delta_{p_2p_2'}(\epsilon_{p_1}+\epsilon_{p_2}) + \langle p_1 p_2|V|p_1'p_2'\rangle\\
B_{p_1p_2,h_1h_2} &=& - \langle p_1 p_2 | V| h_1h_2\rangle\\
C_{h_1h_2,h_1'h_2'} &=& -\delta_{h_1h_1'}\delta_{h_2h_2'}(\epsilon_{h_1} + \epsilon_{h_2}) + \langle h_1 h_2|V|h_1'h_2'\rangle.
\end{eqnarray}
The normalization and closure are very similar to the ph RPA:
\begin{eqnarray}
\sum_{p_1,p_2} (\mathcal{X}^{(pp)m}_{p_1p_2})^*\mathcal{X}^{(pp)m'}_{p_1p_2} - \sum_{h_1,h_2}(\mathcal{Y}^{(pp)m}_{h_1h_2})^*\mathcal{Y}^{(pp)m'}_{h_1h_2} &=& \delta_{mm'}\\
\sum_{p_1,p_2} (\mathcal{Y}^{(pp)n}_{p_1p_2})^*\mathcal{Y}^{(pp)n'}_{p_1p_2} - \sum_{h_1,h_2}(\mathcal{X}^{(pp)n}_{h_1h_2})^*\mathcal{X}^{(pp)n'}_{h_1h_2} &=& -\delta_{nn'}
\end{eqnarray}
and
\begin{eqnarray}
\sum_n \mathcal{X}^{(pp)m}_{\alpha\beta}(\mathcal{X}^{(pp)m}_{\gamma\delta})^* - \sum_m (\mathcal{Y}^{(pp)n}_{\alpha\beta})^*\mathcal{Y}^{(pp)n}_{\gamma\delta} &=& \frac{1}{2}\left(\delta_{\alpha\gamma}\delta_{\beta\delta}-\delta_{\alpha\delta}\delta_{\beta\gamma}\right),\nonumber\\
&&\label{equ:methods:closurepp}
\end{eqnarray}
where $\alpha$, $\beta$, $\gamma$ and $\delta$ are generic indices that label two particles or two holes. The transition to TDA is also made by setting the backward going amplitudes to zero.
							\clearnewpage

\chapter{Faddeev Random Phase Approximation}
\label{cha:frpa}
\begin{figure}
\begin{center}
\includegraphics[scale=0.5,clip=true]{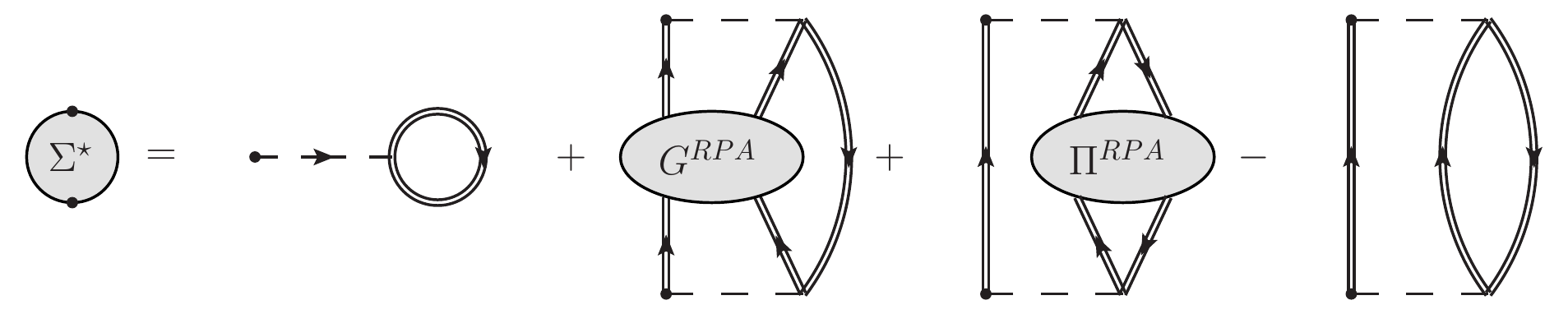}
\caption[Naive resummation of RPA interactions]{
   \label{fig:frpa:naive}
   Naive resummation of RPA interactions between pairs of lines. The third diagram introduces spurious solutions in the Lehmann representation of the propagator.
}
\end{center}
\end{figure}

In this chapter we will demonstrate how a selfenergy of the type \Equ{methods:selfR} can be built, so that it includes interactions between each pair of states that are at the RPA level of theory. This approach will have to go beyond the naive approach that sums the diagrams where the intermediary propagator has been replaced by an RPA propagator. Such a procedure would introduce unphysical solutions of the Dyson equation due to the third diagram in \Fig{frpa:naive}. This diagram is included in both the first and the second diagram and must be subtracted to avoid double counting of diagrams. As a consequence of the minus sign in front of the third diagram, there will be meaningless solutions to the Dyson equation and the normalization of the amplitudes will become problematic due to the erroneous behavior of the derivative of the selfenergy. This naive resummation also lacks the exchange diagram that lets the first particle line interact with the hole line, which is a violation of the Pauli principle. There is no reason why one line should be privileged.

To cure this behavior, it is necessary to perform a partitioning of the interaction into three channels using the Faddeev technique\cite{Faddeev1960}. This allows the pp and ph channels to be mixed, while not making a distinction between any of the three lines. As a result, a better treatment of Pauli correlations is obtained. In contrast to the naive resummation, there is no need to subtract double counting diagrams. The Faddeev procedure takes care of the interaction between two lines separately, which results in a more elegant simultaneous inclusion of pp and ph channels.  

\section{Bethe-Salpeter equation for the irreducible 2p1h/2h1p propagator}
\begin{figure}
\begin{center}
\includegraphics[scale=0.5,clip=true]{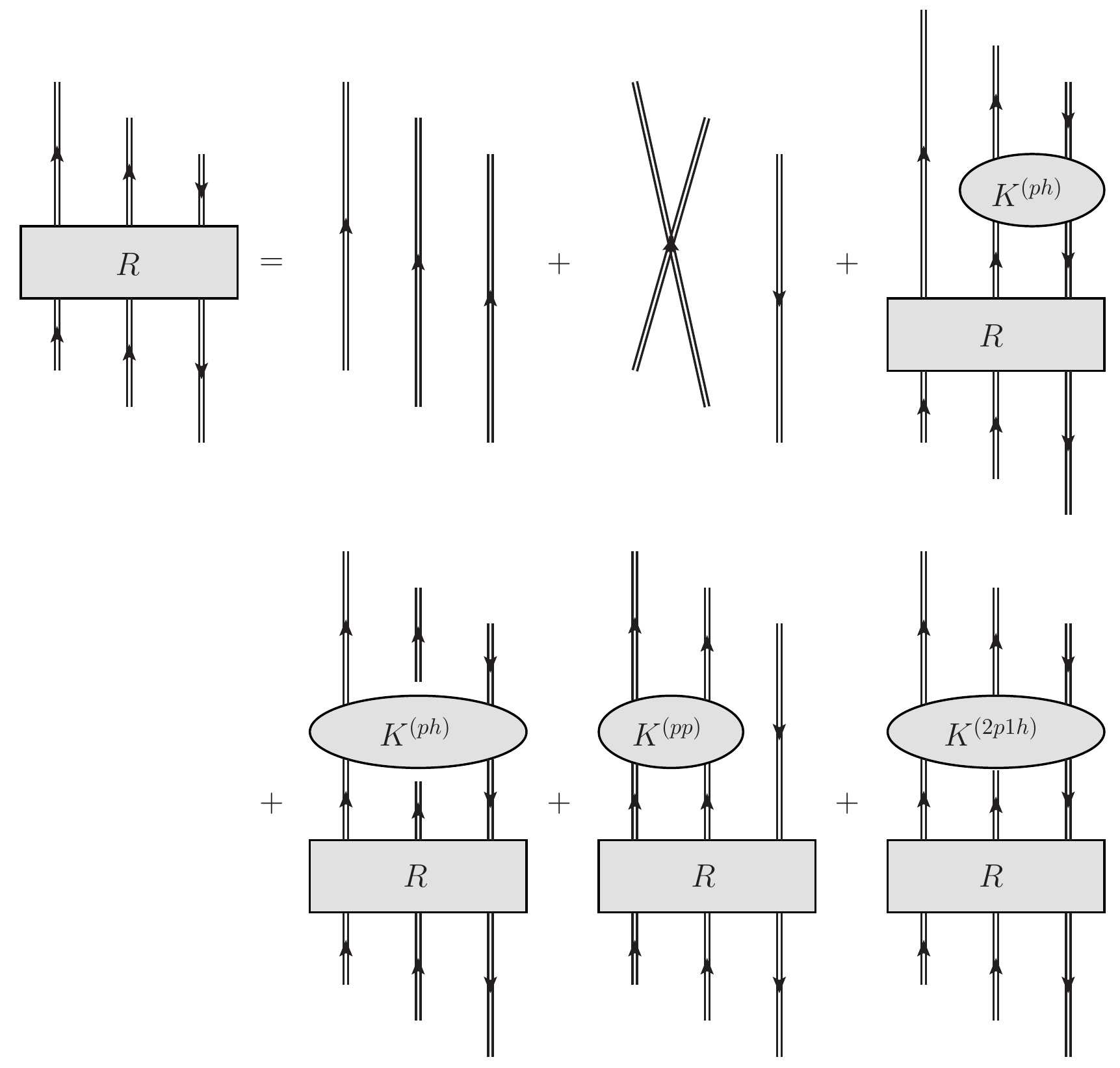}
\caption[Bethe-Salpeter equation for the six-time irreducible 2p1h/2h1p propagator]{
   \label{fig:frpa:Faddeev}
   Bethe-Salpeter equation for the six-time irreducible 2p1h/2h1p propagator. The interaction is split up in separate channels.
}
\end{center}
\end{figure}

It is possible to write a Bethe-Salpeter equation for the irreducible 2p1h/2h1p propagator\cite{Winter1972,Ethofer1969,Ethofer1969b} very similar to the one for the polarization and two-particle propagator. The equation in the time domain
\begin{eqnarray}
R_{\alpha\beta\gamma,\delta\epsilon\zeta}(t_1,t_2,t_3,t_4,t_5,t_6) &=& G_{\alpha\delta}(t_1-t_4)G_{\beta\epsilon}(t_2-t_5)G_{\zeta\gamma}(t_6-t_3) \nonumber\\
&&- G_{\alpha\epsilon}(t_1-t_4)G_{\beta\delta}(t_2-t_5)G_{\zeta\gamma}(t_6-t_3)\nonumber\\
&&+ \sum_{\eta,\theta,\iota,\kappa,\lambda,\mu}\int \mathrm{d}t_1'\int \mathrm{d}t_2'\int\mathrm{d}t_3'\int\mathrm{d}t_4'\int\mathrm{d}t_5'\int\mathrm{d}t_6'\nonumber\\
&& \times G_{\alpha\eta}(t_1-t_1')G_{\beta\theta}(t_2-t_2')G_{\gamma\iota}(t_3'-t_3) \nonumber\\
&&\times K_{\eta\theta\iota,\kappa\lambda\mu}(t_1',t_2',t_3',t_4',t_5',t_6')\nonumber\\
&&\times R_{\kappa\lambda\mu,\delta\epsilon\zeta}(t_4',t_5',t_6',t_4,t_5,t_6)\label{equ:frpa:RK}
\end{eqnarray}
describes the 2p1h propagation as the dressed propagation of the 3 states separately, including exchange and capturing all interactions between the lines in an interaction vertex $K$. This definition of the interaction vertex can be treated using a Faddeev procedure\cite{Faddeev1960}, including a separate interaction vertex for every channel, as is illustrated in \Fig{frpa:Faddeev}
\begin{eqnarray}
K_{\alpha\beta\gamma,\delta\epsilon\zeta}(t_1,t_2,t_3,t_4,t_5,t_6) &=& K^{(ph)}_{\beta\gamma,\epsilon\zeta}(t_2,t_3,t_5,t_6)G_{\alpha\delta}^{-1}(t_1-t_4) \nonumber\\
&& + K^{(ph)}_{\alpha\gamma,\delta\zeta}(t_1,t_3,t_4,t_6)G^{-1}_{\beta\epsilon}(t_2-t_5)\nonumber\\
&& + K^{(pp)}_{\alpha\beta,\delta\epsilon}(t_1,t_2,t_4,t_5)G^{-1}_{\gamma\zeta}(t_6-t_3) \nonumber\\
&&+ K^{(2p1h)}_{\alpha\beta\gamma,\delta\epsilon\zeta}(t_1,t_2,t_3,t_4,t_5,t_6).
\end{eqnarray}

As can be seen in \Equ{frpa:RK}, the full six-time-dependence is needed to write down this Bethe-Salpeter equation. The self-consistency equation can be closed with a four-time reduction of the vertex function, but fewer time arguments are not possible due to the full time dependence of the interaction vertex $K$. There has been no approximation in writing down \Equ{frpa:RK}, so if all objects retain their full time-dependence, this equation solves the 2p1h motion exactly. The $K^{(pp)}$ and $K^{(ph)}$ kernels describe the irreducible interaction of only two lines in the diagram, while the third line propagates freely. It should be noted that $K^{(pp)}$ and $K^{(ph)}$ depend on four times. The $K^{(2p1h)}$ kernel describes the interactions between the three lines that cannot be separated as a succession of pp and ph interactions. In what follows, we will neglect these effective three-body interactions since they do not appear at zeroth order in the problems we are addressing, and will be less important at higher orders. There are no double counting diagrams so it is not necessary to subtract any terms in the equation.

\section{Faddeev procedure in the time domain}
It is customary\cite{Glockle1983} to define three Faddeev components $R^{(i)}$ that will be used to solve \Equ{frpa:RK}. The superscript $(i)$ stands for the number of the line which propagates without interacting with the other lines, while the superscripts $(j)$ and $(k)$ are the lines that are involved in forming the intermediary interaction vertex. The triplet $(i,j,k)$ is always a cyclic permutation of $(1,2,3)$. The Faddeev components are defined as
\begin{eqnarray}
R^{(1)}_{\alpha\beta\gamma,\delta\epsilon\zeta}(t_1,t_2,t_3,t_4,t_5,t_6) &=& \sum_{\eta,\theta,\iota,\kappa} \int \mathrm{d}t_2' \int\mathrm{d}t_3'\int \mathrm{d}t_2''\mathrm{d}t_3'' G_{\beta,\eta}(t_2-t_2')\nonumber\\
&&\times G_{\gamma\theta}(t_3'-t_3) K^{(ph)}_{\eta\theta,\iota\kappa}(t_2',t_3',t_2''t_3'')\nonumber\\
&& \times R_{\alpha\kappa\iota,\delta\epsilon\zeta}(t_1,t_2'',t_3'',t_4,t_5,t_6)\\
R^{(2)}_{\alpha\beta\gamma,\delta\epsilon\zeta}(t_1,t_2,t_3,t_4,t_5,t_6) &=& \sum_{\eta,\theta,\iota,\kappa} \int \mathrm{d}t_1' \int\mathrm{d}t_3'\int \mathrm{d}t_1''\mathrm{d}t_3'' G_{\alpha,\eta}(t_1-t_1')\nonumber\\
&&\times G_{\gamma\theta}(t_3'-t_3) K^{(ph)}_{\eta\theta,\iota\kappa}(t_1',t_3',t_1'',t_3'')\nonumber\\
&&\times R_{\iota\beta\kappa,\delta\epsilon\zeta}(t_1'',t_2,t_3'',t_4,t_5,t_6)\\
R^{(3)}_{\alpha\beta\gamma,\delta\epsilon\zeta}(t_1,t_2,t_3,t_4,t_5,t_6) &=& \sum_{\eta,\theta,\iota,\kappa} \int \mathrm{d}t_1' \int\mathrm{d}t_2'\int \mathrm{d}t_1''\mathrm{d}t_2'' G_{\alpha,\eta}(t_1-t_1')\nonumber\\
&&\times G_{\beta\theta}(t_2-t_2') K^{(pp)}_{\eta\theta,\iota\kappa}(t_1',t_2',t_1'',t_2'')\nonumber\\
&&\times R_{\iota\kappa\gamma,\delta\epsilon\zeta}(t_1'',t_2'',t_3,t_4,t_5,t_6).
\end{eqnarray}
In the current case where the 2p1h kernel $K^{(2p1h)}$ is not taken into account, the full irreducible 2p1h/2h1p propagator is given by
\begin{eqnarray}
R_{\alpha\beta\gamma,\delta\epsilon\zeta}(t_1,t_2,t_3,t_4,t_5,t_6) &=& G_{\alpha\delta}(t_1-t_6)G_{\beta\epsilon}(t_2-t_5)G_{\gamma\zeta}(t_6-t_3) \nonumber\\
&&-G_{\alpha\epsilon}(t_1-t_6)G_{\beta\delta}(t_2-t_5)G_{\gamma\zeta}(t_6-t_3)  \nonumber\\
&& + \sum_{i=1,2,3} R^{(i)}_{\alpha\beta\gamma,\delta\epsilon\zeta}(t_1,t_2,t_3,t_4,t_5,t_6)\label{equ:frpa:R}.
\end{eqnarray}
The meaning of the upper index $(i)$ is clarified by the equations
\begin{eqnarray}
R^{(i)}_{\alpha\beta\gamma,\delta\epsilon\zeta}(t_1,t_2,t_3,t_4,t_5,t_6) &=& \sum_{\eta,\theta,\iota,\kappa,\lambda,\mu}\int \mathrm{d}t_7 \int \mathrm{d}t_8 \int \mathrm{d}t_9\int \mathrm{d}t_{10} \int \mathrm{d}t_{11} \int \mathrm{d}t_{12}\nonumber\\
&& G_{\alpha\eta}(t_1-t_7)G_{\beta\theta}(t_2-t_8)G_{\delta\iota}(t_9-t_3)\nonumber\\
&& \times\Gamma^{(i)}_{\eta\theta\iota,\kappa\lambda\mu}(t_7,t_8,t_9,t_{10},t_{11},t_{12})\nonumber\\
&& \times \left(G_{\kappa\delta}(t_{10}-t_4)G_{\lambda\epsilon}(t_{11}-t_5)G_{\mu\zeta}(t_6-t_{12})\right.\nonumber\\
&& -G_{\lambda\delta}(t_{10}-t_4)G_{\kappa\epsilon}(t_{11}-t_5)G_{\mu\zeta}(t_6-t_{12})\nonumber\\
&& + R^{(j)}_{\kappa\lambda\mu,\delta\epsilon\zeta}(t_{10},t_{11},t_{12},t_4,t_5,t_6)\nonumber\\
&& \left. + R^{(k)}_{\kappa\lambda\mu,\delta\epsilon\zeta}(t_{10},t_{11},t_{12},t_4,t_5,t_6)  \right)\label{equ:frpa:Ri}.
\end{eqnarray}
\begin{figure}
\begin{center}
\includegraphics[scale=0.5,clip=true]{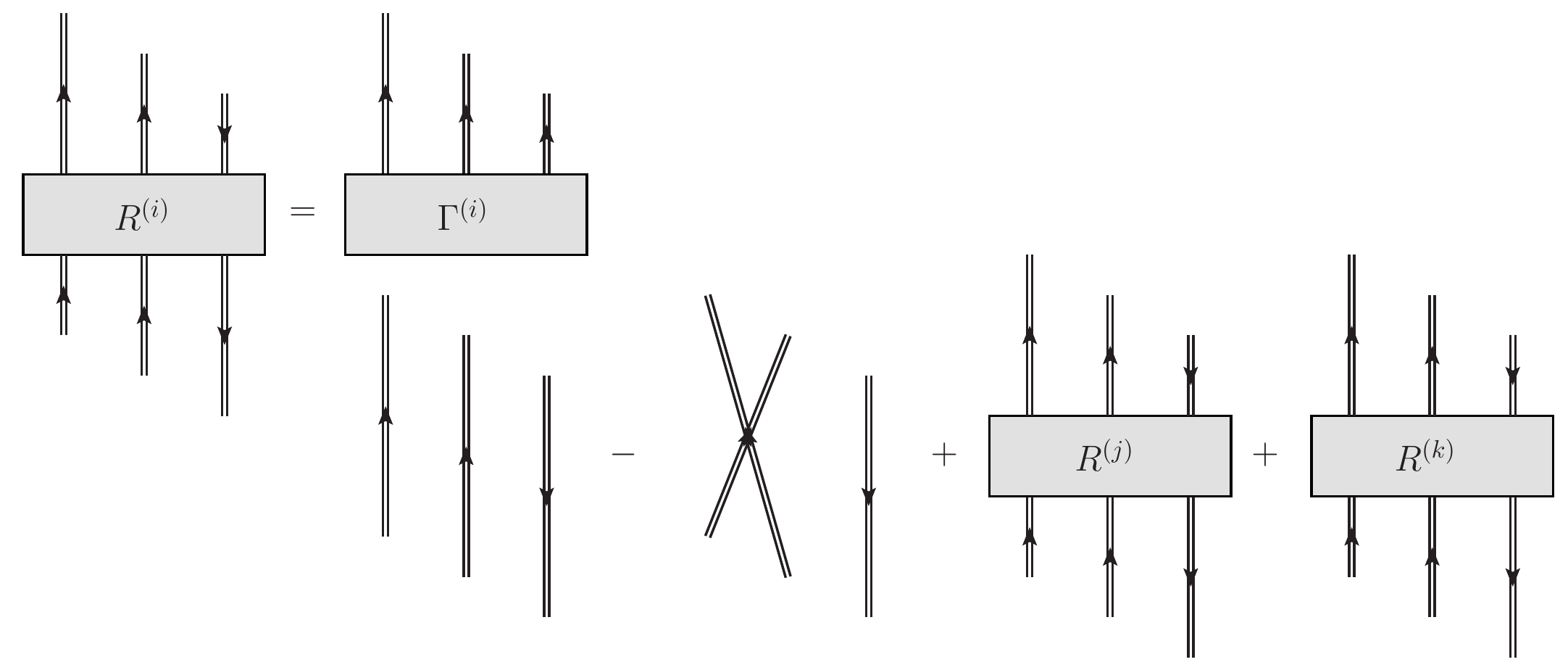}
\caption[Faddeev components]{
   \label{fig:frpa:Ri}
   Every Faddeev component ends in an interaction which lets line $(i)$ go freely. Everything that happens before this interaction is captured in the free propagation and the other two Faddeev components.
}
\end{center}
\end{figure}

Every Faddeev component $R^{(i)}$ ends with an interaction block $\Gamma^{(i)}$, while everything that happens before is captured in the free propagators and the other two components $R^{(j)}$ and $R^{(k)}$. This is illustrated in \Fig{frpa:Ri}. The interaction blocks can be derived from the exact Bethe-Salpeter equations for the pp and ph propagator (\Fig{methods:bethesalpeter}). This Bethe-Salpeter equation can also be written down for the pp and ph four-index interaction blocks. The screened interaction will be equal to the sum of the lowest-order interaction and higher-order diagrams that are generated from a self-consistency procedure. In this way the interaction blocks are derived for both channels
\begin{eqnarray}
\Gamma^{(ph)}_{\alpha\beta,\gamma\delta}(t_1,t_2,t_3,t_4) &=& K^{(ph)}_{\alpha\beta,\gamma\delta}(t_1,t_2,t_3,t_4) \nonumber\\
&& + \sum_{\epsilon,\zeta,\eta,\theta}\int \mathrm{d}t_5 \int \mathrm{d}t_6 \int\mathrm{d}t_7\int\mathrm{d}t_8 \Gamma^{(ph)}_{\alpha\beta,\epsilon\zeta}(t_1,t_2,t_5,t_6)\nonumber\\
&&\times G_{\epsilon\eta}(t_5-t_7)G_{\zeta\theta}(t_8-t_6)K^{(ph)}_{\eta\theta,\gamma\delta}(t_7,t_8,t_3,t_4)\\
\Gamma^{(pp)}_{\alpha\beta,\gamma\delta}(t_1,t_2,t_3,t_4) &=& K^{(pp)}_{\alpha\beta,\gamma\delta}(t_1,t_2,t_3,t_4) \nonumber\\
&& + \sum_{\epsilon,\zeta,\eta,\theta}\int \mathrm{d}t_5 \int \mathrm{d}t_6 \int\mathrm{d}t_7\int\mathrm{d}t_8 \Gamma^{(pp)}_{\alpha\beta,\epsilon\zeta}(t_1,t_2,t_5,t_6)\nonumber\\
&&\times G_{\epsilon\eta}(t_5-t_7)G_{\zeta\theta}(t_6-t_8)K^{(pp)}_{\eta\theta,\gamma\delta}(t_7,t_8,t_3,t_4).
\end{eqnarray}
The interaction blocks $\Gamma^{(i)}$ are six-index interactions composed of the four-index interaction blocks together with the inverse of a single-particle propagator. Two of the $\Gamma^{(i)}$ involve the ph interaction, while one is composed of a pp interaction
\begin{eqnarray}
\Gamma^{(1)}_{\alpha\beta\gamma,\delta\epsilon\zeta}(t_1,t_2,t_3,t_4,t_5,t_6) &=& \Gamma^{(2)}_{\beta\alpha\gamma,\epsilon\delta\zeta}(t_2,t_1,t_3,t_5,t_4,t_6)\\
&=& \Gamma^{(ph)}_{\beta\gamma,\epsilon\zeta}(t_2,t_3,t_5,t_6)G_{\alpha,\delta}^{-1}(t_1-t_4)\\
\Gamma^{(3)}_{\alpha\beta\gamma,\delta\epsilon\zeta}(t_1,t_2,t_3,t_4,t_5,t_6) &=& \Gamma^{(pp)}_{\alpha\beta,\delta\epsilon}(t_1,t_2,t_4,t_5)G_{\gamma\zeta}^{-1}(t_6-t_3).
\end{eqnarray}
\Equ{frpa:Ri} still depends on six times and involves integrals over six more times. Solving this kind of equations is beyond the reach of present-day computers and needs to be avoided at all cost. It should be feasible to circumvent the time-dependence of the Bethe-Salpeter equation and to reduce the time-dependence of $R$ to only two times. The standard method to reduce the time-dependence of this vertex function is by integrating out the excess times. Obviously, this does not offer a proper solution. One part of the solution will be to include a polarization propagator and two-particle propagator that only depend on two times. The other part of the solution is to restrict the class of diagrams that are included to a smaller set.

\section{Faddeev procedure in the energy domain}
\label{sec:frpa:reduction}

The first step is the reduction of the polarization propagator and two-particle propagator to their RPA form, by replacing the kernels $K^{(ph)}$ and $K^{(pp)}$ by their lowest-order approximation, i.e. the bare interaction. In the energy domain, the ph and pp vertices, like the propagators, will depend on only one energy. They are the solution of
\begin{eqnarray}
\Gamma^{(ph)RPA}_{\alpha\beta,\gamma\delta}(E) &=& V_{\alpha\delta,\beta\gamma} + \sum_{\epsilon,\zeta,\eta,\theta} \Gamma^{(ph)RPA}_{\alpha\beta,\epsilon\zeta}(E)\nonumber\\
&& \times \int \frac{\mathrm{d}E_1}{2\pi i} G^{(0)}_{\epsilon\eta}(E+E_1)G^{(0)}_{\zeta\theta}(E_1)V_{\eta\delta,\theta\gamma}\\
\Gamma^{(pp)RPA}_{\alpha\beta,\gamma\delta}(E) &=& V_{\alpha\beta,\gamma\delta} + \frac{1}{2}\sum_{\epsilon,\zeta,\eta,\theta} \Gamma^{(pp)RPA}_{\alpha\beta,\epsilon\zeta}(E)\nonumber\\
&& \times \int \frac{\mathrm{d}E_1}{2\pi i} G^{(0)}_{\epsilon\eta}(E+E_1)G^{(0)}_{\zeta\theta}(-E_1)V_{\eta\zeta,\gamma\delta}\label{equ:frpa:gammapp}.
\end{eqnarray}
The vertices can be derived from the polarization propagator and two-particle propagator. As a result, their Lehmann representation has the same poles as the RPA propagators, but with different amplitudes
\begin{eqnarray}
\Gamma^{(ph)RPA}_{\alpha\beta,\gamma\delta}(E) &=& V_{\alpha\delta,\beta\gamma} + \sum_{\epsilon,\zeta,\eta,\theta} V_{\alpha\zeta,\beta\epsilon}\Pi^{RPA}_{\epsilon\zeta,\eta\theta}(E)V_{\eta\delta,\theta\gamma}\\
&=&  V_{\alpha\delta,\beta\gamma} + \sum_{n\neq 0} \frac{\mathcal{U}^{(ph)n}_{\alpha\delta}(\mathcal{U}^{(ph)n}_{\beta\gamma})^*}{E-\epsilon_n^{(ph)}+i\eta} - \sum_{n\neq 0} \frac{(\mathcal{V}^{(ph)n}_{\alpha\delta})^*\mathcal{V}^{(ph)n}_{\beta\gamma}}{E+\epsilon_n^{(ph)}-i\eta},\nonumber\\
&&
\end{eqnarray}
where
\begin{eqnarray}
\mathcal{U}^{(ph)n}_{\alpha\beta}&=& \sum_{\gamma,\delta}V_{\alpha\delta,\beta\gamma}\mathcal{X}^{(ph)n}_{\gamma\delta}\\
\mathcal{V}^{(ph)n}_{\alpha\beta}&=& \sum_{\gamma,\delta}V_{\alpha\delta,\beta\gamma}\mathcal{Y}^{(ph)n}_{\gamma\delta}
\end{eqnarray}
and
\begin{eqnarray}
\Gamma^{(pp)RPA}_{\alpha\beta,\gamma\delta}(E) &=& V_{\alpha\beta,\gamma\delta} + \sum_{\epsilon,\zeta,\eta,\theta} V_{\alpha\beta,\epsilon\zeta}G^{RPA}_{\epsilon\zeta,\eta\theta}(E)V_{\eta\theta,\gamma\delta}\\
&=& V_{\alpha\beta,\gamma\delta} + \sum_n \frac{\mathcal{U}_{\alpha\beta}^{(pp)n}(\mathcal{U}^{(pp)n}_{\gamma\delta})^*}{E-\epsilon_n^{(pp)+}+i\eta} - \sum_n \frac{(\mathcal{V}^{(pp)n}_{\alpha\beta})^*\mathcal{V}^{(pp)n}_{\gamma\delta}}{E-\epsilon_n^{(pp)-}-i\eta}\nonumber,\\
&&\label{equ:frpa:polepp}
\end{eqnarray}
where
\begin{eqnarray}
\mathcal{U}^{(pp)n}_{\alpha\beta}&=& \sum_{\gamma,\delta}V_{\alpha\beta,\gamma\delta}\mathcal{X}^{(pp)n}_{\gamma\delta}\\
\mathcal{V}^{(pp)n}_{\alpha\beta}&=& \sum_{\gamma,\delta}V_{\alpha\beta,\gamma\delta}\mathcal{Y}^{(pp)n}_{\gamma\delta}.
\end{eqnarray}
The intermediary single-particle propagators are non-interacting. This is a reflection of the quasi-boson approximation in the RPA. In principle dressed propagators\cite{Barbieri2002,Barbieri2001} should be used in the Bethe-Salpeter equations. This would result in different equations for the polarization propagator and two-particle propagator, which are called the dressed RPA (DRPA) equations\cite{Rijsdijk1993,Barbieri2001}.
The inclusion of these effects involves huge matrix sizes as the size of the matrices cubes at every iteration. Therefore, in this work there will be no dressed propagators in the Bethe-Salpeter equations. Since the polarization propagator and two-particle propagator are only calculated with Hartree-Fock propagators, the quality of the vertex functions would be such that it has no advantage to use dressed propagators elsewhere. From now on the non-interacting propagators will be used for the propagation of particles and holes between interaction vertices and vertex functions. This guarantees that at every moment the same level of theory is maintained. In what follows it will also be of some technical importance (see \Equ{frpa:lemma}) that the poles of the intermediary propagation are the same energies as those used in the RPA matrices.

These spectral representations of the vertex functions will be resummed in the irreducible 2p1h/2h1p propagator. As a consequence of the reduction to one energy in the vertex, the irreducible 2p1h/2h1p propagator only depends on three energies
\begin{eqnarray}
R_{\alpha\beta\gamma,\delta\epsilon\zeta}(E_1,E_2,E_3) &=& G^{(0)}_{\alpha\delta}(E_1)G^{(0)}_{\beta\epsilon}(E_2)G^{(0)}_{\gamma\zeta}(-E_3) - G^{(0)}_{\alpha\epsilon}(E_1)G^{(0)}_{\beta\delta}(E_2)G^{(0)}_{\gamma\zeta}(-E_3)\nonumber\\
&& + \sum_{\eta,\theta,\iota,\kappa} G^{(0)}_{\beta\eta}(E_2)G^{(0)}_{\gamma\theta}(-E_3)V_{\eta\kappa,\theta\iota}\nonumber\\
&&\times\int \frac{\mathrm{d}E_4}{2\pi i} R_{\alpha\iota\kappa,\delta\epsilon\zeta}(E_1,E_4,E_2+E_3-E_4)\nonumber\\
&& +\sum_{\eta,\theta,\iota,\kappa} G^{(0)}_{\alpha\eta}(E_1)G^{(0)}_{\gamma\theta}(-E_3)V_{\eta\kappa,\theta\iota}\nonumber\\
&&\times\int \frac{\mathrm{d}E_4}{2\pi i} R_{\iota\beta\kappa,\delta\epsilon\zeta}(E_4,E_2,E_1+E_3-E_4)\nonumber\\
&& +\sum_{\eta,\theta,\iota,\kappa}\frac{1}{2} G^{(0)}_{\alpha\eta}(E_1)G^{(0)}_{\beta\theta}(E_2)V_{\eta\theta,\iota\kappa}\nonumber\\
&&\times\int \frac{(-1)\mathrm{d}E_4}{2\pi i} R_{\iota\kappa\gamma,\delta\epsilon\zeta}(E_4,E_1+E_2-E_4,E_3).
\end{eqnarray}
For the Faddeev components this implies the same decrease in complexity, to only one energy integral
\begin{eqnarray}
R^{(1)}_{\alpha\beta\gamma,\delta\epsilon\zeta}(E_1,E_2,E_3) &=& R^{(2)}_{\beta\alpha\gamma,\epsilon\delta\zeta}(E_2,E_1,E_3)\\
&=& \sum_{\eta,\theta,\iota,\kappa} G_{\beta\eta}^{(0)}(E_2)G_{\gamma\theta}^{(0)}(-E_3)\Gamma^{(ph)RPA}_{\eta\theta,\iota\kappa}(E_2+E_3)\nonumber\\
&&\times\int\frac{\mathrm{d}E_4}{2\pi i} \left(G^{(0)}_{\alpha\delta}(E_1)G^{(0)}_{\iota\epsilon}(E_4)G^{(0)}_{\kappa\zeta}(E_2+E_3-E_4)\right.\nonumber\\
&& -G^{(0)}_{\alpha\epsilon}(E_1)G^{(0)}_{\iota\delta}(E_4)G^{(0)}_{\kappa\zeta}(E_2+E_3-E_4) \nonumber\\
&&+ R^{(2)}_{\alpha\iota\kappa,\delta\epsilon\zeta}(E_1,E_4,E_2+E_3-E_4)\nonumber\\
&&\left.+ R^{(3)}_{\alpha\iota\kappa,\delta\epsilon\zeta}(E_1,E_4,E_2+E_3-E_4)\right)\\ 
R^{(3)}_{\alpha\beta\gamma,\delta\epsilon\zeta}(E_1,E_2,E_3) &=& \sum_{\eta,\theta,\iota,\kappa}G^{(0)}_{\alpha\eta}(E_1)G^{(0)}_{\beta\theta}(E_2)\Gamma^{(pp)RPA}_{\eta\theta,\iota\kappa}(E_1+E_2)\nonumber\\
&&\times \int\frac{\mathrm{d}E_4}{2\pi i} \left(G^{(0)}_{\iota\delta}(E_4)G^{(0)}_{\kappa\epsilon}(E_1+E_2-E_4)G^{(0)}_{\gamma\zeta}(-E_3)\right.\nonumber\\
&& - G^{(0)}_{\iota\epsilon}(E_4)G^{(0)}_{\kappa\delta}(E_1+E_2-E_4)G^{(0)}_{\gamma\zeta}(-E_3)\nonumber\\
&& + R^{(1)}_{\iota\kappa\gamma,\delta\epsilon\zeta}(E_4,E_1+E_2-E_4,E_3)\nonumber\\
&& + R^{(2)}_{\iota\kappa\gamma,\delta\epsilon\zeta}(E_4,E_1+E_2-E_4,E_3).
\end{eqnarray}
While representing a serious decrease in the complexity of the equations, the integral over the single energy still forms a mathematical problem that needs to be avoided. This means a reduction to a single-energy object is mandatory.

The reduction to a single-energy irreducible 2p1h/2h1p propagator is accomplished by restricting the class of diagrams that is resummed in the Faddeev procedure. In the discussion so far, the 2p1h and 2h1p parts of the irreducible 2p1h/2h1p propagator $R$ have still been linked. The propagators in between phonons have both forward and backward propagating parts, potentially connecting a 2p1h state with a 2h1p state. This possibility will now be excluded by restricting the number of phonons that can propagate to one. There will still be an infinite resummation of all possible combinations of ph and pp RPA phonons (including backward propagation), but there is only one phonon propagating at every step in time. In between the phonon propagation, the three states will propagate forward in time. The omitted diagrams will only contribute at higher orders and can safely be discarded. This procedure unlinks the 2p1h and 2h1p spaces and also allows to calculate the irreducible propagator for the two spaces separately. The spaces will still be connected indirectly through the mixing with single-particle space. Even though the 2h1p is more directly linked to experimental results, it is important that $R$ is calculated for both spaces to include the indirect infuence that both spaces have on each other through the coupling with the single-particle states. The derivation of $R$ in the 2p1h and 2h1p space will be very similar, so in what follows the discussion will be restricted to the 2p1h space.

\begin{figure}
\begin{center}
\includegraphics[scale=0.5,clip=true]{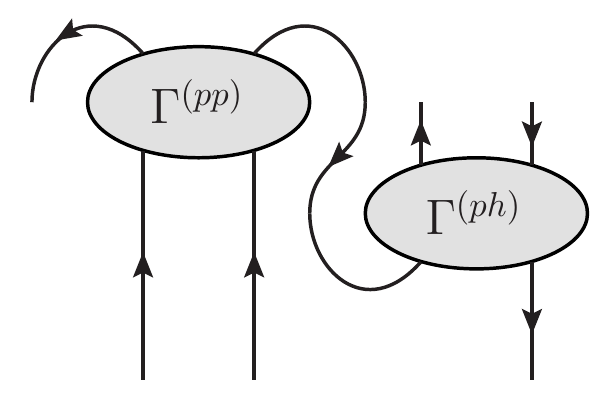}
\caption[An example of a diagram that is not included in the summation of the FRPA]{
   \label{fig:frpa:not}
   An example of a diagram that is not included in the summation of the FRPA. This diagram links the 2p1h space with the 2h1p space, which is prohibited in the present approach.
}
\end{center}
\end{figure}

Limiting the number of phonons to one at every time step is analogous to replacing the factors of three propagators involving forward and backward parts with one propagator depending only on one energy. Since from now on the direction of propagation is fixed, the indices will indicate whether a state is a particle or a hole state. The replacement of the propagators is performed by the substitution
\begin{eqnarray}
G^{(0)}_{p_1p_1'}(E_1)G^{(0)}_{p_2p_2'}(E_2)G^{(0)}_{h_1h_1'}(E_3) \rightarrow G^{(0)>}_{p_1p_2h_1,p_1'p_2'h_1'}(E),
\end{eqnarray}
where the $>$ indicates that only the forward propagating part is retained. The propagator that describes the evolution of the 2p1h state is given by
\begin{eqnarray}
G^{(0)>}_{p_1p_2h_1,p_1'p_2'h_1'}(E) &=& \frac{\delta_{p_1p_1'}\delta_{p_2p_2'}\delta_{h_1h_1'}}{E-(\epsilon_{p_1}+\epsilon_{p_2}-\epsilon_{h_1})+i\eta}.
\end{eqnarray}
Using this propagator for the intermediary evolution of the states results in a new equation for the Faddeev components, depending only on one energy argument:
\begin{eqnarray}
R^{(i)}_{p_1p_2h_1,p_1'p_2'h_1'}(E) &=& \sum_{p_3,p_4,p_5,p_6,h_2,h_3} G^{(0)>}_{p_1p_2h_1,p_3p_4h_2}(E)\Gamma^{(i)}_{p_3p_4h_2,p_5p_6h_3}(E)\nonumber\\
&&\times\left(G^{(0)>}_{p_5p_6h_3,p_1'p_2'h_1'}(E)-G^{(0)>}_{p_5p_6h_3,p_2'p_1'h_1'}(E)\right.\nonumber\\
&&\left.+ R^{(j)}_{p_5p_6h_3,p_1'p_2'h_1'}(E) +R^{(k)}_{p_5p_6h_3,p_1'p_2'h_1'}(E)  \right)\label{equ:frpa:ri}.
\end{eqnarray}

\section{Derivation of the eigenvalue equation}
\begin{figure}
\begin{center}
\includegraphics[scale=0.5,clip=true]{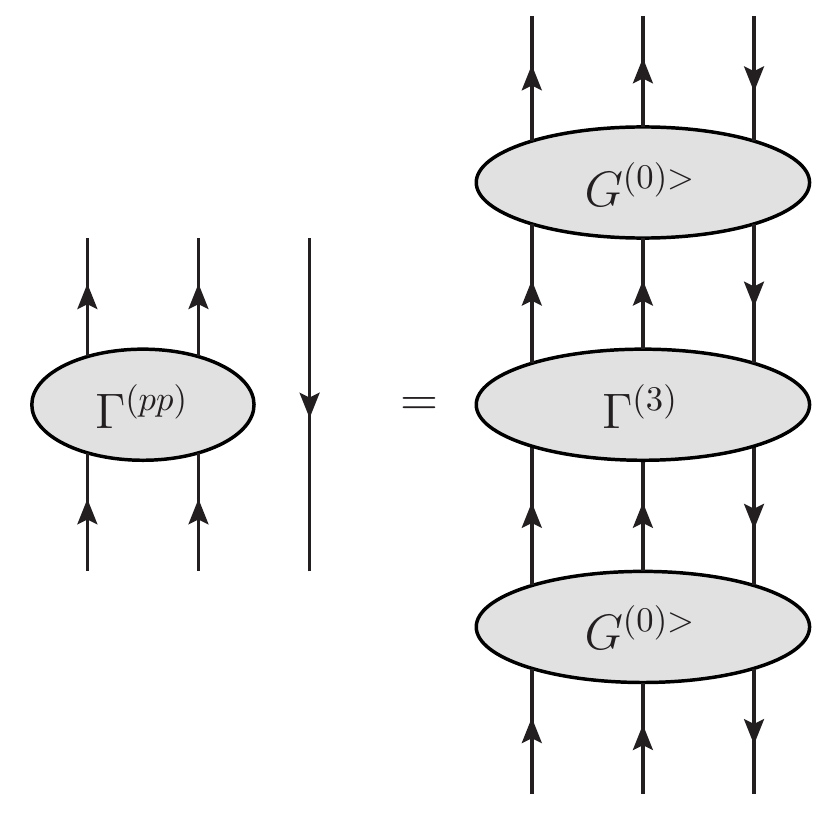}
\caption[Diagrams that illustrate the fixing of the time direction]{
   \label{fig:frpa:gammai}
   Diagrams that illustrate the fixing of the time direction. The left-hand side involves propagators in one direction only. The derivation in \Equ{frpa:gamma3} shows that this is equal to the diagram in the right-hand side.
}
\end{center}
\end{figure}

Fixing the propagation direction of the motion between the phonons has repercussions on the interaction boxes $\Gamma^{(i)}$. There are terms that make the $\Gamma^{(i)}$ different from the normal $\Gamma^{(pp)}$ and $\Gamma^{(ph)}$ vertices. The real structure of these $\Gamma^{(i)}$ is needed to obtain an eigenvalue equation that is formulated in function of the RPA quantities and two-particle interaction matrix elements. As an example the diagram in \Fig{frpa:gammai} is calculated, which will show how the expressions for the $\Gamma^{(i)}$ can be derived. The derivation for channels $(1)$ and $(2)$ is completely analogous but involves the $\Gamma^{(ph)}$ instead of $\Gamma^{(pp)}$.
The integral representation of this diagram is
\begin{eqnarray}
&\sum_{p_3, p_4, p_5, p_6}\frac{1}{2} \int \frac{\mathrm{d}E_1}{2\pi i}\int \frac{\mathrm{d}E_2}{2\pi i} \int \frac{\mathrm{d}E_3}{2\pi i}\, G^{(0)<}_{h_1h_1'}\left(E_1-E\right)G^{(0)>}_{p_1p_3}\left(E_2\right)G^{(0)>}_{p_2p_4}\left(E_1-E_2\right)\nonumber\\
&\quad \times \Gamma^{(pp)}_{p_3p_4,p_5p_6}\left(E_1\right)G^{(0)>}_{p_5p_1'}\left(E_3\right)G^{(0)>}_{p_6p_2'}\left(E_1-E_3\right)\nonumber\\
=& \frac{1}{2} \int \frac{\mathrm{d}E_1}{2\pi i}\frac{\delta_{h_1h_1'}}{E_1-E-\epsilon_{h_1}-i\eta}\Gamma^{(pp)}_{p_1p_2,p_1'p_2'}\left(E_1\right)\frac{1}{E_1-\left(\epsilon_{p_1}+\epsilon_{p_2}\right) + i\eta}\frac{1}{E_1-\left(\epsilon_{p_1'}+\epsilon_{p_2'}\right) + i\eta}\label{equ:frpa:gamma1}.
\end{eqnarray}
Now the pole structure of the pp interaction (\Equ{frpa:polepp}) has to be inserted into \Equ{frpa:gamma1} in such a way that the final result
\begin{eqnarray}
\mathrm{\Equ{frpa:gamma1}} &=& \sum_{p_3,p_4,p_5,p_6,h_2,h_3} G^{(0)>}_{p_1p_2h_1,p_3p_4h_2}\left(E\right)\Gamma^{(3)}_{p_3p_4h_2,p_5p_6h_3}\left(E\right)G^{(0)>}_{p_5p_6h_3,p_1'p_2'h_1'}\left(E\right)\nonumber\\
&&\label{equ:frpa:gamma3}
\end{eqnarray}
defines $\Gamma^{(3)}$ as
\begin{eqnarray}
\Gamma^{(3)}_{p_1p_2h_1,p_1'p_2'h_1'}\left(E\right) & = &\frac{1}{2} \delta_{h_1h_1'} \left( V_{p_1p_2,p_1'p_2'} +\sum_n \frac{\mathcal{U}^{(pp)n}_{p_1p_2}\left(\mathcal{U}^{(pp)n}_{p_1'p_2'}\right)^*}{E_1-\left(\epsilon_n^{(pp)+}-\epsilon_{h_1}\right)+i\eta} \right.\nonumber\\
&& \left. + \sum_{n} \frac{\left(\mathcal{V}^{(pp)n}_{p_1p_2}\right)^*\mathcal{V}^{(pp)n}_{p_1'p_2'}}{\left(\epsilon^{(pp)-}_{n}-\epsilon_{p_1}-\epsilon_{p_2}\right)\left(\epsilon^{(pp)-}_{n}-\epsilon_{p_1'}-\epsilon_{p_2'}\right)} \right.\nonumber\\
&&\left.\times \left(E-\epsilon_{p_1}-\epsilon_{p_2}-\epsilon_{p_1'} - \epsilon_{p_2'} + \epsilon_{h_1} + \epsilon^{(pp)-}_{n}\right)\right)\label{equ:frpa:gamma3}.
\end{eqnarray}
To arrive at an eigenvalue equation, we propose a discrete pole structure for $R^{(i)}$
\begin{eqnarray}
R^{(i)}_{p_1p_2h_1,p_1'p_2'h_1'}(E) &=& \sum_n \frac{\mathcal{X}^{(i)(2p1h)n}_{p_1p_2h_1}(\mathcal{X}^{(2p1h)n}_{p_1'p_2'h_1'})^*}{E-\epsilon^{(2p1h)}_n+i\eta},
\end{eqnarray}
where in this equation
\begin{eqnarray}
\mathcal{X}_{p_1p_2h_1}^{(2p1h)n} &=& \sum_{i=1,2,3} \mathcal{X}_{p_1p_2h_1}^{(i)(2p1h)n}\label{equ:frpa:rri}.
\end{eqnarray}
Substituting this form of $R^{(i)}$ into \Equ{frpa:ri} and multiplying by $(E-\epsilon_n^{(2p1h)})$ while simultaneously taking the limit for $E\rightarrow\epsilon_n^{(2p1h)}$ and assuming that the correlated $R$ has no poles in common with the free 2p1h propagator $G^{(0)>}$ gives, again for $(i)=(3)$,
\begin{eqnarray}
\mathcal{X}^{(3)(2p1h)n}_{p_1p_2h_1} &=& \sum_{p_3,p_4}G^{(0)>}_{p_1p_2h_1',p_1p_2h_1}(\epsilon^{(2p1h)}_n)\Gamma^{(3)}_{p_1p_2h_1,p_3p_4h_1}(\epsilon^{(2p1h)}_n)\nonumber\\
&& \times \left(\mathcal{X}^{(1)(2p1h)n}_{p_3p_4h_1}+\mathcal{X}^{(2)(2p1h)n}_{p_3p_4h_1}\right)\label{equ:frpa:x3}.
\end{eqnarray}
To go any further in this derivation, it is necessary to postulate a property of the pp vertex. When the integral in \Equ{frpa:gammapp} is worked out and the limit procedure is applied, but now for $E\rightarrow \epsilon_{\alpha} + \epsilon_{\beta}$, the result is a crucial property of the vertex, namely
\begin{eqnarray}
\Gamma^{(pp)}_{\alpha\beta,\gamma\delta}(\epsilon_{\alpha}+\epsilon_{\beta}) &=& 0\label{equ:frpa:lemma}.
\end{eqnarray}
This property is of great importance to reduce the complexity of \Equ{frpa:x3} together with \Equ{frpa:gamma3}. An analogous relation holds for the ph vertex $\Gamma^{(ph)}$. After some algebra the equations can be transformed into
\begin{eqnarray}
\mathcal{X}^{(3)(2p1h)m}_{p_1p_2h_1} & = &\frac{1}{2} \left(\sum_n \frac{\mathcal{U}^{(pp)n}_{p_1p_2}\left(\mathcal{U}^{(pp)n}_{p_3p_4}\right)^*}{\left(\epsilon_n^{(pp)+} -\epsilon_{p_1} - \epsilon_{p_2}\right)\left(\epsilon_m^{(2p1h)} - \epsilon_n^{(pp)+} + \epsilon_{h_1}\right)} \right.\nonumber\\
&&\left.+ \sum_{n} \frac{\left(\mathcal{V}^{(pp)n}_{p_1p_2}\right)^*\mathcal{V}^{(pp) n}_{p_3p_4}}{\left(\epsilon^{(pp)-}_{n} - \epsilon_{p_1} - \epsilon_{p_2}\right)\left(\epsilon_{n}^{(pp)-} - \epsilon_{p_3} - \epsilon_{p_4}\right)}\right)\nonumber\\
&&\times \left(\mathcal{X}^{(1)(2p1h)m}_{p_3p_4h_1}+\mathcal{X}^{(2)(2p1h)m}_{p_3p_4h_1}\right)\label{equ:frpa:eigenvalue1}.
\end{eqnarray}
This is a non-linear eigenvalue equation with the Faddeev energies $\epsilon^{(2p1h)}_m$ as eigenvalues and the amplitudes of the Faddeev components as eigenvectors. \Equ{frpa:eigenvalue1} allows us to define the following vectors
\begin{eqnarray}
\left(U^{(3)}\right)_{p_1p_2h_1,nh_1'} &=& \delta_{h_1h_1'}\frac{1}{\sqrt{2}}\mathcal{X}^{(pp)n}_{p_1p_2}\\
\left(H^{(3)}\right)_{p_1p_2h_1,nh_1'} &=& \delta_{h_1h_1'}\frac{1}{\sqrt{2}}\mathcal{Y}^{(pp)n}_{p_1p_2}\\
\left(T^{(3)}\right)_{p_1p_2h_1,nh_1'} &=& \delta_{h_1h_1'}\frac{1}{\sqrt{2}}\left(\epsilon^{(pp)+}_n-\epsilon_{p_1}-\epsilon_{p_2}\right)\mathcal{X}^{(pp)n}_{p_1p_2}\\
\left(D^{(3)}\right)_{nh_1,n'h_1'} &=& \delta_{nn'}\delta_{h_1h_1'}\left(\epsilon^{(pp)+}_n-\epsilon_{h_1}\right).
\end{eqnarray}
The definitions for $(i)=(1),(2)$ are analogous but involve the eigenvalues and eigenvectors of the ph RPA. The second index of $U^{(i)}$, $H^{(i)}$, $T^{(i)}$ and both indices of $D^{(i)}$ are a combination of an index labeling the RPA eigenvalues and the line propagating freely. This guarantees that both indices include the whole 2p1h space. In these vectors, the formulation of the non-linear eigenvalue equation for any $(i)$ is given by
\begin{eqnarray}
\mathcal{X}^{(i)(2p1h)} &=& \left(U^{(i)}\frac{1}{\epsilon^{(2p1h)} - D^{(i)}}T^{(i)\dagger} + H^{(i)}H^{(i)\dagger}\right)\nonumber\\
&&\times\left(\mathcal{X}^{(j)(2p1h)}+\mathcal{X}^{(k)(2p1h)}\right)\label{equ:frpa:eigenvalue2}.
\end{eqnarray}
The index $(i)$ can also be captured in a vector formulation by introducing
\begin{eqnarray}
X &=&\left(\begin{array}{c}\mathcal{X}^{(1)(2p1h)}\\\mathcal{X}^{(2)(2p1h)}\\\mathcal{X}^{(3)(2p1h)}\end{array}\right)
\end{eqnarray}
and the matrix $M$ in this space of three-component vectors
\begin{eqnarray}
M &=& \left(\begin{array}{c c c} 0 & I & I \\ I & 0 & I \\ I & I & 0\end{array}\right).
\end{eqnarray}
These definitions together with the convention that the matrices $U$, $H$, $T$ and $D$ are the matrices with their respective components on the diagonal allow the reformulation of all three components of \Equ{frpa:eigenvalue2} into one matrix equation
\begin{eqnarray}
X = \left( U \frac{1}{\epsilon^{(2p1h)}-D}T^{\dagger}+HH^{\dagger}\right)MX.
\end{eqnarray}
After some algebra the problem then reduces to a linear non-Hermitian eigenvalue equation:
\begin{eqnarray}
\omega X = F X, \quad F=A^{-1}B,\label{eq:frpa:F}
\end{eqnarray}
with 
\begin{eqnarray}
A &=& \left(\begin{array}{ccc}   I                       & -H^{(1)}H^{(1)\dagger}   & -H^{(1)}H^{(1)\dagger}\\
                                 -H^{(2)}H^{(2)\dagger}  & I                        & -H^{(2)}H^{(2)\dagger}\\
                                 -H^{(3)}H^{(3)\dagger}  & -H^{(3)}H^{(3)\dagger}   & I                        \end{array}\right)\\
B &=& \left(\begin{array}{ccc}   0                       &  U^{(1)}T^{(1)\dagger}   & U^{(1)}T^{(1)\dagger}\\
                                 U^{(2)}T^{(2)\dagger}   & 0                        & U^{(2)}T^{(2)\dagger}\\
                                 U^{(3)}T^{(3)\dagger}   & U^{(3)}T^{(3)\dagger}    & 0                        \end{array}\right)\\
&&                                 +
      \left(\begin{array}{ccc}   U^{(1)}D^{(1)}U^{(1)-1} &  0                       &0 \nonumber\\
                                 0                       & U^{(2)}D^{(2)}U^{(2)-1}  & 0 \\
                                 0                       &  0                       & U^{(3)}D^{(3)}U^{(3)-1} \end{array}\right)\nonumber\\
&&\times\left(\begin{array}{ccc}   I                       & -H^{(1)}H^{(1)\dagger}   & -H^{(1)}H^{(1)\dagger}\\
                                 -H^{(2)}H^{(2)\dagger}  & I                        & -H^{(2)}H^{(2)\dagger}\\
                                 -H^{(3)}H^{(3)\dagger}  & -H^{(3)}H^{(3)\dagger}   & I                        \end{array}\right)
\end{eqnarray}

\section{Elimination of spurious solutions}
Through the definition of the three Faddeev components, the dimension of the matrices that have to be diagonalized has tripled. The number of physical solutions to the Dyson equation has stayed the same, so two thirds of these solutions have to be discarded. This can be seen from the way in which the final amplitudes $\mathcal{X}^{(2p1h)}$ are retrieved from the Faddeev components (\Equ{frpa:rri}). Combinations of Faddeev components that sum to zero have a vanishing amplitude in the irreducible 2p1h propagator and will not contribute to the selfenergy.

The solutions to the three-component eigenvalue equation can be divided into two classes according to their eigenvalue with respect to the operator
\begin{eqnarray}
P &=& \left(\begin{array}{ccc} 0 & I^{ex} & 0 \\ I^{ex} & 0 & 0 \\ 0 & 0 & I^{ex}\end{array}\right),
\end{eqnarray}
where $I^{ex}$ is the matrix that exchanges the two particle indices
\begin{eqnarray}
I^{ex}_{p_1p_2h_1,p_1'p_2'h_1'} &=& \delta_{p_1p_2'}\delta_{p_2p_1'}\delta_{h_1h_1'}.
\end{eqnarray}
$P$ commutes with $F$ and the common set of eigenvectors of these operators are
\begin{eqnarray}
X_{-1} = \left(\begin{array}{c} \mathcal{X}^{(2p1h)}_a \\ -I^{ex}\mathcal{X}^{(2p1h)}_a \\ \mathcal{X}^{(2p1h)}_b - I^{ex}\mathcal{X}^{(2p1h)}_b\end{array}\right) \mathrm{and~} X_{1} = \left(\begin{array}{c} \mathcal{X}^{(2p1h)}_a \\ I^{ex}\mathcal{X}^{(2p1h)}_a \\ \mathcal{X}^{(2p1h)}_b + I^{ex}\mathcal{X}^{(2p1h)}_b\end{array}\right)
\end{eqnarray}
with $\mathcal{X}^{(2p1h)}_a$ and $\mathcal{X}^{(2p1h)}_b$ arbitrary vectors in the Faddeev component space. It can be shown that $X_{-1}$ and $X_1$ are also eigenvectors of $F$. $X_{-1}$ gives rise to anti-symmetric amplitudes to the irreducible 2p1h propagator, while $X_1$ results in symmetric amplitudes under the exchange of the two particle indices. Since we are dealing with electrons, the eigenvectors of interest are of the type $X_{-1}$. They consist of the symmetric and anti-symmetric part of $\mathcal{X}^{(2p1h)}_a$ and only the anti-symmetric part of $\mathcal{X}^{(2p1h)}_b$. Now the unphysical solutions can be selected. Defining the bases
\begin{eqnarray}
Y^{Ph} = \frac{1}{\sqrt{6}}\left(\begin{array}{c} I-I^{ex} \\ I-I^{ex} \\ I-I^{ex} \end{array}\right) \mathrm{and~} Y^{Sp} = \frac{1}{\sqrt{6}}\left(\begin{array}{c} -I \\ I^{ex} \\ I-I^{ex}\end{array}\right)\label{eq:frpa:vectors},
\end{eqnarray}
it can immediately be seen that the vector space spanned by $Y^{Sp}$ concerns eigenvectors that sum to zero amplitude and has to be projected out. Since we only need the physical solutions of the eigenvalue equation for the spectral amplitudes, it suffices to project both matrices $A$ and $B$ from the right-hand side with the physical and spurious solutions and take the matrix block $\langle Y^{Ph}|F|Y^{Ph}\rangle$ to be diagonalized\cite{Barbieri2001,Navratil1999}.

The second row and second column of the projected matrix can be left out because of the exchange symmetry. The right-hand side projection of $A$ results in
\begin{eqnarray}
\tilde{A} &=& A\left(\begin{array}{cc} Y^{Ph} & Y^{Sp}\end{array}\right)\label{eq:frpa:atilde1}\\
&=& \frac{1}{\sqrt{6}}\left(\begin{array}{cc}   I-I^{ex} - 2H^{(1)}H^{(1)\dagger}\left(I-I^{ex}\right) & -I - H^{(1)}H^{(1)\dagger}\\
                              I-I^{ex} - 2H^{(3)}H^{(3)\dagger}\left(I-I^{ex}\right) & I-I^{ex} + H^{(3)}H^{(3)\dagger}\left(I-I^{ex}\right)\end{array}\right)\nonumber\\
                              &&\label{eq:frpa:atilde2}\\
&=& \frac{1}{\sqrt{6}}\left(\begin{array}{cc}   I-I^{ex} - 2H^{(1)}H^{(1)\dagger}\left(I-I^{ex}\right) & -I -H^{(1)}H^{(1)\dagger}\\
                              I-I^{ex} - 4H^{(3)}H^{(3)\dagger}                      & I-I^{ex} + 2H^{(3)}H^{(3)\dagger}\end{array}\right)\label{eq:frpa:atilde3}\\
&=& \frac{1}{\sqrt{6}}\left(\begin{array}{cc}   \left(3I-2U^{(1)}U^{(1)\dagger}\right)\left(I-I^{ex}\right)   & -U^{(1)}U^{(1)\dagger}\\
                              3\left(I-I^{ex}\right) - 4 U^{(3)}U^{(3)\dagger}       & 2U^{(3)}U^{(3)\dagger} \end{array}\right),\label{eq:frpa:atilde4}\\
&=& \left(\begin{array}{cc} Q & R \\ P & S\end{array}\right)
\end{eqnarray}
where the transition between the last three lines in the equation are accomplished by using the anti-symmetry in the first two indices and by inserting the closure of the RPA equations \Equ{methods:closureph} and \Equ{methods:closurepp}.
The projection of $B$ goes along the same lines:
\begin{eqnarray}
\tilde{B} 
&=& \frac{1}{\sqrt{6}}\left(\begin{array}{cc}  \left(-2U^{(1)}U^{(1)\dagger}E^f + 3 U^{(1)}D^{(1)}U^{(1)-1}\right)\left(I-I^{ex}\right) &  -U^{(1)}U^{(1)\dagger}E^f\\
                                       -4U^{(3)}U^{(3)\dagger}E^f + 6 U{(3)}D^{(3)}U^{(3)-1}  &  2U^{3}U^{(3)\dagger}E^f \end{array}\right)\nonumber\\
                                       && \label{eq:frpa:btilde2},\\
&=& \left(\begin{array}{cc} L & N\\ M & K\end{array}\right)
\end{eqnarray}
where the diagonal matrix $E^f$ was introduced as
\begin{eqnarray}
E^f_{p_1p_2h_1,p_1'p_2'h_1'} = \left(\epsilon_{p_1}+\epsilon_{p_2}-\epsilon_{h}\right)\delta_{p_1p_1'}\delta_{p_2p_2'}\delta_{h_1h_1'}.
\end{eqnarray}
The matrix $\tilde{A}$ can be inverted by using block-inversion\cite{Press2007}
\begin{eqnarray}
\tilde{A}^{-1} &=& \left( \begin{array}{cc} \tilde{Q} & \tilde{R} \\ \tilde{P} & \tilde{S} \end{array}\right)\\
&=& \left(\begin{array}{cc} \left(Q-RS^{-1}P\right)^{-1} & -\tilde{Q}RS^{-1} \\ -S^{-1}P\tilde{Q} & S^{-1} + S^{-1}P\tilde{Q}RS^{-1}\end{array}\right).
\end{eqnarray}
The matrix block of interest is
\begin{eqnarray}
\langle Y^{Ph} | F | Y^{Ph} \rangle &=& \tilde{Q}L + \tilde{R}M\\
&=& \tilde{Q}\left(L - RS^{-1}M\right).
\end{eqnarray}
The calculation of $\tilde{Q}$ can be avoided by solving the general eigenvalue equation
\begin{eqnarray}
\tilde{P}^{-1}\mathcal{X}^{(2p1h)} \epsilon^{(2p1h)}&=& (L-RS^{-1}M)\mathcal{X}^{(2p1h)}.
\end{eqnarray}
The left-hand side is given in terms of the Faddeev component matrices as
\begin{eqnarray}
\tilde{Q}^{-1} &=& \frac{1}{2} \left(I-I^{ex}\right) - 2U^{(3)}U^{(3)\dagger} \nonumber\\
&&+ U^{(3)}U^{(3)\dagger}U^{(1)\dagger-1}U^{(1)-1}(I-I^{ex})
\end{eqnarray}
while the right-hand side is given by
\begin{eqnarray}
L-RS^{-1}M &=& U^{(3)}D^{(3)}U^{(3)-1} -2U^{(3)}U^{(3)\dagger}E_f \nonumber\\
&&+ U^{(3)}U^{(3)\dagger}U^{(1)\dagger-1}D^{(1)}U^{(1)-1}(I-I^{ex})
\end{eqnarray}
This is the non-Hermitian eigenvalue problem that has to be solved and which results in the poles and amplitudes of the irreducible 2p1h propagator in the present approximation.
\section{Coupling to the single-particle space}
After the calculation of the excitation energies $\epsilon^{(2p1h)}$ and $\epsilon^{(2h1p)}$ together with their respective amplitudes, these excitations have to be coupled with the single-particle space. The coupling happens through the procedure introduced in \Equ{methods:selfR}, applied in the energy domain. Instead of building the selfenergy matrix explicitly from the Faddeev eigenvalues and eigenvectors and diagonalizing it afterwards, it is preferable to reformulate the problem in the space of the single-particle states and the 2p1h and 2h1p solutions. In this space the hamiltonian matrix becomes
\begin{eqnarray}
H &=& \left(\begin{array}{ccc} \epsilon -U +\Sigma^{HF} & \tilde{V}^{(2p1h)} & \tilde{V}^{(2h1p)} \\ \tilde{V}^{(2p1h)\dagger} & \epsilon^{(2p1h)} & 0\\ \tilde{V}^{(2h1p)\dagger} & 0 & \epsilon^{(2h1p)}\end{array}\right)\label{equ:frpa:hamiltonian}.
\end{eqnarray}
The two blocks $\epsilon^{(2p1h)}$ and $\epsilon^{(2h1p)}$ are diagonal in the space of the Faddeev solutions. Note that there is no explicit coupling between the 2p1h and 2h1p states. Their mixing blocks are equal to zero and these spaces are only linked indirectly through the single-particle space. The two blocks $V^{(2p1h)}$ and $V^{(2h1p)}$ take into account this coupling. These blocks are not simply the bare interactions. Due to the procedure introduced in \Sec{frpa:reduction}, where the class of diagrams was restricted, we have neglected a class of diagrams. These are, however, not energy dependent and can be reintroduced a posteriori by replacing the bare interaction with the interaction leading to the description that sums all diagrams up to third order. As was derived by Schirmer et al.\cite{Schirmer1983}, the interaction that gives the correct behavior is
\begin{eqnarray}
V^{(2p1h)}_{\alpha,p_1p_2h_1} &=& V_{\alpha h_1,p_1p_2} + \frac{1}{2}\sum_{h_2,h_3} \frac{V_{\alpha h_1,h_2h_3} V_{h_2 h_3, p_1 p_2 }}{\epsilon_{h_2} + \epsilon_{h_3} - \epsilon_{p_1} - \epsilon_{p_2}}\nonumber\\
&& + \sum_{p_3, h_2}\frac{V_{\alpha p_3, p_1 h_2}V_{h_1 h_2,p_2 p_3}}{\epsilon_{h_1}+\epsilon_{h_2} - \epsilon_{p_2} - \epsilon_{p_3}} - \sum_{p_3,h_2} \frac{V_{\alpha p_3, p_2 h_2}V_{h_1 h_2,p_1p_3}}{\epsilon_{h_1}+\epsilon_{h_3} - \epsilon_{p_1} - \epsilon_{p_3}}.\nonumber\\
&&
\end{eqnarray}
An analogous expression can be found for the coupling between the single-particle states and 2h1p states. As the diagonal blocks are expressed in the basis of the physical states $Y^{Ph}$ from \Equ{frpa:vectors}, a sum has to be carried out over the 2p1h indices to change to the basis of Faddeev solutions
\begin{eqnarray}
\tilde{V}^{(2p1h)}_{\alpha,n} = \sum_{p_1,p_2,h_1} \mathcal{X}^{(2p1h)n}_{p_1p_2h_1}V_{\alpha,p_1p_2h_1}.
\end{eqnarray}

The diagonalization of the Hamiltonian matrix \Equ{frpa:hamiltonian} has to be done iteratively. In order to resum all diagrams correctly up to third order, the Hartree-Fock selfenergy has to be calculated using the correlated density matrix. At every step of the iteration, a new density matrix is calculated from the eigenvectors of \Equ{frpa:hamiltonian}. Based on this density matrix, a new Hartree-Fock term is generated and inserted in the Hamiltonian matrix. This procedure has to be repeated until convergence. Note that the coupling blocks and diagonal Faddeev blocks stay the same during this iterative process. Therefore, they have to be calculated only once.

This partial selfconsistency procedure is also used in the FRPA calculations for nuclei\cite{Barbieri2002}. However the procedure here is somewhat more involved. In order to include the correct description of the short range tensor correlations, a presummation of diagrams is carried out. This is done by solving the Brueckner-Goldstone equations for the effective nucleon-nucleon interaction or $G$ matrix\cite{Muther1993}. In this procedure, the Pauli operator excludes the states from the space to which the many body perturbation theory is applied. This prevents double countings of ladder-diagrams, while at the same time adding the intermediate states that are necessary to obtain correct results. This presummation of the Brueckner-Hartree-Fock propagator constitutes the main difference between the processes of calculation for nuclei and molecules. If one wanted to apply the same routine as for molecules, one could use renormalized low-momentum interactions. These interactions have the short-range physics included and a prediagonalization of ladder diagrams is no longer needed. The low momentum interactions do however require the use of three-nucleon forces in the formalism.

\section{FTDA as a special case of FRPA}
The following section presents the proof that when interactions at the TDA level of theory are used in the Faddeev procedure, the Faddeev TDA corresponds to ADC(3)\cite{Schirmer1983}. The single-particle selfenergy block and coupling matrices in the Hamiltonian \Equ{frpa:hamiltonian} are identical to what is found in ADC(3). The question is whether the diagonal matrices reduce to the standard ADC(3) form after elimination of the spurious solutions. The reduction can be done by setting the matrices that involve backward propagating amplitudes to zero, expressing the lack of such amplitudes in the TDA. The corresponding eigenvalue equation for the 2p1h space is
\begin{eqnarray}
\epsilon^{(2p1h)}\mathcal{X}^{(2p1h)}&=& \frac{1}{6}\left(I-I^{ex}\right)\left(4U^{(1)}T^{(1)\dagger}+2U^{(1)}D^{(1)}U^{(1)-1} \right.\nonumber\\
&&\left.+ 2U^{(3)}T^{(3)\dagger} + U^{(3)}D^{(3)}U^{(3)-1}\right)\left(I-I^{ex}\right)\mathcal{X}^{(2p1h)},\nonumber\\
&&
\end{eqnarray}
where the matrices $U^{(i)}$, $T^{(i)}$ and $D^{(i)}$ are now built up with the RPA quantities replaced by TDA quantities. If one does the same substitution of zero backward going amplitudes in the closure relations \Equ{methods:closureph} and \Equ{methods:closurepp}, the first two terms reduce to
\begin{eqnarray}
\left(4U^{(1)}T^{(1)\dagger}+2U^{(1)}D^{(1)}U^{(1)-1} \right)_{p_1p_2h_1,p_1'p_2'h_1'} &=& 2\delta_{p_1p_1'}\delta_{p_2p_2'}\delta_{h_1h_1'}\nonumber\\
&&\times\left(\epsilon_{p_1}+\epsilon_{p_2}-\epsilon_{h_1}\right)\nonumber\\
&&+ \delta_{p_1p_1'}6V_{p_2'h_1,h_1'p_2}\nonumber,\\
&&
\end{eqnarray}
while the last two terms give
\begin{eqnarray}
\left( 2U^{(3)}T^{(3)\dagger} + U^{(3)}D^{(3)}U^{(3)-1}\right)_{p_1p_2h_1,p_1'p_2'h_1'} &=& \frac{1}{2}\delta_{h_1h_1'}\left(\delta_{p_1p_1'}\delta_{p_2p_2'}-\delta_{p_1p_2'}\delta_{p_2p_1'}\right)\nonumber\\
&&\times\left(\epsilon_{p_1}+\epsilon_{p_2}-\epsilon_{h_1}\right)\nonumber\\
&&+ \frac{3}{2}\delta_{h_1h_1'}V_{p_1p_2,p_1'p_2'}.
\end{eqnarray}
Together with the anti-symmetrization of both particle indices this yields for the eigenvalue problem
\begin{eqnarray}
\epsilon^{(2p1h)}_n\mathcal{X}^{(2p1h)n}_{p_1p_2h_1} &=& \sum_{p_1',p_2',h_1'}\left(\delta_{h_1h_1'}\left(\delta_{p_1p_1'}\delta_{p_2p_2'}-\delta_{p_1p_2'}\delta_{p_2p_1'}\right)\left(\epsilon_{p_1}+\epsilon_{p_2}-\epsilon_{h_1}\right)\right.\nonumber\\
&&+ \delta_{h_1h_1'} V_{p_1p_2,p_1'p_2'} + \delta_{p_1p_1'} V_{p_2'h_1,h_1'p_2} - \delta_{p_1p_2'} V_{p_1'h_1,h_1'p_2}\nonumber\\
&&\left.+ \delta_{p_2p_2'} V_{p_1'h_1,h_1'p_1} - \delta_{p_2p_1'} V_{p_2'h_1,h_1'p_1}\right)\mathcal{X}^{(2p1h)n}_{p_1'p_2'h_1'},
\end{eqnarray}
which is exactly the equation that is solved in the ADC(3)\cite{Schirmer1983}. In what follows, results obtained with ADC(3) will be labeled as FTDA, seeing that both methods are completely equivalent.

There have been efforts to make a connection between Green's function theories and CC theories. The equivalence of the ring CC with doubles (CCD) and RPA\cite{Scuseria2008} shows that in some cases it is possible to find such a connection, although this correspondence holds only for the RPA with direct interaction and is based on the equality of the correlation energies. When the level of theory becomes higher, the equivalence relations become more and more intricate. It is possible to link an approximate CC method with doubles (CC2) to the ADC(2) matrix by replacing the CC2 ground-state amplitudes by those obtained in first-order perturbation theory and taking the Hermitian average\cite{Hattig2005}. There is no obvious systematic mechanism for how to extend these correspondences to higher orders or more complex theories. Given the very involved diagrammatic content of FRPA, it should be no surprise that no equivalent CC theory has yet been found. However, hybrid methods using concepts of Green's function theory combined with coupled cluster theory have been devised\cite{Nooijen1992} and have proven to give accurate results for small molecules\cite{Nooijen1993}.
							\clearnewpage
\chapter{Results}
\label{cha:results}

In this chapter the method developed in the previous chapters is applied to some physical systems. First an application to atoms is discussed. Considering the spherically symmetric nature of closed-shell atoms, it is possible to exploit the symmetry to the maximal extent and to push the calculations towards very large basis sets. Theoretical extrapolation to the basis set limit makes precise comparison with experiment possible.

The application to small molecules is another example where quantum mechanical correlation plays an important role in the description of the physical properties. The most common quantum chemistry methods like HF and density functional theory (DFT) fail to simultaneously describe correct ground-state energies and ionization energies, even at equilibrium geometry. First the equilibrium properties of these molecules are studied, after which the dissociation is studied in some detail. The physical implications of the separation of the atoms in a diatomic molecule are catastrophic for the RPA. Triplet instabilities prohibit the calculation of properties in the FRPA.

The Hubbard molecule is used as a toy model to investigate the this further and to search for possible solutions.

\section{Accurate atomic calculations}

The first application of the FRPA to electronic structure was a calculation for neon\cite{Barbieri2007}. This calculation showed that the main ionization energies obtained with the FRPA are at least of the same level, and even somewhat better than those obtained with ADC(3). Two basis sets were employed: the augmented correlation-consistent polarization valence triple-zeta (aug-cc-pVTZ) basis, a standard quantum-chemical Gaussian basis set and the HF+continuum, a numerical basis set based on Hartree-Fock and subsequent discretization of the continuum. Comparing results obtained in a finite basis set to experimental values may lead to faulty conclusions. Therefore, it is of importance to investigate the behavior of the results in the basis set limit, i.e. to systematically enlarge the basis set until the result converges. We considered a set of neutral atoms and ions with closed shells and performed calculations using Gaussian basis sets. The smallest systems (He, Be and Be$^{2+}$) were described in correlation-consistent polarization valence basis sets (cc-pVXZ) of increasing quality from double to quintuple zeta (X=2-5). For the larger systems, it is necessary to include core-valence functions to describe the correlation correctly. For these systems the correlation-consistent polarization core-valence (cc-pCVXZ) basis sets with increasing X were used. The basis sets for the charged species were obtained from the neutral-atom basis sets by rescaling according to
\begin{eqnarray}
\phi^i_{X^{y}}(r) \propto \phi^i_{X}\left(r\frac{Z}{Z-y}\right),
\end{eqnarray}
where $Z$ is the central charge of the nucleus and $y$ is the total charge of the ion. All calculations were performed without frozen-core approximation, in a full active space. Since the FRPA calculations are done in a non-relativistic framework, the experimental values need to be corrected for relativistic effects. This has been studied in Ref.~\cite{Pernpointner2004}. Since both ADC(3) and FRPA are of comparable level of theory, we assume that the relativistic corrections of ADC(3) are also applicable to the FRPA values. Both the FTDA and FRPA results were calculated with self-consistency on the Hartree-Fock diagram. This means that all results in this section are in fact equivalent to the FTDAc and FRPAc results defined in the next section.

\subsection{Convergence of the atomic calculations}

\Tab{results:atom1} shows total binding energies for Ne in the cc-pCVDZ basis and for Be in four increasingly large basis sets. The results obtained in FRPA are shown together with values calculated in the FTDA, CC with singles and doubles (CCSD) and FCI for comparison. For the Ne atom, all three methods are in agreement with the FCI result, showing only a small correction of less than $2~\mathrm{mH}$. The discrepancy from the exact result of FTDA is halved by going to FRPA. The Be atom represents a hard test case for electronic structure methods. Due to the near degeneracy of the $2s$ and $2p$ orbitals, its ground state is very badly described by a single Slater determinant. As a result the RPA will be driven to an instability in the $S=1$ channel. For this channel alone, the RPA eigenvalues and eigenvectors are replaced with their TDA counterparts. All other channels are calculated at the RPA level of theory. The results show that the values calculated with FTDA and FRPA are very close. This means that in the case of the Be atom, it is not crucial to use RPA for the screening of the interactions. This should not be surprising as the RPA is not the optimal method for few-body methods since the violation of the Pauli-principle that is inherent to RPA plays a more important role in these systems. It is reassuring that the FRPA and FTDA results are close in this case, implying that one can safely use the FRPA even for smaller systems. Both Green's function methods are systematically off by $9~\mathrm{mH}$ compared to CCSD and FCI. This means that the description of the excitations at the level of RPA is not enough to describe all correlations. More involved methods for the excitations could cure this behavior.

\begin{table}[H]
\begin{tabular*}{\textwidth}{@{\extracolsep{\fill}}lcccccc}
\toprule
$E_{\mbox{tot}}$      &    Ne        &&   \multicolumn{4}{c}{Be} \\
\cmidrule{2-2} \cmidrule{4-7}
&   cc-pCVDZ   &&  cc-pVDZ  & cc-pVTZ  & cc-pVQZ  & cc-pV5Z \\
\midrule
ADC(3)/FTDA         &   -128.7191  &&    -14.6089  &	-14.6154  &   -14.6314  &   -14.6375 \\
FRPA                  &   -128.7210  &&    -14.6084  &	-14.6150  &   -14.6310  &   -14.6371 \\
CCSD                  &   -128.7211  &&    -14.6174  &	-14.6236  &   -14.6396  &   -14.6457 \\
\\
full CI               &   -128.7225  &&    -14.6174  &	-14.6238 &    -14.6401  &   -14.6463\\
\bottomrule
\end{tabular*}
\caption[Total binding energies for Ne and Be]{Total binding energies (in Hartree) for Ne and Be obtained for cc-p(C)VXZ bases
of different sizes. The results obtained with ADC(3) and FRPA (with self-consistency
in the HF diagram) and with CCSD are compared to FCI calculations.} 
\label{tab:results:atom1}
\end{table}

The data in  \Tab{results:atom2} involve the extrapolation of two results according to the formula
\begin{eqnarray}
E_X = E_{\infty} + AX^{-3}\label{equ:results:extrapol},
\end{eqnarray}
where $X$ is the number of zeta functions in the basis set and $E_{\infty}$ is the extrapolated basis set limit. This relation is known to yield good extrapolations for correlation energies\cite{Helgaker2004}. \Tab{results:atom2} indicitates that the extrapolated result changes by less than $2~\mathrm{mH}$ when going from extrapolation between X=T,Q to extrapolation between X=Q and 5. This leads us to decide that the current accuracy suffices to calculate the basis set limit and the deviation serves as an error estimate of the extrapolation. The convergence for the larger atom Mg is slower but the error does not exceed $10~\mathrm{mH}$.

The convergence of ionization energies (IEs) and electron affinities (EAs) tends to be faster than the total binding energies because they represent differences of energies. Inaccuracies are comparable for the $N\pm1$ and the $N$ particle system and they tend to cancel out. This means that \Equ{results:extrapol} still holds for IEs and EAs. This relation is not always right. When shake-up configurations are important, the basis set limit will not obey this relation. Also for smaller basis sets the behavior of the extrapolation may be prone to larger instabilities. An example of both possibilities is given in \Tab{results:atom3}. The $3p$ orbit of Ar has a strong one-hole character and therefore has smooth convergence behavior. The $3s$ orbit of Ar is a notable counterexample. Here the calculated values show an oscillatory behavior when increasing basis set size. This does not exclude smoother convergence when one goes beyond X=5, but this kind of calculations are beyond our current capabilities. In what follows we apply \Equ{results:extrapol} and estimate the errors for the IEs to be of the order of $2~\mathrm{mH}$.

\subsection{Extrapolated results for simple atoms}

\Tab{results:atom4} presents the extrapolated ground-state energies for a series of simple atoms. The formula from \Equ{results:extrapol} was applied using the data from the calculations for the X=Q and X=5 basis sets. For the lighter atoms (He and Be) the cc-pVXZ basis sets were used, while for the larger atoms the cc-pCVXZ basis sets were used. The experimental values have been corrected for relativistic effects along the lines of Ref.~\cite{Pernpointner2004}. The data obtained at the CCSD level for He and Be$^{2+}$ can be regarded as FCI results. Still there is no guarantee that the extrapolation to the basis set limit has fully converged. This explains why these values can be below the empirical values. For these two species, we see that FRPA and FTDA lack about $1~\mathrm{mH}$ of the correlation energy, which corresponds to $2\%$ of the total correlation energy. Be forms a special case as explained earlier. The bad description of the near degeneracy precludes the correct description with the Green's function methods. It should be noted that the error for the CCSD calculation is also largest for Be. This may lead to the conclusion that the basis set used in this calculation does not allow for the correct description of the correlation effects in this case. The FRPA describes at least $99\%$ of the correlation for the larger atoms. Both FRPA and FTDA agree with the experimental results within the error estimate of the extrapolation procedure. It can generally be said that FRPA predicts the ground-state energies at least as well as FTDA ($\sigma_{rms} = 9.5~\mathrm{mH}$). When the deviating results for Be are kept out of the calculation of $\sigma_{rms}$, the FRPA ($\sigma_{rms}=3.4~\mathrm{mH}$) achieves even slightly better accuracy than CCSD ($\sigma_{rms}=4.2~\mathrm{mH}$). 

The extrapolated ionization energies are presented in \Tab{results:atom5}. Again the difference between the light (He and Be) atoms and larger atoms shows. The FTDA results are approximately $1~\mathrm{mH}$ closer to the experimental results than the FRPA results. This is due to the bad few-body behavior of the RPA. The FRPA overestimates the correlations, which results in ionization energies that are too low. The ionization energies of the larger atoms are systematically better reproduced by the FRPA than in the FTDA. This results in a $\sigma_{rms}$ that is $3~\mathrm{mH}$ lower for FRPA than for FTDA. The FTDA and FRPA are drastically better than the second-order results, which is recovered when $R$ is the free propagator. This indicates that one has to describe the selfenergy at least up to third order in perturbation theory correctly to account for the correct behavior of the ionization energies. The FTDA is the lowest level of theory that takes into account mixing of 2h1p states, which is required to describe the higher ionization energies correctly. One such example is the $3s$ orbital of Ar. In second order, there is one peak at an energy $36~\mathrm{mH}$ away from the experimental value. This peak has a spectral strength of $0.81$. The second smaller peak carries a strength of $0.10$ and lies at a higher energy. The inclusion of configuration mixing between 2h1p states shifts both energies downwards and redistributes the spectral strength of the main peak to $0.61$, which corresponds better to the experimental value of $0.51$.

\begin{sidewaystable}[H]
\begin{tabular*}{\textwidth}{@{\extracolsep{\fill}} llccccr}
\toprule
& $E_{\mbox{tot}}$                  & cc-p(C)VDZ  & cc-p(C)VTZ  & cc-p(C)VQZ  & cc-p(C)V5Z  & Experiment \\
\midrule
Be & calc.            &   -14.6084  &   -14.6150  &     -14.6310  &    -14.6371  &    -14.6674 \\
& extrap.          &             &   -14.6178  &     -14.6427  &    -14.6436   \\
\\
Ne & calc.            &   -128.7210 &   -128.8643 &    -128.9079  &   -128.9226  &   -128.9383   \\
& extrap.          &             &   -128.9246 &    -128.9397  &   -128.9381   \\
\\
Mg & calc.            &   -199.8147 &   -199.9507 &    -200.0033  &  -200.0271   &    -200.054  \\
& extrap.          &             &   -200.0080 &    -200.0417  &  -200.0519    \\
\bottomrule
\end{tabular*}
\caption[Convergence of total binding energies in the FRPA]{Convergence of total binding energies (in Hartree) in the FRPA approach. First lines: energies calculated using double (X=D) to quintuple (X=5)
valence orbits basis sets. Second lines: results extrapolated from two consecutive
sets.
The Be atom was calculated with the cc-pVXZ bases, while Ne and Mg were done
using cc-pCVXZ.
The experimental values are from Refs.~\cite{Davidson1991,Chakravorty1996,Martin,McCarthy1989}}
\label{tab:results:atom2}
\end{sidewaystable}

\begin{sidewaystable}[H]
\begin{tabular*}{\textwidth}{@{\extracolsep{\fill}} llccccr}
\toprule
&IE                   & cc-pCVDZ  & cc-pCVTZ  & cc-pCVQZ  & cc-pCV5Z  & Experiment \\
\midrule
Ar (3p) & calc.       &      0.5623 &      0.5695 &       0.5751  &     0.5770	 &   0.579  \\
& extrap.          &             &      0.5725 &       0.5792  &     0.5788    \\
\\
Ar (3s) & calc.       &      1.0985 &      1.0616 &       1.0599  &     1.0622   &   1.075  \\
& extrap.          &             &      1.0461 &       1.0586  &     1.0646    \\
\bottomrule
\end{tabular*}
\caption[Convergence of ionization energies in the FRPA]{Convergence of IEs (in Hartree) in the FRPA approach.
First lines: energies calculated using double (X=D) to quintuple (X=5)
valence orbits basis sets. Second lines: results extrapolated from two consecutive
sets.
The experimental values are from Refs.~\cite{Davidson1991,Chakravorty1996,Martin,McCarthy1989}.} 
\label{tab:results:atom3}
\end{sidewaystable}

\begin{sidewaystable}[H]
\begin{tabular*}{\textwidth}{@{\extracolsep{\fill}} lcccccccccc}
\toprule
&  & Hartree-Fock
&  & FTDA
&  & FRPA
&  & CCSD
&  & Experiment  \\
\midrule
He         & &    -2.8617 (+42.0)  & &  -2.9028 (+0.9)   & &   -2.9029 (+0.8)  & &   -2.9039  (-0.2) & &   -2.9037 \\
Be$^{2+}$  & &   -13.6117 (+43.9)  & &  -13.6559 (-0.3)  & &  -13.6559 (-0.3)  & &  -13.6561  (-0.5) & &  -13.6556 \\
Be         & &   -14.5731 (+94.3)  & &  -14.6438 (+23.6) & &  -14.6436 (+23.8) & &  -14.6522 (+15.2) & &  -14.6674 \\
Ne         & &  -128.5505 (+387.8) & &  -128.9343 (+4.0) & & -128.9381 (+0.2)  & & -128.9353  (+3.0) & & -128.9383 \\
Mg$^{2+}$  & &  -198.83 7 (+444)   & &  -199.226  (-5)   & & -199.228  (-7)    & & -199.225 (-4)     & & -199.221 \\
Mg         & &  -199.616  (+438)   & &  -200.048  (+6)   & & -200.052  (+2)    & & -200.050 (+4)     & & -200.054 \\
Ar         & &  -526.820  (+724)   & &  -527.543  (+1)   & & -527.548  (-4)    & & -527.536 (+8)     & & -527.544 \\
\\
$\sigma_{rms}$ [mH] & &    392     & &     9.5 (3.6)     & &    9.5 (3.4)      & &    6.9 (4.2)      & & \\
\bottomrule
\end{tabular*}
\caption[Extrapolation of total binding energies]{Hartree-Fock, ADC(3)/FTDA, FRPA and CCSD total binding energies (in Hartree) extrapolated from the cc-p(C)VQZ and cc-p(C)V5Z basis sets. He, Be$^{2+}$ and Be where calculated with the cc-pVXZ bases, while cc-pCVXZ bases were used for the remaining atoms. The deviations from the experiment are indicated in parentheses (in mHartree). The experimental energies are from Refs.~\cite{Davidson1991,Chakravorty1996,Martin}. The rms errors in parentheses are calculated by neglecting the Be results.
} 
\label{tab:results:atom4}
\end{sidewaystable}

\begin{sidewaystable}[H]
\begin{tabular*}{\textwidth}{@{\extracolsep{\fill}} llcccccccccc}
\toprule
& & & Hartree-Fock
& & 2$^{nd}$  order
& & FTDA/
& & FRPA
& & Experiment   \\
& & &  & &  & & ADC(3)  & &   & &   \\
\midrule
He         & 1s &  &  0.918 (+14)   &  &  0.9012 (-2.5) &  &  0.9025 (-1.2)  &  &  0.9008 (-2.9)   &  &  0.9037 \\
\\
Be$^{2+}$: & 1s &  &  5.6672 (+116) &  &  5.6542 (-1.4) &  &  5.6554 (-0.2)  &  &  5.6551 (-0.5)   &  &  5.6556 \\
\\
Be         & 2s &  &  0.3093 (-34)  &  &  0.3187 (-23.9) & &  0.3237 (-18.9) &  &  0.3224 (-20.2) &  &  0.3426 \\
& 1s &  &  4.733 (+200)  &  &  4.5892 (+56)  &  &  4.5439 (+11)   &  &  4.5405 (+8)    &  &  4.533 \\
\\
Ne         & 2p &  &  0.852 (+57)   &  &  0.752 (-41)   &  &  0.8101 (+17)   &  &  0.8037 (+11)   &  &  0.793 \\
& 2s &  &  1.931 (+149)  &  &  1.750 (-39)   &  &  1.8057 (+24)   &  &  1.7967 (+15)   &  &  1.782 \\
\\
Mg$^{2+}$: & 2p &  &  3.0068 (+56.9)&  &  2.9217 (-28.2) & &  2.9572 (+7.3)  &  &  2.9537 (+3.8)  &  &  2.9499 \\
& 2s &  &  4.4827        &  &  4.3283        &  &  4.3632         &  &  4.3589         &  &   \\
\\
Mg         & 3s &  &  0.253 (-28)   &  &  0.267 (-14) &  &  0.272 (-9)     &  &  0.280 (-1)     &  &  0.281 \\
& 2p &  &  2.282 (+162)  &  &  2.117 ( -3) &  &  2.141 (+21)    &  &  2.137 (+17)    &  &  2.12  \\
\\
Ar         & 3p &  &  0.591 (+12)   &  &  0.563 (-16)  &  &  0.581 (+2)     &  &  0.579 ($\approx$0)&& 0.579 \\
& 3s &  &  1.277 (+202)  &  &  1.111 (+36)  &  &  1.087 (+12)    &  &  1.065 (-10)    &  &  1.075 \\
& 3s &  &                &  &  1.840        &  &  1.578          &  &  1.544          &  &     \\
\\
$\sigma_{rms}$ [mH] & & &  81.4  &  & 29.3 &  &     13.7        &  &     10.6    &  & \\ 
\bottomrule
\end{tabular*}
\caption[Extrapolation of ionization energies]{Ionization energies obtained with Hartree-Fock, second-order perturbation theory for the selfenergy (plus the cHF term), FTDA and with the full Faddeev-RPA (in Hartree). All results are extrapolated from the cc-p(C)VQZ and cc-p(C)V5Z basis sets. The deviations from the experiment (indicated in parentheses) and the rms errors are given in mHartree. The experimental energies are from Ref.~\cite{McCarthy1989,NIST, Thompson2001}.
} 
\label{tab:results:atom5}
\end{sidewaystable}

It can be concluded that both FTDA and FRPA yield very similar results for the lightest atoms, while for the larger systems, going from TDA to RPA for the pp and ph channel leads to a systematic improvement of the ionization energies. The accuracy of the FRPA for small systems gives confidence that the method can be used even for few-body systems. The extrapolation to the basis set limit shows that the FRPA yields the same accuracy as in smaller basis sets and does not lead to overcorrelation.

\section{Molecular results}

After calculations for nuclei and atoms, the next logical step in the study of the FRPA are calculations for molecules. The molecules that were chosen all have a closed shell structure and their ground state is a spin singlet. Due to this singlet ground state, the results are independent of the third component of the spin. Spin angular momentum algebra can be applied to all matrices, which improves the computational scaling. The calculations were again done in Gaussian basis sets\cite{BSE}. With the exception of H${_2}$O, all molecules studied here are linear molecules. This means that there is only one parameter fixing the external potential: the nuclear separation.  The calculation is facilitated through the conservation of the rotational quantum number around the nuclear axis. The resulting reduction of matrix size yields a considerable speed-up of the calculations and a lower memory usage. The results are compared to CCSD(T) and to FCI or experiments where available.

\subsection{Ground-state properties}

\Tab{results:molecule1} and \Tab{results:molecule2} present the ground-state properties of a series of molecules which are ordered according to increasing correlation energy. These ground-state properties were obtained by performing a scan of different separation lengths (and angles in the case of H$_2$O) around the equilibrium geometry for each molecule. A third-order polynomial was then fitted to these data points, after which the minimum was located. It is this minimum that is tabulated here. The calculations were performed in a cc-pVDZ basis set. The FCI results were calculated with the geometry from the FRPAc fit, which within the accuracy obtained here also corresponds to the CCSD(T) fit. The CCSD(T) ionization energies were derived by performing two calculations, one for the $N$ particle system and one for the $N-1$ particle system, subsequently subtracting both results. The CCSD(T) results for H$_2$ are equivalent to FCI. The FTDA and FRPA columns present the results of the procedures without performing the iterative calculation of the HF-diagram. This corresponds to the first iteration in the procedure for the calculation of FTDAc and FRPAc, which do include the Hartree-Fock diagram with the correlated density matrix. 

The ground-state energies for \Tab{results:molecule1} are very similar for FTDAc and FRPAc, differing by at most $4~\mathrm{mH}$. The deviations are such that the FRPAc is closer to the FCI results where available. The FRPAc results are generally close to the values obtained with CCSD(T), with a maximum deviation of $5~\mathrm{mH}$ for H$_2$O. The bond lengths obtained with FTDAc, FRPAc and CCSD(T) are of the same quality when compared to the experimental values. The Green's function methods are slightly better at describing the first ionization energies. The importance of the iterative calculation of the HF-diagram is clear from this table. The results are systematically driven closer to the reference values by adding this diagram, for both FTDA and FRPA.

The deviations between FTDAc and FRPAc for the ground-state energies are of the order of $10~\mathrm{mH}$. The FRPAc ground-state energies are generally closer to the CCSD(T) results, which will be regarded as the reference here. Calculating the geometry of C$_2$H$_2$ forms a problem for the Green's function methods, while in the other cases they predict geometries close to the experiment. It should be emphasized that for these molecules the presence of the iterative HF-diagram plays an even greater role. For the calculation of C$_2$H$_2$, N$_2$ and CO$_2$ it is even the case that no minimum is found in the potential energy curve without this diagram. This does not mean that there are no results, only that the dissociation curve has the wrong shape. For the other two molecules, self consistency tends to adjust the results toward experiment.

\subsection{Ionization energies}

To make the comparison with previously published ADC(3) calculations\cite{Trofimov2005}, \Tab{results:molecule3} has an extra column with these results. This column agrees well with the values obtained with the FTDAc. The small differences are already present in the reference HF calculations. The basis set used for this table was the augmented-cc-pVDZ (aug-cc-pVDZ) basis set. The FRPAc and FTDAc are of comparable accuracy. The $1\sigma_u$ level of N$_2$ is an exception, FRPA and FRPAc are off by about $0.8~\mathrm{eV}$, indicating that in this case the RPA is driven close to instability for this channel, which affects the results negatively.

\subsection{Basis set convergence}

In \Tab{results:molecule4} the results for different Gaussian basis sets are shown for the HF molecule. The differences in ionization energies between the basis sets with double zeta functions and those with triple zeta functions are of the order of $0.75~\mathrm{eV}$ for the non-augmented and $0.25~\mathrm{eV}$ for the augmented basis sets. The convergence behavior of the ground-state energies calculated with the FRPA and FRPAc is very similar to that of CCSD(T) and MP3. The convergence of the FRPAc is faster, again indicating that the self-consistency for the HF-diagram is desirable. The ground-state energy in the largest basis set aug-cc-pVTZ obtained with the FRPAc is very close to the CCSD(T) result and outperforms the value from FRPA and MP3.

\begin{sidewaystable}[H]
\begin{tabular*}{\textwidth}{@{\extracolsep{\fill}} c  c  r  r  r  r  r  r  r  }
\toprule
&  & FTDA & FTDAc & FRPA & FRPAc & CCSD(T) & FCI &  \multicolumn{1}{c}{Expt.}\\ 
\midrule

$\mathrm{H}_2$ \\
& $E_0$ & $-1.170$ & $-1.161$ & $-1.170$ & $-1.161$ & $-1.164$ & $-1.164$ & $-1.175$\\
& $r_{H-H}$ & $0.769$ & $0.757$ & $0.770$ & $0.757$ & $0.761$ &  & $0.741$ \\
& I & $16.16$ & $16.03$ & $16.16$ & $16.03$ & $ 16.12 $ & & $16.08$\\

$\mathrm{BeH}_2$\\
& $E_0$ & $-15.855$ & $-15.831$ & $-15.856$ & $-15.832$ & $-15.835$ & $-15.836$ & -\\
& $r_{Be-H}$ & $1.374$ & $1.337$ & $1.383$ & $1.337$ & $1.339$ & & $1.340$ \\
& I & $11.89$ & $11.78$ & $11.84$ & $11.76$ & $11.89$ & & -\\

$\mathrm{HCl}$\\
& $E_0$ & $-460.295$ & $-460.256$ & $-460.293$ & $-460.255$ & $-460.254$ & & -\\
& $r_{H-Cl}$ & $1.314$ & $1.297$ & $1.314$ & $1.293$ & $1.290$ & & $1.275$ \\
& I & $12.44$ & $12.24$ & $12.44$ & $12.24$ & $12.26$ & & -\\

$\mathrm{HF}$\\
& $E_0$ & $-100.175$ & $-100.224$ & $-100.173$ & $-100.228$ & $-100.228$ & $-100.231$ & -\\
& $r_{H-F}$ & $0.904$ & $0.916$ & $0.897$ & $0.913$ & $0.920$ & & $0.917$ \\
& I & $15.70$ & $15.70$ & $15.56$  & $15.54$ & $ 15.42 $ & & $16.12$\\

$\mathrm{H_2O}$ \\
& $E_0$ & $-76.248$  & $-76.240$ & $-76.243$ & $-76.236$ & $-76.241$ & & -\\
& $r_{H-O}$ & $0.986$ & $0.964$ & $0.981$ & $0.962$ & $0.967$ & & $0.958$\\
& $\Lambda_{O-H-O}$ & $101$ & $102$ & $100$ & $102$ & $102$ & & $104$\\
& I & $12.07$ & $12.15$ & $12.25$ & $12.21$ & $11.94$ & & $12.61$\\

\bottomrule
\end{tabular*}
\caption[Ground state properties of small molecules with correlation energy up to $200~\mathrm{mH}$]{FRPA results for a set of small molecules with a correlation energy up to $200~\mathrm{mH}$ in a cc-pVDZ basis set. The ground-state energy $E_0$ is given in Hartree, the ionization energy $\mathrm{I}$ in electronvolt, equilibrium bond distances are in Angstrom and the equilibrium angles in degrees. FRPA and FTDA refer to the calculations after the first iteration, while FRPAc and FTDAc refer to the calculations where consistency at the Hartree-Fock level was applied. The calculated data are compared to the Coupled Cluster method at the level of CCSD(T) and to experimental data or exact calculations taken from Ref. \cite{CCCBDB}. The FCI energies were calculated at the FRPAc geometry.}
\label{tab:results:molecule1}
\end{sidewaystable}

\begin{sidewaystable}[H]
\begin{tabular*}{\textwidth}{@{\extracolsep{\fill}} c  c  r  r  r  r  r  r  r  }
\toprule
&  & FTDA & FTDAc & FRPA & FRPAc & CCSD(T) & FCI &  \multicolumn{1}{c}{Expt.}\\ 
\midrule

$\mathrm{BF}$\\
&$E_0$ & $-124.331$ & $-124.365$ & $-124.332$ & $-124.368$ & $-124.380$ & & -\\
& $r_{B-F}$ & $1.285$ & $1.284$ & $1.305$ & $1.285$ & $1.295$ & & $1.267$ \\
& I & $11.35$ & $10.75$ & $11.73$ & $10.94$ & $11.01$ & & -\\

$\mathrm{C_2H_2}$\\
& $E_0$ & - & $-77.102$ & - & $-77.093$ & $-77.111$ & & -\\
& $r_{C-C}$ & - & $1.298$ & - & $1.298$ & $1.232$ & & $1.203$\\
& $r_{C-H}$ & - & $1.083$ & - & $1.080$ & $1.081$ & & $1.063$\\
& I & - & $11.26$ & - & $11.14$ & $11.08$ & & $11.49$\\

$\mathrm{CO}$\\
& $E_0$ & $-113.096$  & $-113.037$ & $-113.100$ & $-113.048$ & $-113.055$ & & -\\
& $r_{C-O}$ & $1.140$ & $1.130$ & $1.133$ & $1.123$ & $1.145$ & & $1.128$\\
& I & $14.39$ & $13.69$ & $14.23$ & $14.44$ & $13.64$ & & $14.01$\\

$\mathrm{N}_2$\\
& $E_0$ & - & $-109.258$ & - & $-109.272$ & $-109.276$ & & -\\
& $r_{N-N}$ & - & $1.104$ & - & $1.106$ & $1.119$ & & $1.098$ \\
& I & - & $15.37$ & - & $14.80$ & $15.05$ & & $15.58$ \\

$\mathrm{CO_2}$\\
& $E_0$ & -  & $-188.139$ & - & $-188.134$ & $-188.148$ & & -\\
& $r_{C-O}$ & - & $1.162$ & - & $1.162$ & $1.175$ & & $1.162$\\
& I & - & $13.25$ & - & $13.42$ & $13.26$ & & $13.78$\\

\bottomrule
\end{tabular*}
\caption[Ground state properties of small molecules with correlation energy higher than $200~\mathrm{mH}$]{FRPA results for a set of small molecules with a correlation energy higher than $200~\mathrm{mH}$ in a cc-pVDZ basis set. The ground-state energy $E_0$ is given in Hartree, the ionization energy $\mathrm{I}$ in electronvolt, equilibrium bond distances are in Angstrom and the equilibrium angles in degrees. FRPA and FTDA refer to the calculations after the first iteration, while FRPAc and FTDAc refer to the calculations where consistency at the Hartree-Fock level was applied. The calculated data are compared to the Coupled Cluster method at the level of CCSD(T) and to experimental data or exact calculations taken from Ref. \cite{CCCBDB}. The FCI energies were calculated at the FRPAc geometry.}
\label{tab:results:molecule2}
\end{sidewaystable}

\begin{sidewaystable}[H]
\begin{tabular*}{\textwidth}{@{\extracolsep{\fill}} l c c c c c c c c}
\toprule
& & HF & FTDA & FTDAc & ADC(3) & FRPA & FRPAc & Expt.\\
&  & & & & & & \\
\midrule
$\mathrm{HF}$ & & & & & & &\\
& 1$\pi$ &                 17.17 & 16.22 &  16.46 & 16.48 &  16.05 &  16.35 &  16.05\\
& 3$\sigma$ &              20.98 & 20.14 &  20.33 & 20.36 &  20.03 &  20.24 &  20.0\\
$\mathrm{CO}$ & & & & & & & \\
& 5$\sigma$ &              15.10 & 14.48 &  13.88 & 13.94 &  14.37 &  13.69 &  14.01\\
& 1$\pi$ &                 17.44 & 17.02 &  16.93 & 16.98 &  16.95 &  16.84 &  16.91\\
& 4$\sigma$ &              21.99 & 20.05 &  20.11 & 20.19 &  19.46 &  19.59 &  19.72\\
$\mathrm{N}_2$ & & & & & & &\\
& 3$\sigma_g$ &            17.25 & 16.14 &  15.65 & 15.72 &  15.76 &  15.18 &  15.60\\
& 1$\pi_u$ &               16.73 & 17.20 &  16.82 & 16.85 &  17.71 &  17.14 &  16.98\\
& 2$\sigma_u$ &            21.25 & 19.35 &  18.99 & 19.06 &  18.29 &  17.90 &  18.78\\
$\mathrm{H_2O}$ & & & & & & &\\
& 1$b_1$ &                 13.86 & 12.80 &  12.83 & 12.86 &  12.62 & 12.67  & 12.62\\
& 3$a_1$ &                 15.93 & 15.06 &  15.11 & 15.15 &  14.91 & 14.98  & 14.74\\
& 1$b_2$ &                 19.56 & 19.15 &  19.19 & 19.21 &  19.06 & 19.13  & 18.51\\
\\
& $\bar{\Delta}$ (eV) &   1.26 (1.14) &  0.34 (0.31) & 0.27 (0.28) & 0.30 (0.30) & 0.25 (0.23)  &  0.31 (0.26)\\
& $\Delta_{max}$ (eV)&          2.47 (2.27) &  0.64 (0.64) & 0.68 (0.68) & 0.70 (0.70) & 0.73 (0.73)  &  0.88 (0.62)\\
\bottomrule
\end{tabular*}
\caption[Ionization energies of small molecules]{Ionization energies in electronvolt calculated in the aug-cc-pVDZ basis set. The geometry was taken at the experimental value (see \Tab{results:molecule1} and \Tab{results:molecule2}). The last two rows show the mean absolute deviation and maximum absolute deviation compared to experiment. The values between braces are calculated without the $2\sigma_u$-level of N$_2$. The column labeled ADC(3) represents the ADC(3) results from Ref.~\cite{Trofimov2005}. Experimental values are from Ref.~\cite{Trofimov2005,Kimura1981}.}
\label{tab:results:molecule3}
\end{sidewaystable}

\begin{table}[H]
\begin{tabular*}{\textwidth}{@{\extracolsep{\fill}} l c c c c c c}
\toprule
& & cc-pVDZ & aug-cc-pVDZ & cc-pVTZ & aug-cc-pVTZ & Expt.\\ 
Method & Level \\
\midrule
FRPA \\
&$E_0$ &                   -100.172 &  -100.106 &  -100.335 &  -100.305 & -\\
& 1$\pi$ &                 15.46 &  16.05    &  16.19    &  16.33    &  16.11\\
& 3$\sigma$ &              19.56 &  20.03    &  20.05    &  20.22    &  20.00\\
\\
FRPAc\\
&$E_0$  &                  -100.228 &  -100.261 &   -100.346   &  -100.357 & -\\
& 1$\pi$ &                 15.54 &  15.35    &  16.16       &  16.41    &  16.11\\
& 3$\sigma$ &              19.54 &  20.24    &  20.00       &  20.27    &  20.00\\
\\
CCSD(T)\\
&$E_0$ &                   -100.228 & -100.264  & -100.338  & -100.350 & -   \\
& 1$\pi$ &                 15.44    &  16.06    & 15.96     &  16.16&  16.11\\
MP3\\
&$E_0$ &                   -100.224 & -100.256  & -100.330  & -100.340 & - \\
& 1$\pi$ &                 15.42    & 15.99     & 15.88     & 16.04 & 16.11\\
\bottomrule
\end{tabular*}
\caption[Basis set convergence of molecular ionization energies]{Ground-state energies in Hartree and vertical ionization energies in electronvolt for $\mathrm{HF}$, calculated in different basis sets. The geometry was taken at the experimental value of $0.917$~Angstrom. Experimental values are from Ref.~\cite{Trofimov2005,Kimura1981}, CCSD(T) and MP3 calculations were performed at the same geometry.}
\label{tab:results:molecule4}
\end{table}

\subsection{Example of the pole structure}

\begin{figure}
\begin{center}
\includegraphics[scale=0.65,clip=true]{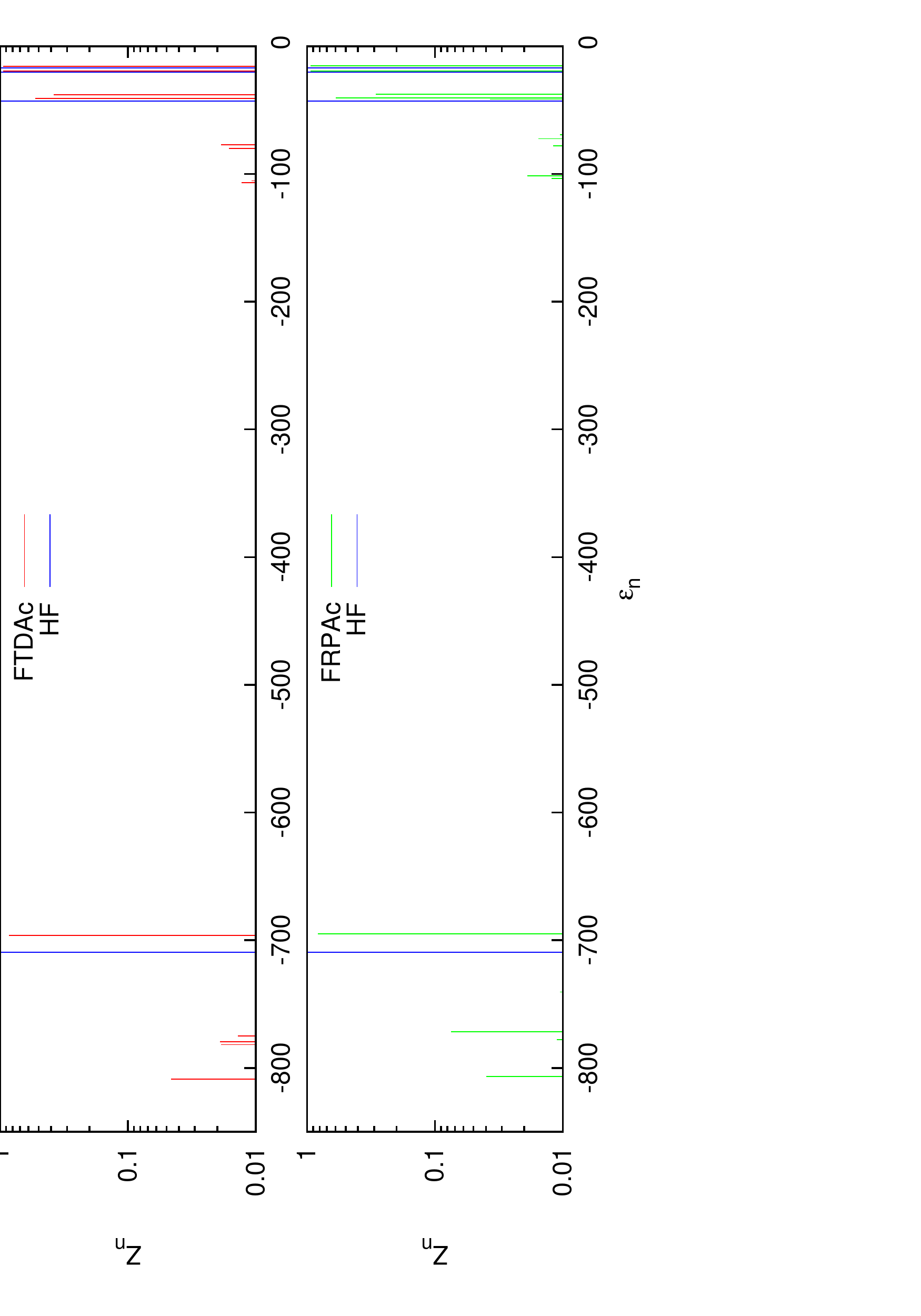}
\caption[Pole structure of the Green's function for HF]{
   \label{fig:results:poles}
   The pole structure of the Green's function for HF in Hartree-Fock, FTDAc and FRPAc. The energy axis is in eV.
}
\end{center}
\end{figure}

\Fig{results:poles} presents a plot of the spectral strength of the Green's function for the negative energy region of HF, calculated in the cc-pVDZ basis set. Only peaks with a spectral strength higher than $0.05$ are retained. This plot shows the fragmentation of the deeper lying peak corresponding with the core orbitals of F. There are two peaks close to the Fermi-level that are almost unaltered compared to Hartree-Fock. Their spectral strength is around $95\%$ in both FRPAc and FTDAc. The main difference with Hartree-Fock occurs for the peak around $-40~\mathrm{eV}$. This peak is split up into two separate peaks with reduced strength. For FRPA this is a peak of $0.60$ and one of $0.30$, while FTDA gives a peak of $0.50$ and $0.40$. There are also some satellites at lower energies.

The calculations at equilibrium geometry indicate that FRPAc yields results that are of similar quality as those obtained with FTDAc, just as for atoms. While RPA was developed for extended systems, it seems that FRPA is can be applied to both small and more extended molecules. The self-consistent treatment of the HF-diagram is mandatory to obtain good results.

\begin{figure}[h]
\begin{center}
\includegraphics[scale=1.0,clip=true]{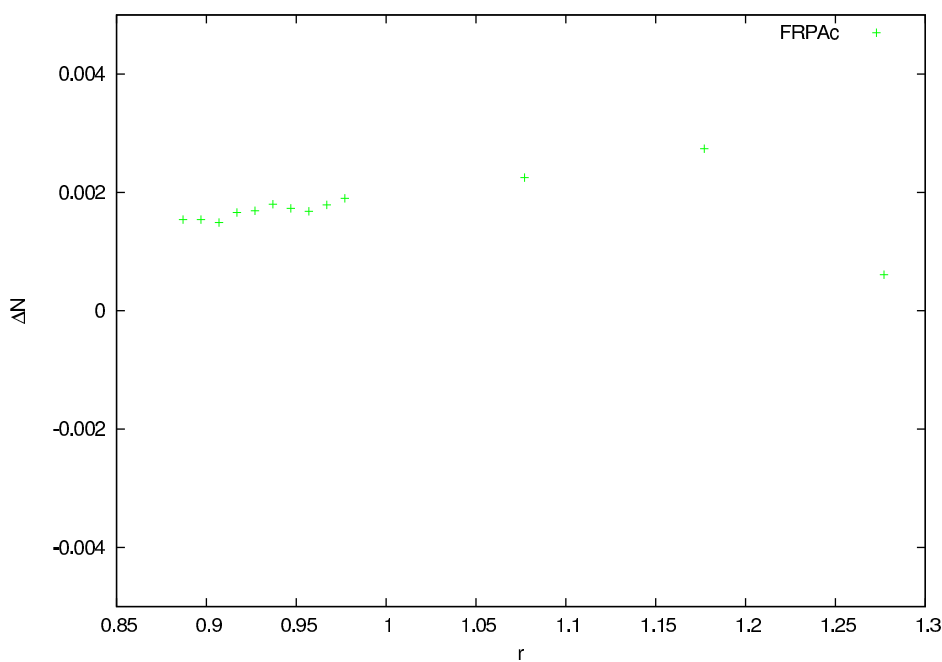}
\caption[Dissociation curve for HF]{
   \label{fig:results:partnumb}
   The deviation of the particle number for HF in the cc-pVDZ basis set. The deviation $\Delta N$ is given relative to the physical particle number and the bond length $r$ is in Angstrom.
}
\end{center}
\end{figure}

A point that deserves some attention is the conservation of particle number as predicted by the Baym-Kadanoff theorem. Since we are using only partial self consistency, it is to be expected that the particle number will not be exactly equal to the physical value. This can be seen from \Fig{results:partnumb}, where the deviation from the real particle number is plotted in function of the separation distance of the nuclei. The deviation is of the order of $10^{-3}$ and is not varying much in the neighborhood of the equilibrium. The deviations are more strongly oscillating in the dissociation limit where the behavior of the RPA is getting worse. Surprisingly, there is almost no difference between FRPA and FRPAc in this respect and hence only the FRPAc results are shown in the figure.

\section{Dissociation limit}

\begin{figure}[h]
\begin{center}
\includegraphics[scale=1.0,clip=true]{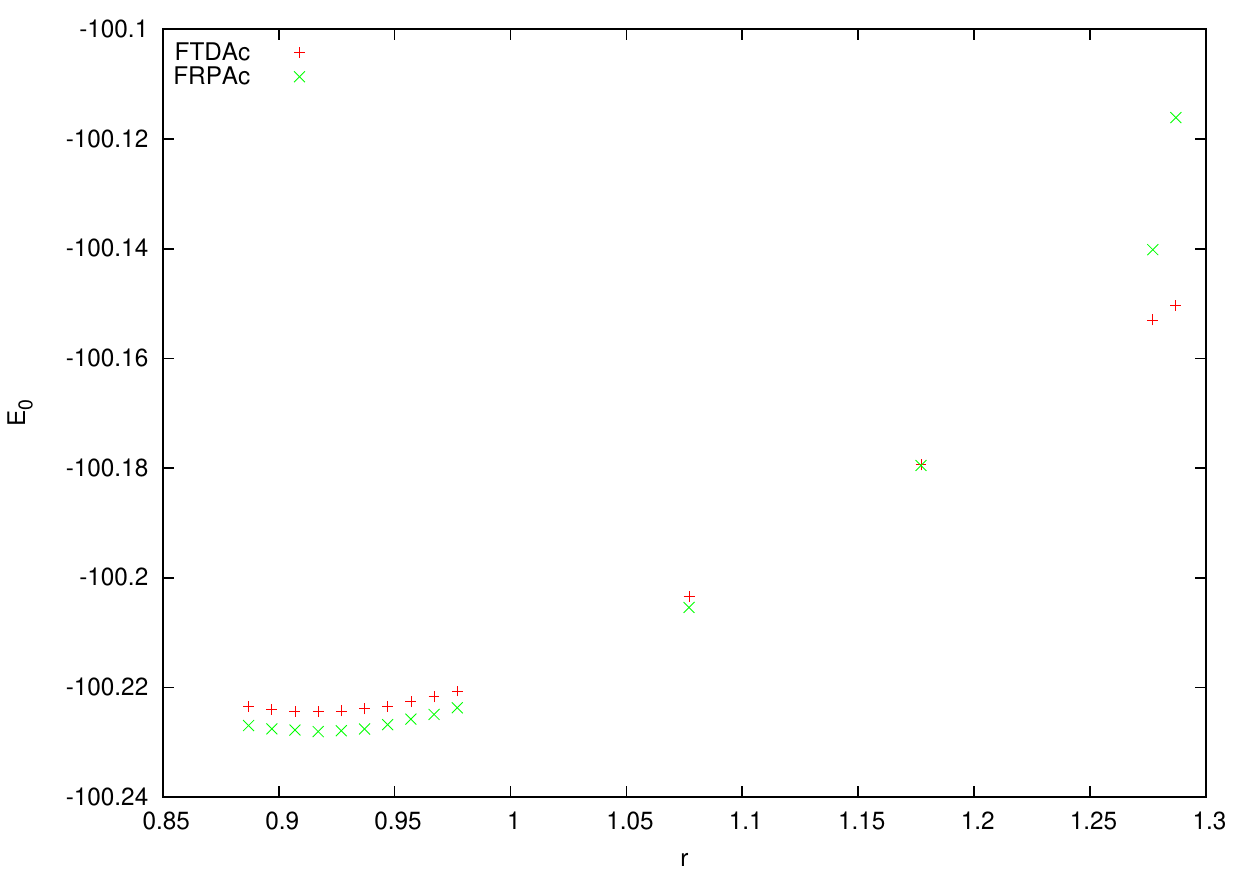}
\caption[Dissociation curve for HF]{
   \label{fig:results:dissoHF}
   The dissociation curve for HF in the cc-pVDZ basis set. The total energy $E_0$ is in Hartree and the bond length $r$ is in Angstrom.
}
\end{center}
\end{figure}

For chemical purposes, it is interesting to look at the dissociation behavior of a method. In the disscociation limit, the constituents of the molecule are placed far apart. A size-extensive method yields the same result in this limit as when the two constituents are calculated separately. \Fig{results:dissoHF} shows the dissociation curve of HF calculated with FRPAc and FTDAc in the cc-pVDZ basis set. Around the experimental equilibrium geometry, the FRPA and FTDA results are very similar as seen from \Tab{results:molecule1}. At larger bond lengths, however, the FRPA energy starts to diverge from the FTDA result. At a bond length of $1.287~\mathrm{\AA}$, the FRPA result even stops to give a result. This is not due to a numerical instability in the calculation of the $\epsilon^{(2p1h)}$  or $\epsilon^{(2h1p)}$ but to an instability in the ph RPA. The closer to this point of instability of the RPA, the less the RPA describes what physically happens. As a result the FRPA, that heavily depends on the quality of the intermediary states, starts to diverge and predicts wrong ground-state energies. FTDA continues to give qualitatively correct results. This is a trend that occurs for every molecule in the test set. Moving away from equilibrium, there is always an RPA instability at some point that prohibits the calculation of FRPA values.

This behavior can be understood from calculations for H$_2$ in a minimal basis set and is known as the "triplet instability"\cite{Thouless1961,Cizek1967}. It is not so much a problem of RPA as of the HF wave function. In the dissociation limit, the HF ground state becomes unstable with respect to ph excitations. The Hartree-Fock stability matrix is nothing more than the matrix in the left-hand side of \Equ{methods:rpaab}. When one of the eigenvalues of this matrix becomes zero, the RPA eigenvalues reach a bifurcation and become complex. The inadequacy of the HF ground state can be easily seen from an analysis of H$_2$ in a minimal basis set. The spatial wave functions are $1s$ functions centered on the H-atoms A and B. These can be put in a bonding and anti-bonding combination for the molecular orbitals. This is also the definition of the particle and hole state of the Hartree-Fock solutions
\begin{eqnarray}
|h) = |b) &=& \frac{1}{\sqrt{2}}\left(|A) + |B)\right)\\
|p) = |a) &=& \frac{1}{\sqrt{2}}\left(|A) - |B)\right).
\end{eqnarray}
These states are normalized to one at great separation, which is the case of interest to us. The restricted Hartree-Fock ground state is the spatially symmetric state with positive parity and spin $S=0$
\begin{eqnarray}
|\Phi_0\rangle &=& |bb) \frac{1}{\sqrt{2}}\left(|\uparrow\downarrow)-|\downarrow\uparrow)\right)\\
&=& \left[a_{b}^{\dagger} \otimes a_{b}^{\dagger}\right]^{S=0}_{M_S=0}|0\rangle.
\end{eqnarray}
Particle-hole excitations can only be formed by removing a particle from a bonding state and replacing it with a particle in an anti-bonding state. The excitations can be split up in a spin triplet and a spin singlet. The energy of the triplet state
\begin{eqnarray}
|\Phi\rangle = \left[a^{\dagger}_{a} \otimes a_{b}^{\dagger}\right]^{S=1}_{M_S} | 0 \rangle
\end{eqnarray}
is found to be
\begin{eqnarray}
\langle \Phi | H | \Phi \rangle &=& (ab|H|ab)-(ab|H|ba).
\end{eqnarray}
When the A and B H-atoms are separated by a large distance, the overlap between the wave functions becomes neglegible and the energy of the triplet state reduces to
\begin{eqnarray}
\langle \Phi | H | \Phi \rangle &=& (AB|H|AB)-(AB|H|BA).
\end{eqnarray}
This is the energy one would expect for the dissociation state where each H-atom receives one electron. It is the physically correct solution which restricted HF fails to describe. So the energy of the triplet state will always be lower than that of the restricted HF singlet ground state.

\begin{figure}[h]
\begin{center}
\includegraphics[scale=1.0,clip=true]{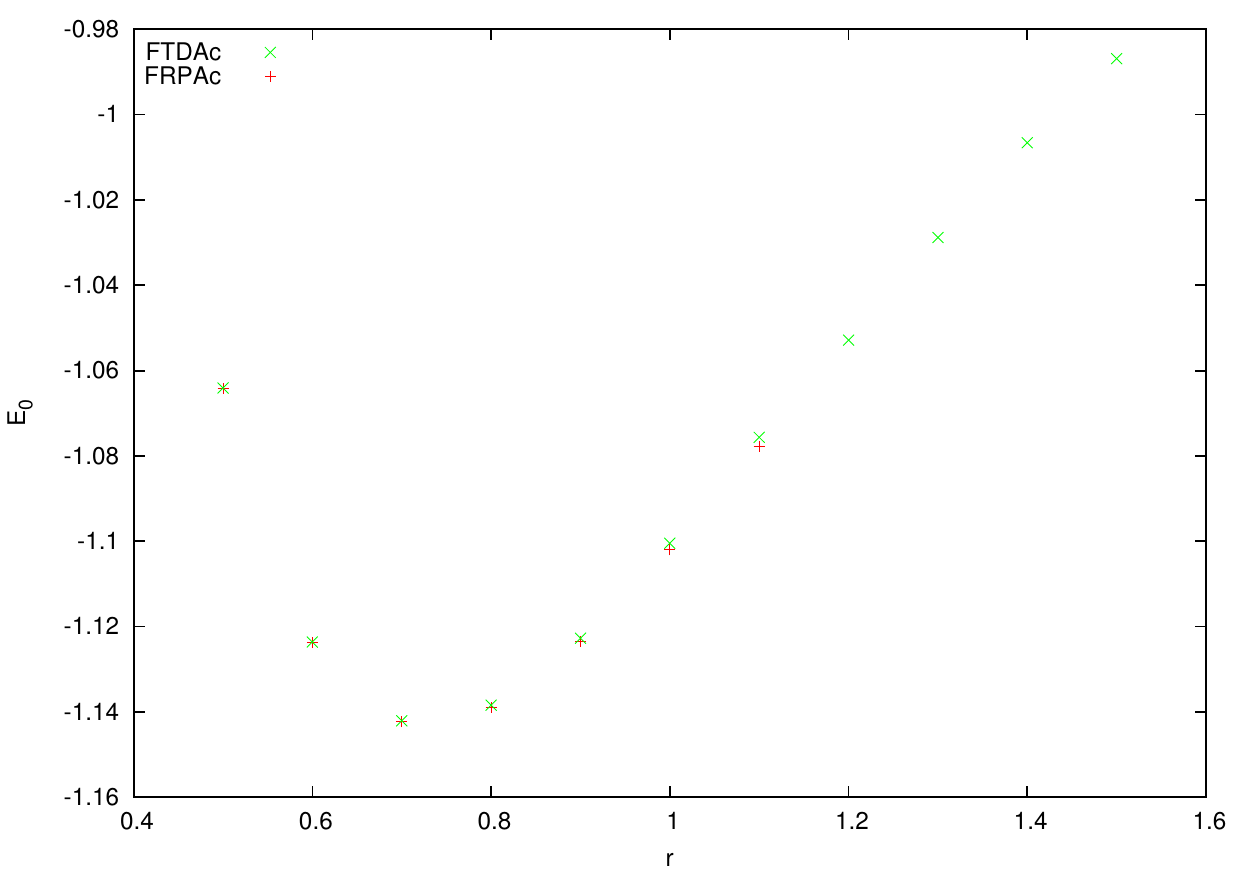}
\caption[Dissociation curve for H$_2$ in a minimal basis set]{
   \label{fig:results:gsenergy}
   The dissociation curve for H$_2$ in the STO-6G basis set. The total energy $E_0$ is in Hartree and the bond length $r$ is in Angstrom.
}
\end{center}
\end{figure}

\begin{figure}[h]
\begin{center}
\includegraphics[scale=1.0,clip=true]{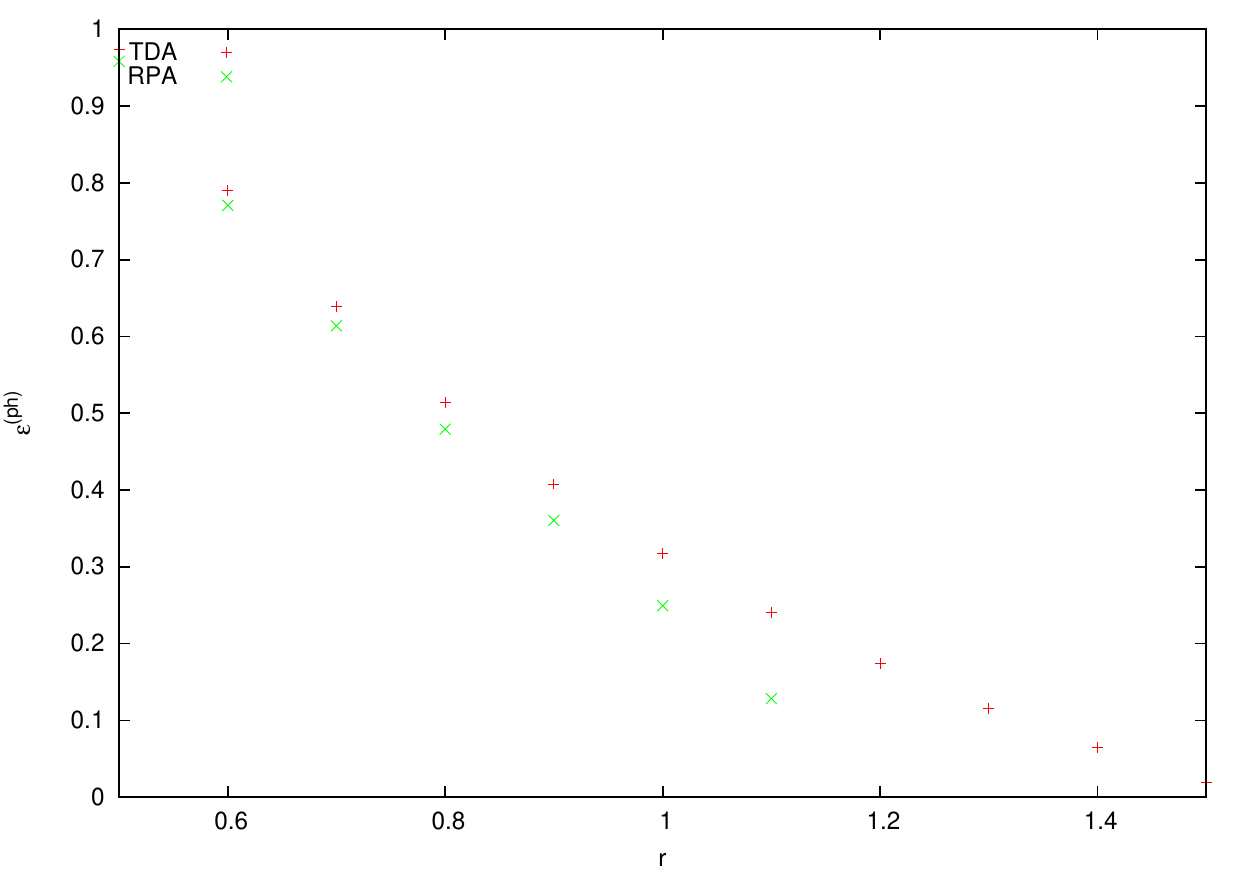}
\caption[Excitation energy for H$_2$]{
   \label{fig:results:excitation}
   The lowest $S=1$ excitation energy as a function of the nuclear separation distance $r$ for H$_2$ in the STO-6G basis set. The excitation energy $\epsilon^{(ph)}$ is in Hartree and the bond length $r$ is in Angstrom.
}
\end{center}
\end{figure}

As a result of the instability of the HF ground state to a ph excitation, the RPA matrix no longer is positive semi-definite. As the separation of the two atoms grows, the lowest $S=1$ RPA eigenvalue approaches zero and then becomes complex. This behavior is plotted in \Fig{results:excitation}. At a separation of $1.2~\mathrm{\AA}$, the excitation energy crosses zero and becomes complex. As a result, there are no FRPAc results beyond this separation distance in \Fig{results:gsenergy}. The TDA does not have this problem. Because of the unconnected positive and negative energy diagonalization problem, a TDA eigenvalue can always be found. However, \Fig{results:excitation} shows that the TDA eigenvalue in the same channel also becomes negative. This means that the Hartree-Fock state should not be used as a reference. It is possible to calculate FTDA values beyond this point even though the HF ground state becomes an invalid starting point for perturbation theory.

\section{The Hubbard molecule}

In condensed matter physics, many efforts have been made to account for the behavior of spin systems. They form a model for the great variety of types of magnetic arrangements that occur in experiments with real solids. Since its introduction\cite{Hubbard1963}, the Hubbard model in particular has been the subject of extensive research. Although it was originally postulated for the study of correlations of d-electrons in transition metals, it is often used for the description of effects such as itinerant magnetism and metal-insulator transitions. At the same time, the repulsive Hubbard model for two sites ("Hubbard molecule") can be used to study the competition between localization and delocalization of electrons in a schematic model. It is this case of strong correlations where the RPA fails. This tool has been used to cure some of the deficiencies of the GW approximation\cite{Romaniello2009} and to study an approach that resums pp and ph interactions\cite{Romaniello2012} with a screened interaction. The physical interpretation, together with the access to analytical results, makes the Hubbard molecule the perfect background to study the dissociation behavior of the FRPA.

The Hubbard molecule is a special case of the general Hubbard Hamiltonian for two sites. The Hamiltonian in the site basis is given by
\begin{eqnarray}
H &=& \sum_{i,\sigma_i} \epsilon_{0} a^{\dagger}_{i\sigma_i}a_{i\sigma_i} - \sum_{i\neq j,\sigma_i} t a^{\dagger}_{i\sigma_i}a_{j\sigma_i}\nonumber\\
&&+ \frac{1}{2}\sum_{i,\sigma_i} U a^{\dagger}_{i\sigma_i}a^{\dagger}_{i-\sigma_i}a_{i-\sigma_i}a_{i\sigma_i},
\end{eqnarray}
where $i$ and $j$ run over the two sites $1$ and $2$ and $\sigma_i$ is the third component of the spin with possibilities $\uparrow$ and $\downarrow$. Note that in the repulsive Hubbard model both $U$ and $t$ are positive. It is advantageous to go over to states with a definite parity and third component of the spin by applying the rotation
\begin{eqnarray}
a_{1\sigma} &=& \frac{1}{\sqrt{2}}\left(a_{+\sigma}+a_{-\sigma}\right)\\
a_{2\sigma} &=& \frac{1}{\sqrt{2}}\left(a_{+\sigma}-a_{-\sigma}\right).
\end{eqnarray}
The matrix elements of the interaction conserve parity and spin projection, which gives the interaction a block-diagonal structure. The matrix of the single-particle part of the Hamiltonian is even diagonal in this basis. We are interested in the H$_2$ molecule-like behavior of the Hamiltonian, so there are two electrons in the system. This corresponds to the Hubbard model at half filling, two electrons for four states. In this sector, the exact wave function is given by
\begin{eqnarray}
|\Psi_0^2\rangle &=& \frac{1}{\sqrt{\left(4t-\sqrt{16t^2+U^2}\right)^2+U^2}}\left(U|+\uparrow+\downarrow\rangle\right.\nonumber\\
&& \left.+ \left(4t-\sqrt{16t^2+U^2}\right)|-\uparrow-\downarrow\rangle\right)\label{equ:results:hubbardgs},
\end{eqnarray}
which has an exact $N=2$ ground-state energy of 
\begin{eqnarray}
E_0^2 &=& 2\epsilon_0 +\frac{U}{2}-\frac{\sqrt{16t^2+U^2}}{2}.
\end{eqnarray}
For the exact Green's function, it is important to have the difference of the ground-state energy with the energy of the excited states. The $N+1$ and $N-1$ systems are the $3$ and $1$ electron systems. The two-site Hubbard model with only one electron is a trivial problem. The ground state is the positive parity state with spin projection $\uparrow$ or $\downarrow$. This degeneracy has to be lifted by choosing one of the two projections. The excited states are listed in \Tab{results:hubbard1}.\begin{table}[h]
\begin{center}
\begin{tabular}{c c c}
\toprule
$i$ & $\Psi_i^1$ & $E_i^1$\\
\midrule
1 & $|+\uparrow\rangle$ & $\epsilon_0-t$\\
2 & $|+\downarrow\rangle$ & $\epsilon_0-t$\\
3 & $|-\uparrow\rangle$ & $\epsilon_0+t$\\
4 & $|-\downarrow\rangle$ & $\epsilon_0+t$\\
\bottomrule
\end{tabular}
\end{center}
\caption[Hubbard molecule $N=1$]{Ground and excited states of the Hubbard molecule in the $1$ electron sector} 
\label{tab:results:hubbard1}
\end{table}

The three-electron system forms the dual problem: one hole has to be located in one of four states. In contrast to the $N=1$ system, the ground state here is formed by a negative parity state. The hole state that has to be accommodated gets a minus sign for its parity so that it is a positive parity state. The excited states for $N=3$ are given in \Tab{results:hubbard2}.

\begin{table}[h]
\begin{center}
\begin{tabular}{c c c}
\toprule
$i$ & $\Psi_i^3$ & $E_i^3$\\
\midrule
1 & $|+\uparrow+\downarrow-\uparrow\rangle$ & $3\epsilon_0-t + U$\\
2 & $|+\uparrow+\downarrow-\downarrow\rangle$ & $3\epsilon_0-t + U$\\
3 & $|-\uparrow-\downarrow+\uparrow\rangle$ & $3\epsilon_0+t + U$\\
4 & $|-\uparrow-\downarrow+\downarrow\rangle$ & $3\epsilon_0+t +U$\\
\bottomrule
\end{tabular}
\end{center}
\caption[Hubbard molecule $N=3$]{Ground and excited states of the Hubbard molecule in the $3$ electron sector} 
\label{tab:results:hubbard2}
\end{table}

\begin{figure}
\begin{center}
\includegraphics[scale=0.5,clip=true]{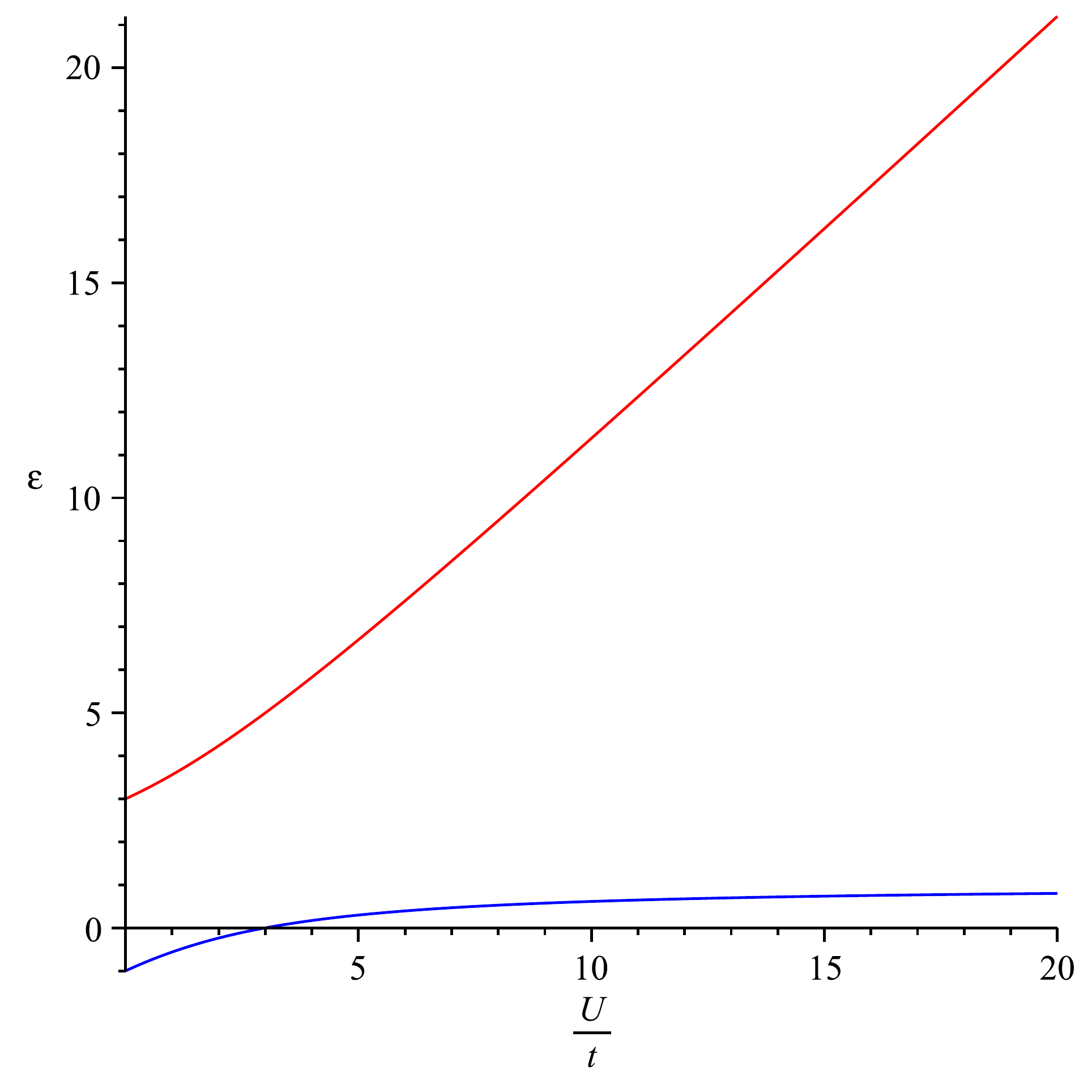}
\caption[Poles of the exact Green's function for the Hubbard molecule]{
   \label{fig:results:hubbardpoles}
   The pole energies for the exact $G_{+\sigma}$ are shown as a function of the reduced interaction strength $\frac{U}{t}$. The offset $\epsilon_0$ is chosen to be zero. The red line represents the forward propagating pole, while the blue line represents the backward propagating pole.
}
\end{center}
\end{figure}

\begin{figure}
\begin{center}
\includegraphics[scale=0.5,clip=true]{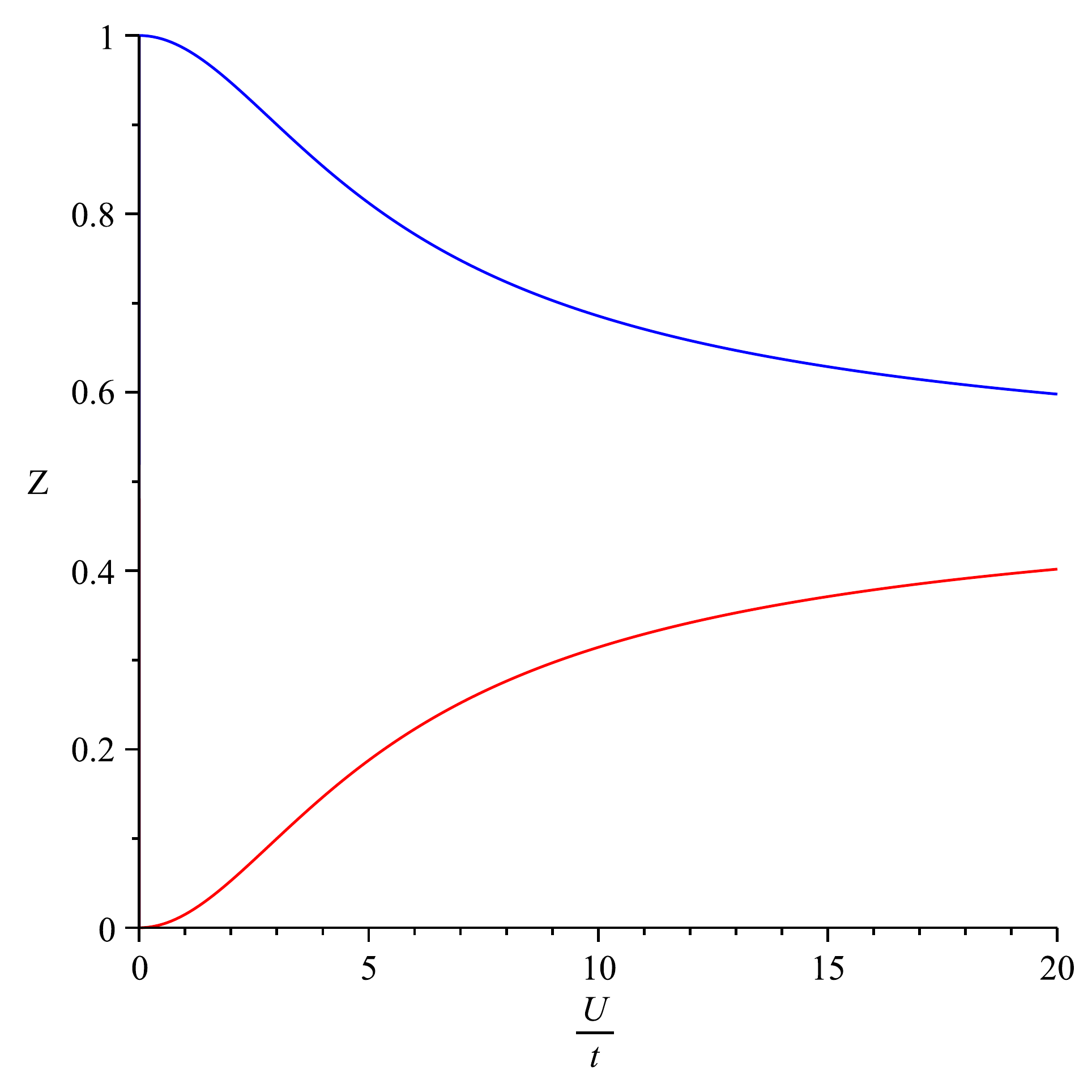}
\caption[Spectral strengths of the exact Green's function for the Hubbard molecule]{
   \label{fig:results:hubbardstrength}
   The spectral strength of the poles from \Fig{results:hubbardpoles} are plotted as a function of the reduced interaction strength $\frac{U}{t}$ for $G_{+\sigma}$. The red line represents the forward propagating pole, while the blue line represents the backward propagating pole. Both strenghts converge to $0.5$ in the limit of infinite interaction. The pole strengths of $G_{-\sigma}$ are the same, but with the forward and backward propagating pole switched.
   }
\end{center}
\end{figure}

After close inspection, it can be seen that the Green's function is diagonal in both parity and spin. The four only components of the Green's function are defined by
\begin{eqnarray}
G_{\pm \sigma} &=& \sum_{n} \frac{|\langle\Psi_0^2|a_{\pm \sigma}|\Psi_n^3\rangle|^2}{E-(E_n^3-E_0^2)+i\eta} + \sum_n  \frac{|\langle\Psi_n^1|a_{\pm \sigma}|\Psi_0^2\rangle|^2}{E-(E_0^2-E_n^1)-i\eta}.
\end{eqnarray}
The numerators are the squares of the coefficients in the definition of the ground-state wave function in \Equ{results:hubbardgs}. The resulting Green's functions are
\begin{eqnarray}
G_{+\sigma} &=& \frac{1}{\left(4t-\sqrt{16t^2+U^2}\right)^2+U^2}\left(\frac{\left(4t-\sqrt{16t^2+U^2}\right)^2}{E-\left(\epsilon_0+t+\frac{U}{2}+\frac{\sqrt{16t^2+U^2}}{2}\right)+i\eta}\right.\nonumber\\
&& \left. + \frac{U^2}{E-\left(\epsilon_0+t+\frac{U}{2}-\frac{\sqrt{16t^2+U^2}}{2}\right)-i\eta}\right)\\
G_{-\sigma} &=&  \frac{1}{\left(4t-\sqrt{16t^2+U^2}\right)^2+U^2}\left(\frac{U^2}{E-\left(\epsilon_0-t+\frac{U}{2}+\frac{\sqrt{16t^2+U^2}}{2}\right)+i\eta}\right.\nonumber\\
&&\left. + \frac{\left(4t-\sqrt{16t^2+U^2}\right)^2}{E-\left(\epsilon_0-t+\frac{U}{2}-\frac{\sqrt{16t^2+U^2}}{2}\right)-i\eta}\right).
\end{eqnarray}
The resulting pole energies are plotted in \Fig{results:hubbardpoles}. In the limit of infinite interaction strength $\frac{U}{t}\rightarrow \infty$, the forward propagating pole exhibits linear behavior, while the backward propagating pole converges to $U$. \Fig{results:hubbardstrength} shows the corresponding spectral strengths. At small interaction, only the backward propagating pole carries any strength. In the limit of higher interaction, both poles converge to a spectral strength of $0.5$. This means that in this limit, they are equally important and the system can only be described by a fragmented single-particle propagator.  

\begin{figure}
\begin{center}
\includegraphics[scale=0.5,clip=true]{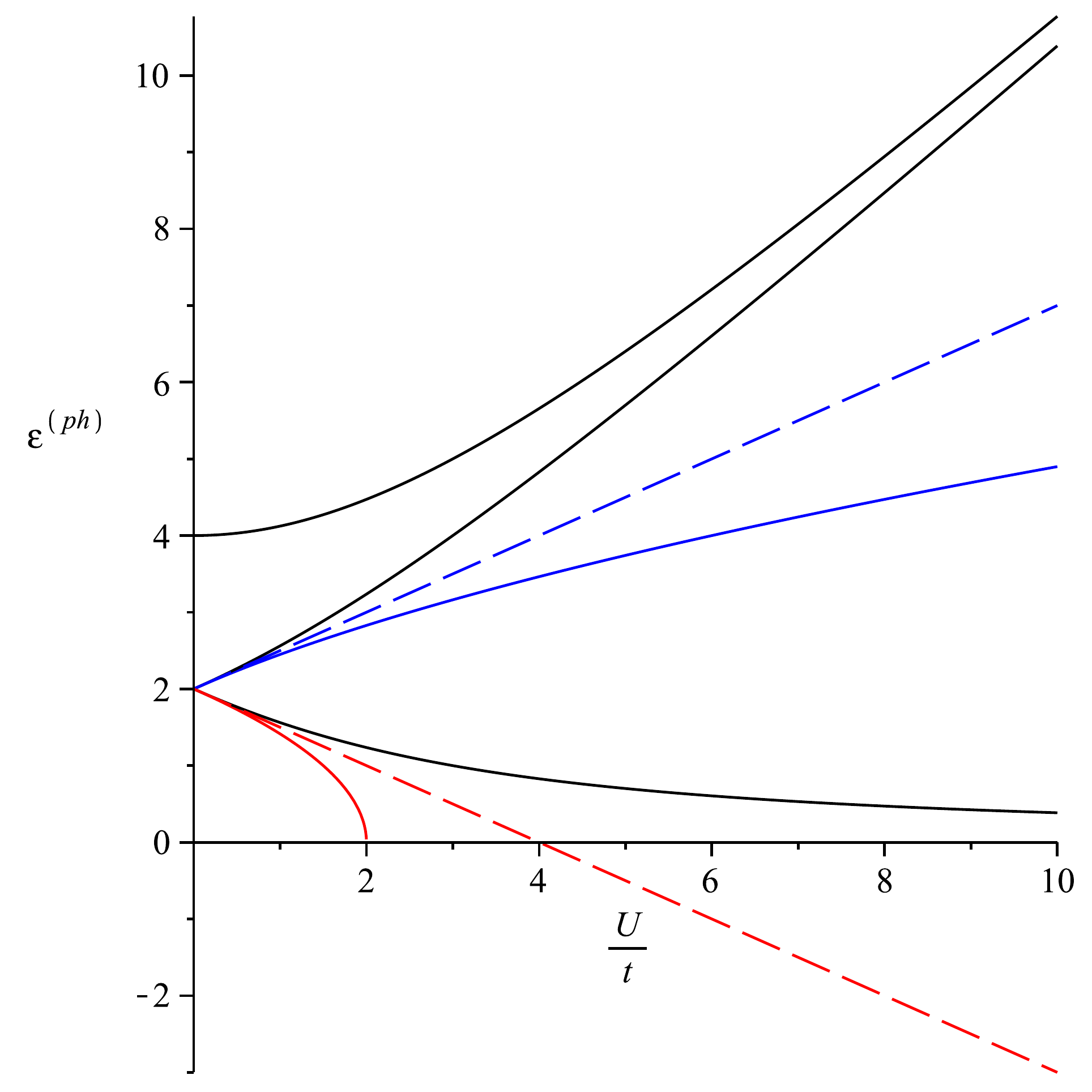}
\caption[ph energies for the Hubbard molecule]{
   \label{fig:results:hubbardph}
   The ph excitation energies for the Hubbard molecule are plotted. The red curves correspond with the $S=1$ solution, while the blue curves are the $S=0$ solutions. Dashed lines are TDA, full lines are RPA. The black lines are the exact excitation energies. It should be noted that the upper exact solution has no counterpart in TDA nor RPA. This is because the parity of this excitation is negative, while this impossible to reach with a TDA or an RPA ph excitation which removes a hole with positive parity and adds a particle with negative parity to the HF ground state.
   }
\end{center}
\end{figure}

To build the RPA matrices, it is necessary to dispose of the Hartree-Fock solution for the $N=2$ system. The hole states are given by the positive parity states, while the negative parity states are found to be the particle states. The HF energies are 
\begin{eqnarray}
\epsilon_h &=& \epsilon_0-t+\frac{U}{2}
\end{eqnarray}
for the hole states and 
\begin{eqnarray}
\epsilon_p &=& \epsilon_0+t+\frac{U}{2}
\end{eqnarray}
for the particle states. With the HF reference state, the ph RPA matrix (\Equ{methods:rpaab}) that needs to be diagonalized is
\begin{eqnarray}
&&\left(\begin{array}{cc}A& B\\-B^* & -A^*\end{array}\right)\nonumber\\
&=&  \left(\begin{array}{cccccccc} 2t-\frac{U}{2} & 0 & & & 0 & -\frac{U}{2} & &  \\ 0 & 2t-\frac{U}{2} & & & -\frac{U}{2} & 0& & \\ & & 2t & \frac{U}{2} & & & 0 & \frac{U}{2}\\ & & \frac{U}{2} & 2t & & & \frac{U}{2} & 0 \\ 0 & \frac{U}{2} & & & -2t+\frac{U}{2} & 0 & & \\\frac{U}{2} & 0& & & 0& -2t+\frac{U}{2} & & \\ & & 0 & -\frac{U}{2} & & & -2t & -\frac{U}{2} \\ & & -\frac{U}{2} & 0 & & & -\frac{U}{2} & -2t\end{array}\right)\nonumber.\\
&&
\end{eqnarray}
The eigenvalues of this matrix are
\begin{eqnarray}
\epsilon^{(ph)>}_0 = \sqrt{4t^2+2tU}\\
\epsilon^{(ph)<}_0 = -\sqrt{4t^2+2tU}\\
\epsilon^{(ph)>}_1 = \sqrt{4t^2-2tU}\\
\epsilon^{(ph)<}_1 = -\sqrt{4t^2-2tU},
\end{eqnarray}
where the solutions are ordered according to their total spin in a singlet and a triplet. It is clear that for the triplet, an imaginary eigenvalue is possible. When the interaction $\frac{U}{t}$ is greater than two, the eigenvalue becomes complex. This does not manifest itself in the case of the TDA eigenvalues
\begin{eqnarray}
\epsilon^{(ph)>}_0 = 2t+\frac{U}{2}\\
\epsilon^{(ph)<}_0 = -2t-\frac{U}{2}\\
\epsilon^{(ph)>}_1 = 2t-\frac{U}{2}\\
\epsilon^{(ph)<}_1 = -2t+\frac{U}{2},
\end{eqnarray}
but at an interaction $\frac{U}{t}$ of four, the excitation energy for the forward propagating pole becomes negative. This means that the excited state is lower in energy than the ground state of the system. This situation is illustrated in \Fig{results:hubbardph}. For reference the exact excitations to the $N$ particle system are given in black. It is clear that the lowest two exact excitations are reproduced in the TDA and RPA. The spin singlet has no instability, while the RPA spin triplet becomes complex at an interaction $\frac{U}{t}=2$ and the TDA triplet becomes negative at an interaction $\frac{U}{t}=4$. The highest excitation has no counterpart in RPA or TDA because these two models do not allow the correct parity. The pp RPA channel does not exhibit this kind of critical behavior. Over the whole parameter range, the eigenvalues stay well behaved.

\begin{figure}
\begin{center}
\includegraphics[scale=0.5,clip=true]{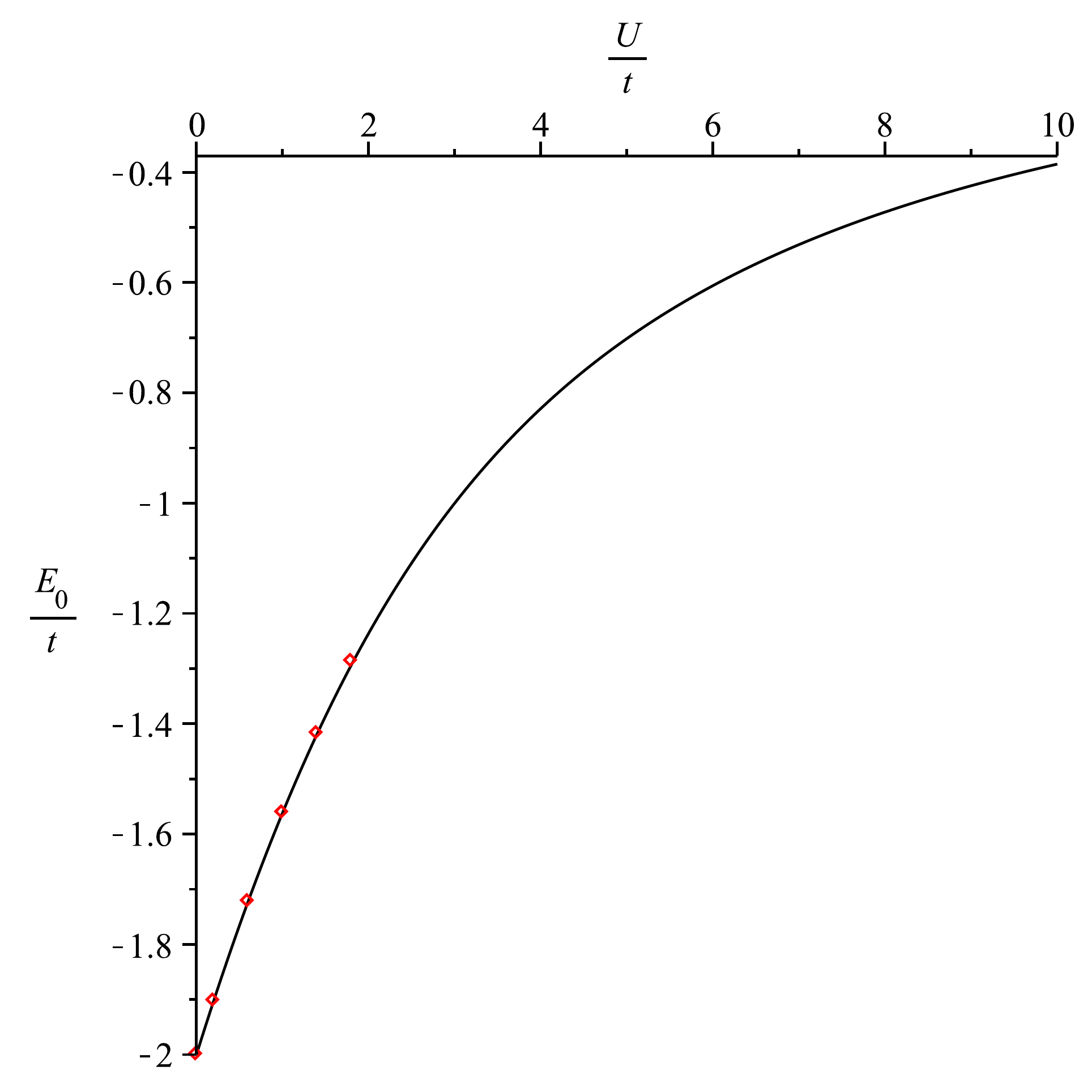}
\caption[Ground-state energies of the Hubbard molecule in FRPA]{
   \label{fig:results:hubbardE0}
   Ground-state energies of the Hubbard molecule versus reduced interaction strength. The red squares are the FRPA results with self-consistency on the level of the HF diagram; the black line is the exact result which coincides with the FTDA result. At $\frac{U}{t}=2$ the RPA-instability sets in.
 }
\end{center}
\end{figure}

The pp and ph RPA can now be used as building blocks for the FRPA. The FRPA matrices get too involved to be calculated analytically. The results of the simulation can be seen in \Fig{results:hubbardE0}. The ground-state energy of the Hubbard molecule is plotted as a function of the reduced interaction strength. Just as with the real H$_2$ molecule, the results of the FRPA follow the exact values very closely until the breakdown point at $\frac{U}{t}=2$. So unlike the HF molecule, the results are trustworthy even for this system with a very strong few-body character. After the instability occurs, the calculation becomes impossible. The exact results coincide with the FTDA results. Even the poles and strengths are reproduced exactly. So for this problem, although the excitation energies are not exact in the TDA (see \Fig{results:hubbardph}), the FTDA Green's function \textit{is} the exact result.

\begin{figure}
\begin{center}
\includegraphics[scale=0.5,clip=true]{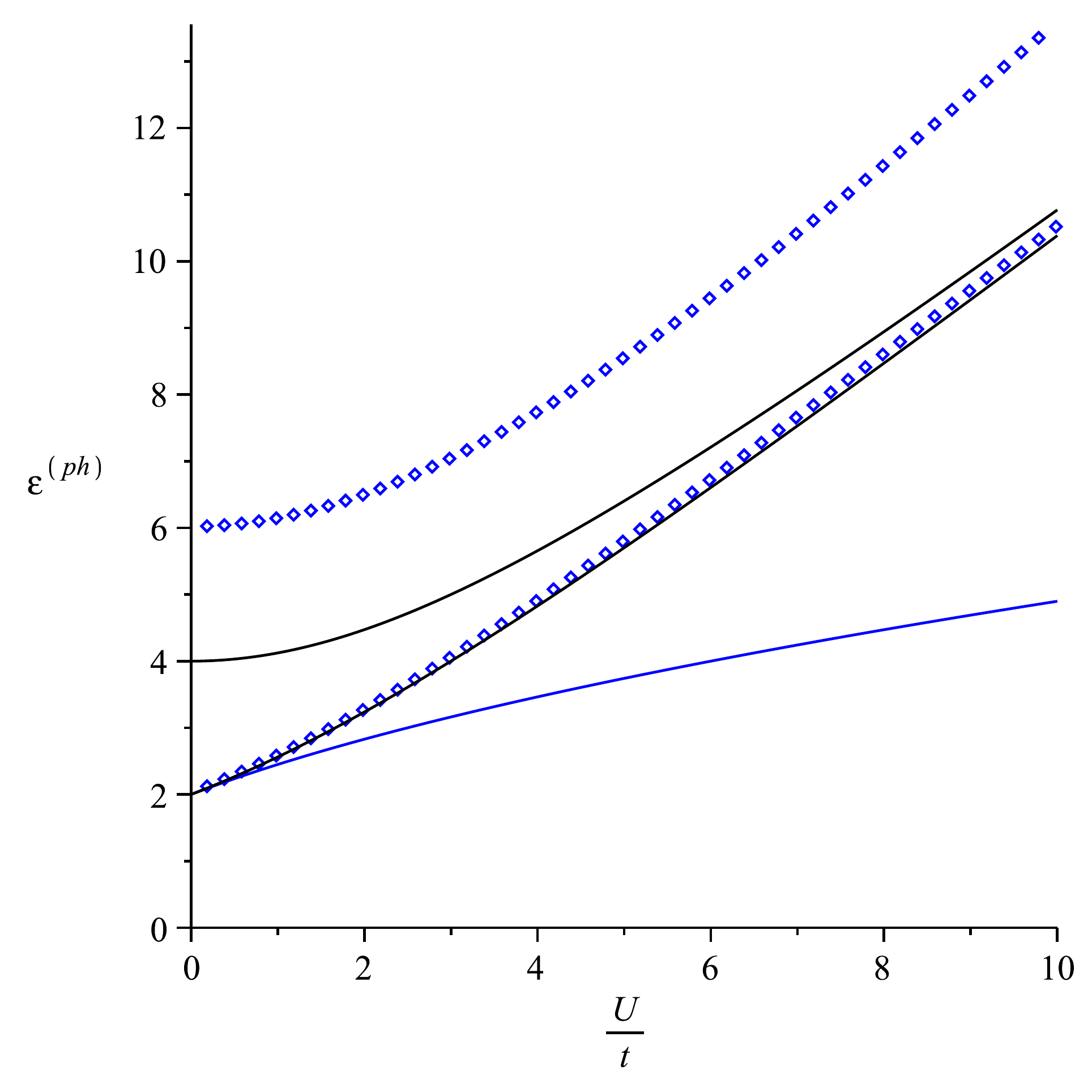}
\caption[ph energies for the Hubbard molecule with the exact propagator for $S=0$]{
   \label{fig:results:hubbards=0}
   The ph excitation energies for the Hubbard molecule calculated with the exact propagator are plotted for the $S=0$ channel. The squares are the values calculated with the exact propagator as a reference, while the solid red line are the normal RPA values. The black lines give the exact $S=0$ excitation. The highest exact excitation is the positive parity excitation which is not produced in either the normal RPA or the RPA based on the exact propagator.
   }
\end{center}
\end{figure}

\begin{figure}
\begin{center}
\includegraphics[scale=0.5,clip=true]{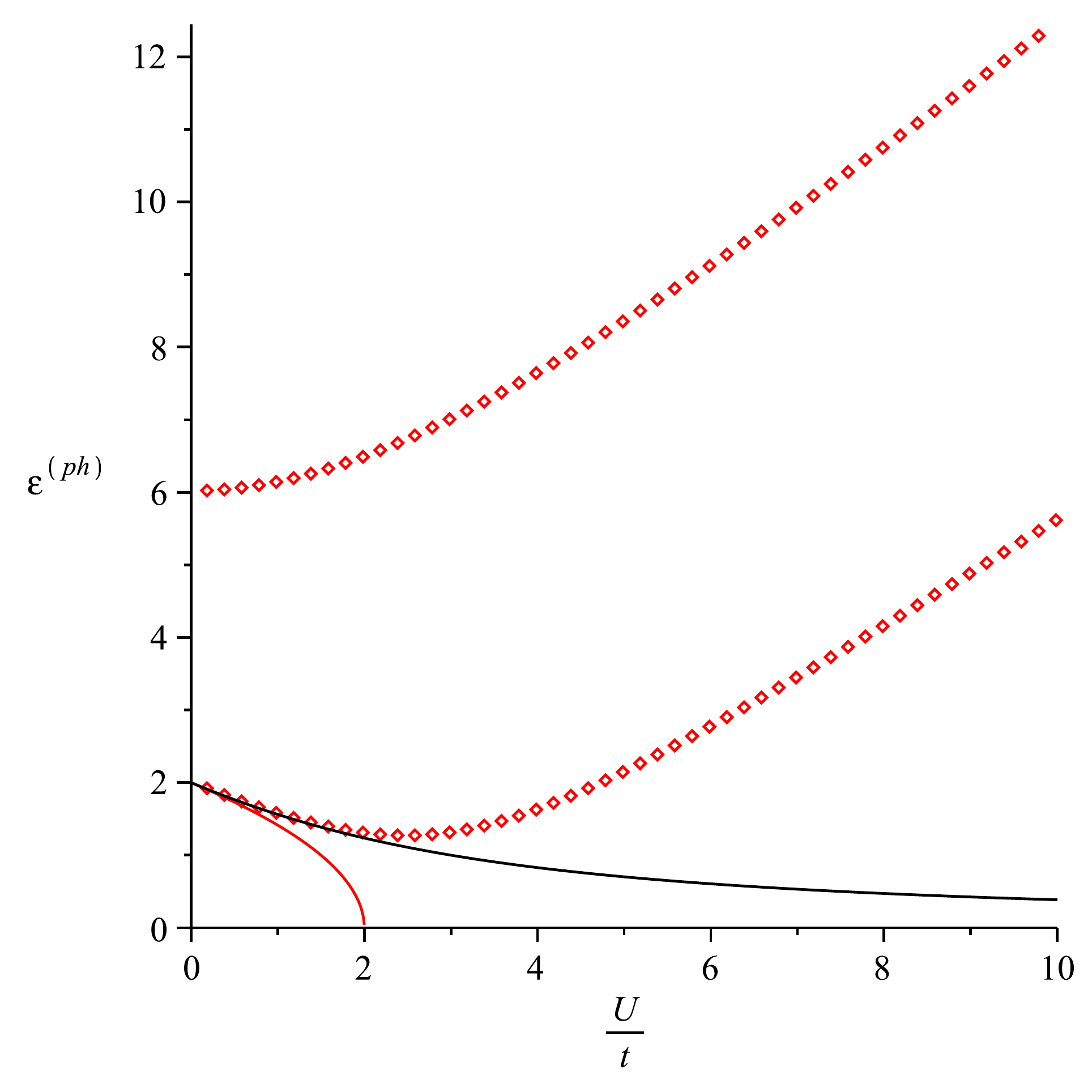}
\caption[ph energies for the Hubbard molecule with the exact propagator for $S=1$]{
   \label{fig:results:hubbards=1}
   The ph excitation energies for the Hubbard molecule calculated with the exact propagator are plotted for the $S=1$ channel. The squares are the values calculated with the exact propagator as a reference, while the solid red line are the normal RPA values. The black line gives the exact $S=1$ excitation.
   }
\end{center}
\end{figure}

One solution for this problem is to replace the unstable channel with its TDA counterpart, i.e. replacing one single energy and amplitude with the values calculated from a separate TDA calculation. This allows for values to be calculated beyond the instability, but the results obtained there are very close to FTDA. So no real advantage is gained. Another idea is to see if added self-consistency can cure this problem, as was done for example in Ref.~\cite{Barbieri2002}. This approach, however, does not eliminate the possibility of complex RPA eigenvalues. To accomplish a higher level of consistency, we use the exact Green's function to construct the $\Pi^{(0)}$ in \Equ{methods:rpapi}. Seeing the small dimensions of this problem, the diagonal nature of the exact Green's function and the division of the matrices into blocks with definite spin, it is possible to directly invert \Equ{methods:rpapi} to obtain the RPA polarization propagator $\Pi$. The resulting poles are shown in \Fig{results:hubbards=0} and \Fig{results:hubbards=1}. A striking difference is the occurence of two times two branches per spin. This is due to the fragmentation of the exact single-particle Green's function, this doubles the matrix dimension of the RPA problem. Of course it is still impossible to reach the excitation with different parity, so there are two excitation energies too many. We consider the higher two excitations to be spurious and discard them in further calculations. It should be noted that the $S=0$ excitation becomes exact up to numerical errors. The $S=1$ solution differs from the exact result, but it is striking that no instability appears at any interaction strength due to the use of a fragmented single-particle propagator. Therefore, it is possible to do calculations beyond $\frac{U}{t}=2$ with the FRPA using these eigenvalues and eigenvectors.

\begin{figure}
\begin{center}
\includegraphics[scale=0.5,clip=true]{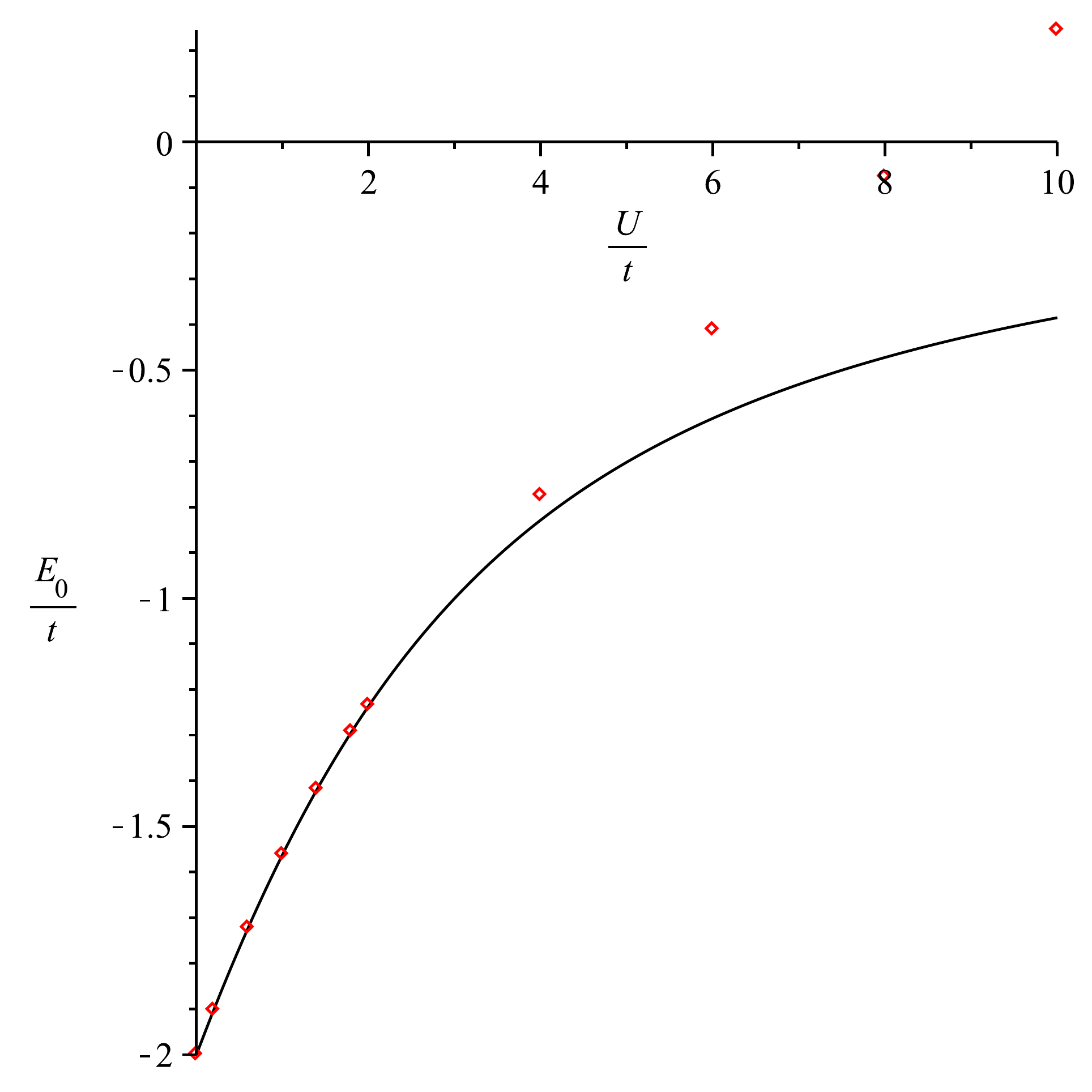}
\caption[Ground-state energies of the Hubbard molecule in FRPA with exact propagator]{
   \label{fig:results:hubbardE02}
   Ground-state energies of the Hubbard molecule versus reduced interaction strength. The red squares are the FRPA results based on the ph RPA with the exact propagator with self-consistency on the level of the HF diagram, the black line shows the exact result. There is no instability in this case.
   }
\end{center}
\end{figure}

\Fig{results:hubbardE02} shows the results of this calculation. At low interactions, just as with the normal FRPA, this calculation based on the exact single-particle propagator follows the exact result very closely. Even up to a reduced interaction of four, the general trend of the exact result is reproduced. Afterwards the values become detached from the exact curve very quickly. At $\frac{U}{t}=10$, the system is not even bound anymore. It is unlikely that adding further self-consistency would positively influence the results, as the RPA is already calculated with the best propagator possible. This could mean that the RPA misses a crucial diagram that cannot be cured by self-consistency, but only by making the transition to a ph interaction that is exact beyond first order in perturbation theory.

Schuck and coworkers\cite{Schuck1973b,Dukelsky1990} have developed the so-called SCRPA that goes beyond the first-order RPA. The SCRPA is a theory that can be applied to both the pp and ph interactions and differs from the standard RPA mainly in the use of a correlated ground state for the calculation of its expectation values. This means that the first-order approach in perturbation theory of RPA is upgraded to a nonperturbative variational theory, although not of the Rayleigh-Ritz type. The SCRPA has been applied to some test models and also to the Hubbard model\cite{Schaefer1999,Jemai2005}. In Ref.\cite{Jemai2005} it is found that the SCRPA solves the two-body problem exactly, while conserving the matrix size and characteristic structure of the RPA eigenvalue problem. Due to the flexible nature of the FRPA, the SCRPA holds the promise of making the FRPA exact for two-body systems and allowing the calculations to go beyond the point where the classical RPA experiences unstable behavior. However, the SCRPA has not been applied to realistic systems yet because of the appearance of the single-particle density matrix in the equations. The RPA amplitudes give access to only part of the density matrix, while all matrix elements are needed to build the SCRPA matrices. This problem could be solved by incorporating the SCRPA into the FRPA framework. As it is a single-particle Green's function method, the FRPA gives easy access to the single-particle density matrix. A procedure could be developed in which the FRPA density matrix is used to update the SCRPA matrices, after which the SCRPA amplitudes and eigenvalues are used to update the FRPA matrix, and to repeat this until convergence. Taking a suitable starting point for the single-particle density matrix to be used in the first RPA iteration could circumvent the occurence of instabilities.
							\clearnewpage
\chapter{Conclusion}
\label{cha:conclusion}

In this work we have introduced a Green's function technique for the calculation of ground-state energies and ionization energies in quantum many-body systems. The elementary building blocks of the theory are the RPA (random phase approximation) excitations. These are used to construct an approximation for the selfenergy in a specific manner, called the FRPA (Faddeev RPA). The FRPA holds the promise to be a Green's function method with a wide applicability, from finite systems like atoms and molecules to extended systems like the uniform electron gas and nuclear matter. The fully self-consistent FRPA is conserving in the Baym-Kadanoff sense and consequently obeys important conservation laws.

In \Cha{frpa} we have presented the FRPA technique as a Green's function method that is able to couple pp (particle-particle) and ph (particle-hole) excitations of the RPA type. The procedure involves the extension of the space in which the selfenergy is expanded from the single-particle space, to the single-particle space together with the 2p1h (two-particle-one-hole) and 2h1p (two-hole-one-particle) spaces. The consecutive inclusion of ph and pp excitations is governed by the Faddeev equations. The eigenvalues and amplitude from the RPA problems form the building blocks for the Faddeev components. The Faddeev technique is needed when including excitations at the RPA level. In this respect, the current method differs from the ADC(3) (algebraic diagrammatic construction method of third order) which only involves excitations of the TDA (Tamm-Dancoff approximation) level to be exchanged. Both methods are closely related: the ADC(3) can be rederived from the FRPA by the exclusion of backward going diagrams in the pp and ph interaction vertices. Diagrammatically, this close relation was already clear, but now it has been shown rigorously. Replacing the TDA intermediary excitations with their RPA counterparts implies that the FRPA behaves better in the limit of extended systems. It also means that the matrix equations become non-Hermitian, which can lead to the occurence of complex eigenvalues. This should always be kept in mind when performing an FRPA calculation. At first sight, the matrix dimension of the FRPA seems to triple compared to ADC(3), but after carefully studying the nature of the solutions, we can discard two thirds as unphysical. The separation of physical and spurious solutions can be carried out by a straightforward projection, which reduces the matrix problem to its orginal size. 

First we have performed calculations of the electronic structure for a series of atoms. These calculations showed that even for very small systems, where the RPA violations of the Pauli principle are expected to be important, the FRPA obtained satisfactory results. In general, FRPA and FTDA yield very similar results for the lightest systems, while the additional diagrams in RPA lead to a small but systematic improvement as the atomic number increases. Both Green's function methods agree well with CCSD (coupled cluster with singles and doubles) in the basis set limit and deviations from experimental results are within the estimated extrapolation error. The near-degeneracy of the Be atom seems harder to describe, although the errors are systematic, appearing for all methods, and seem to be mostly due to deficiencies in the basis set.
The conclusions for the ionization energies follow the same trend. For the smaller two-electron atoms, the FRPA does not improve on the FTDA results. For all other atoms, the FRPA pushes the calculated values in the direction of the experiment. The Be atom also forms a problematic case for the ionization energies.

The next logical step is to calculate properties of small molecules. The diatomic molecules of interest have the nuclear separation as an extra parameter. Around equilibrium geometry, we find results that are in line with the atomic calculations. For smaller molecules, FRPA and FTDA are very close to each other, while for larger molecules the differences are more pronounced. In any case, the scan of nuclear separations clearly shows that self-consistency on the Hartree-Fock diagram should always be performed. The ionization energies are of comparable quality for FTDA and FRPA. In these calculations, there is also a problematic case: the third ionization energy of the N$_2$ molecule produces results that deviate from the experiment. An attempt was made to push these results to the basis set limit, but this is for now still out of reach of our computer resources.

The molecular calculations in the dissociation limit show the "Achilles' heel" of the FRPA. Being highly dependent on the quality of the RPA solution, the applicability of the FRPA only stretches so far as the applicability of the RPA. It is known that the ph RPA exhibits triplet instabilities when the bond in diatomic molecules is stretched. For the smaller systems, we do not find that the FRPA results are adversely influenced by the imminent RPA instability. However, the results for the HF molecule show that, for molecules with higher correlation energies, the results of the FRPA start to deviate in the vicinity of the instability. This means that the FRPA values can only be trusted around the equilibrium geometry.

The Hubbard molecule allows us to look at this instability in its most simple form. The possibility to obtain exact results serves a guideline to improve on the current theory. We have demonstrated that by adding fragmentation to the single-particle propagator, it is possible to go beyond the point of instability in the RPA. 

\section{Outlook}

Although the FRPA is a promising method, there is still room for improvement, not only on the level of the numerical methods used, but also with respect to the physical content of the theory. We will list some of the possible pathways to a more refined Green's function method:
\begin{itemize}
\item The computational cost for solving the FRPA problem can be drastically reduced by using Arnoldi techniques in the 2p1h and 2h1p diagonalizations. This approach has been previously applied in Ref.~\cite{Barbieri2009} and it was found that a limited number of Arnoldi vectors guarantee correct converged results for total energies and ionization potentials.
\item The SCRPA (self-consistent RPA) may offer a viable alternative for the calculation of ph and pp interactions. This scheme has the property that it solves the two-body problem exactly. It goes beyond the quasi-boson approximation by taking into account the killing condition for the excitation operators. From the formulation in Ref.~\cite{Schaefer1999}, it can be seen that the SCRPA procedure has the same diagrammatical structure as the standard RPA and can also be translated in the diagrammatical language used to describe the FRPA. The occurence of single-particle density matrices in the SCRPA matrix forms an obstacle for the implementation. This problem can be tackled in two ways, either by resorting to perturbation theory of the density matrix in function of the SCRPA amplitudes or by using matrix relations that are system specific. For realistic systems, another solution has to be found. One possibility is to use the single-particle density matrix from a Faddeev calculation and to insert this into the SCRPA pp and ph equations, the solutions of which can then be used to build the Faddeev matrices. The matrix dimensions of this problem should stay the same as in the current implementation of the FRPA. The additional cost would be the iterative solution of the SCRPA equations, which should be small compared to the diagonalization in 2p1h and 2h1p space.
\item The current approach would be more elegant if it could be translated into an EOM-approach along the lines of Ref.~\cite{Rowe1968}. However, the existence of an excitation operator that has the current formalism as a result has not yet been found. In Ref.~\cite{Jemai2011}, the excitation operators as used for the normal RPA equations were extended to incorporate higher excitations. This was done in such a way that the excitation operator destroys the correlated vacuum state. One can apply this approach for the polarization propagator to the case of the single-particle propagator by extending the excitation operators from the single-particle space to the 2p1h and 2h1p space. The final matrix problem can be made Hermitian by symmetrizing the anti-commutators.
\item An ultimate test for robustness of the FRPA would be to apply it to extended systems. For this to be possible a method must be devised to handle the continuous variables that are characteristic to this sort of systems. Their are two possible paths that were previously applied to nuclear matter\cite{Dewulf1997}, the BIN\cite{VanNeck1991,VanNeck1993} and BAGEL\cite{Müther1988,Müther1993} methods. In the BIN method, the energy axis is discretized and all quantities of interest are represented in these equidistant energy bins. In the BAGEL method, the Green's function is represented by a few characteristic poles, chosen in such a way that the lowest order moments of the exact distribution are reproduced. A more difficult path is to reformulate the theory where all the quantities are kept in their analytical form. Success is not guaranteed when pursuing this path.
\end{itemize}
							\clearnewpage

\appendix

\chapter{Uniform electron gas}
\label{app:electrongas}

In this chapter we derive the expression for the Lindhard function of the uniform electron gas in the TDA. The derivation goes along the lines of the derivation of the RPA polarization propagator\cite{Dickhoff2005}, with the exception that the backward propagating amplitude is not present. The expression for the Lindhard function is used to calculate the discrete pole which arises in the limit of vanishing total momentum. This so-called plasmon-pole can be derived classically and is described correctly by the RPA, but not by the TDA.

\section{Polarization propagator for infinite systems}
The Bethe-Salpeter equation for the RPA and TDA polarization propagator for continuous systems is given by
\begin{eqnarray}
\Pi_{SM_S}(\mathbf{p},\mathbf{p}';\mathbf{Q},E) &=& \delta_{\mathbf{p},\mathbf{p}'}\Pi^{(0)}(\mathbf{p};\mathbf{Q},E) + \Pi^{(0)}(\mathbf{p};\mathbf{Q},E)\nonumber\\
&&\times\sum_{\mathbf{p''}}\langle \mathbf{Q},\mathbf{p}|V^{(ph)}_{S M_S}|\mathbf{Q},\mathbf{p}''\rangle \Pi_{S M_S}(\mathbf{p}'',\mathbf{p}';\mathbf{Q},E).\nonumber\\
&&
\end{eqnarray}
Here the $\mathbf{Q}$ represents total momentum of the ph state, while $\mathbf{p}$, $\mathbf{p}'$ and $\mathbf{p}''$ are relative momenta. The interaction $V$ corresponds to a local, central interaction without spin dependence, wich resulting in a polarization propagator which is diagonal in total momentum and in spin degrees of freedom. The superscript $(ph)$ indicates that the interaction is written in a basis that emphasizes the physical content of the interaction: it connects ph pairs
\begin{eqnarray}
\langle \alpha \beta^{-1} | V^{(ph)} | \gamma \delta^{-1} \rangle &=& \langle \alpha\bar{\delta}|V|\bar{\beta}\gamma\rangle,
\end{eqnarray}
where the time-reversal relation is given in \Equ{methods:timereversal}. With the current assumptions for the interaction, we arrive at matrix elements
\begin{eqnarray}
\langle\mathbf{Q},\mathbf{p}|V_{S M_S}^{(ph)}|\mathbf{Q},\mathbf{p}'\rangle &=& \frac{1}{V}\left(\delta_{S,0}V(\mathbf{Q})-V(\mathbf{p}-\mathbf{p}')\right).
\end{eqnarray}
Since the momentum $\mathbf{Q}$ is a conserved quantity, the direct term will completely dominate the ph interaction at small values of $\mathbf{Q}$, which is the plasmon limit we are interested in. As a consequence, the sum over the momentum differences can be carried out. This leads to a new Bethe-Salpeter equation
\begin{eqnarray}
\Pi_{S M_S}(\mathbf{Q},E) &=& \Pi^{(0)}(\mathbf{Q},E) + \Pi^{(0)}(\mathbf{Q},E) V_{S M_S}^{(ph)}(\mathbf{Q})\Pi_{S M_S}(\mathbf{Q},E),
\end{eqnarray}
where the definitions
\begin{eqnarray}
\Pi_{S M_S}(\mathbf{Q},E) &=& \frac{1}{V} \sum_{\mathbf{p}}\sum_{\mathbf{p}'} \Pi_{S M_S}(\mathbf{p},\mathbf{p}';\mathbf{Q},E)
\end{eqnarray}
and
\begin{eqnarray}
\Pi^{(0)}(\mathbf{Q},E) &=& \frac{1}{V} \sum_{\mathbf{p}} \Pi^{(0)}(\mathbf{p};\mathbf{Q},E)
\end{eqnarray}
have been used. Neglecting the exchange term and including a factor two for the spin degeneracy, the polarization propagator in the TDA or RPA becomes
\begin{eqnarray}
\Pi_{S=0}(Q,E) = \frac{\Pi^{(0)}(Q,E)}{1-2V(Q)\Pi^{(0)}(Q,E)}\label{equ:electrongas:rpa}.
\end{eqnarray}
Assuming that the polarization propagator has a discrete pole at energy $E_p$, the following Lehmann representation can be proposed
\begin{eqnarray}
\Pi_{S=0}(Q,E) &=& \frac{A_p(Q)}{E-E_p + i\eta} - \frac{B_p(Q)}{E+E_p-i\eta} + \text{continuum}.
\end{eqnarray}
Replacing the polarization propagator in \Equ{electrongas:rpa} and searching for the pole $E_p$ results in the equation
\begin{eqnarray}
1-2V(Q)\Re\Pi^{(0)}(Q,E_p) = 0
\end{eqnarray}
Finding a value for the plasmon pole is now reduced to finding an expression for the real part of the non-interacting polarization propagator. The simplest way of doing this, is by using the dispersion relation
\begin{eqnarray}
\Pi^{(0)}(\mathbf{Q},E) &=& -\frac{1}{\pi}\int \frac{\Im\Pi^{(0)}(\mathbf{Q},E')}{E-E'+i\eta}\mathrm{d}E'\nonumber\\
&& + \frac{1}{\pi} \int \frac{\Im\Pi^{(0)}(\mathbf{Q},E')}{E-E'-i\eta}\mathrm{d}E'\label{equ:electrongas:integral},
\end{eqnarray}
where the second and backward propagating term is retained only in the RPA and not in the TDA. The imaginary part of the Lindhard function for positive energy values can be calculated as 
\begin{eqnarray}
\Im\Pi^{(0)}(Q,E) &=& \left\{\begin{array}{l}  -\frac{m^2E}{4\pi\hbar^3Q}\\\quad\quad\text{when } \frac{Q}{p_F} < 2 \text{ and }0 < E < \frac{Qp_F}{m}-\frac{Q^2}{2m} \\ -\frac{m}{8\pi\hbar^3Q}\left(p_F^2-\left(\frac{mE}{Q}-\frac{Q}{2}\right)^2\right)\\\quad\quad\text{when }\frac{Q}{p_F} < 2 \text{ and } \frac{Qp_F}{m}-\frac{Q^2}{2m} < E\\ -\frac{m}{8\pi\hbar^3Q}\left(p_F^2-\left(\frac{mE}{Q}-\frac{Q}{2}\right)^2\right)\\\quad\quad\text{when } 2 < \frac{Q}{p_F}. \end{array}\right.\nonumber\\
&&
\end{eqnarray}
\section{Tamm-Dancoff approximation}
Calculating the first term of \Equ{electrongas:integral} results in the function
\begin{eqnarray}
\Re\Pi^{(0)}(Q,E) &=& \left\{\begin{array}{l}  
\frac{m}{8\pi^2\hbar^3Q} \left(mE-Qp_F+2mE\ln\left|\frac{2mE}{2mE-2Qp_F+Q^2}\right| \right.\\
\left. + \left(p^2_F-\left(\frac{mE}{Q}-\frac{Q}{2}\right)^2\right)\ln\left|\frac{2mE-2Qp_F+Q^2}{2mE-2Qp_F-Q^2}\right|\right)\\
\quad\quad\text{when } \frac{Q}{p_F} < 2\\
\frac{m}{8\pi^2\hbar^3Q} \left(-Qp_F+\frac{2mEp_F}{Q} \right. \\
\left. + \left(p^2_F-\left(\frac{mE}{Q}-\frac{Q}{2}\right)^2\right)\ln\left|\frac{2mE+2Qp_F-Q^2}{2mE-2Qp_F-Q^2}\right|\right)\\
\quad\quad\text{when }2 < \frac{Q}{p_F}. \end{array}\right.\nonumber\\
&&
\end{eqnarray}
Taking the limit for vanishing momentum and writing this expression as a series expansion in function of the momentum, one arrives at
\begin{eqnarray}
\lim_{Q\rightarrow 0}\Re\Pi^{(0)}(Q,E) = \frac{mp_F}{4\pi^2\hbar^3}\left(\frac{p_FQ}{mE}+\frac{1}{3}\frac{p_F^2Q^2}{m^2E^2} + \mathcal{O}(Q^3)\right).
\end{eqnarray}
When substituted in \Equ{electrongas:rpa} this does not have the correct $Q$ dependence to cancel the $\frac{1}{Q^2}$ behavior of the Coulomb-interaction. As a consequence the plasmon pole will diverge and the TDA does not reproduce the classical behavior.
\section{Random Phase approximation}
For the RPA derivation, both positive and negative energy parts of the $\Im\Pi^{(0)}$ are needed. It suffices to see that the $\Im\Pi^{(0)}(Q,E)$ is an even function of E, so that the integral in \Equ{electrongas:integral} can be carried out. The resulting real part of the Lindhard function is given by a single expression, valid for all momenta
\begin{eqnarray}
\Re\Pi^{(0)}(Q,E) &=& \frac{mp_F}{4\pi^2\hbar^3}\left\{-1+\frac{p_F}{2Q}\left(1-\left(\frac{mE}{Qp_F}-\frac{Q}{2p_F}\right)^2\right)\right.\nonumber\\
&&\times \left.\ln\left|\frac{2mE+2Qp_F-Q^2}{2mE-2Qp_F-Q^2}\right|-\frac{p_F}{2Q}\left(1-\left(\frac{mE}{Qp_F}+\frac{Q}{2p_F}\right)^2\right)\right.\nonumber\\
&&\left.\times\ln\left|\frac{2mE+2Qp_F+Q^2}{2mE-2Qp_F-Q^2}\right| \right\}
\end{eqnarray}
The backward propagating term cancels the terms that are linear in the energy. The low-momentum limit is given by
\begin{eqnarray}
\lim_{Q\rightarrow 0}\Re\Pi^{(0)}(Q,E) = \frac{mp_F}{4\pi^2\hbar^3}\left(\frac{2}{3}\frac{p_F^2Q^2}{m^2E^2}+\mathcal{O}(Q^3)\right),
\end{eqnarray}
which exactly cancels the $\frac{1}{Q^2}$ of the interaction and results in the correct classical plasmon pole.
 \clearnewpage
\chapter{Pictures}
\label{app:pictures}

We will introduce two pictures of quantum mechanics in which we can analyze the Schr\"odinger equations.

\section{Schr\"odinger picture}
Most of the time we deal with a description of quantum mechanics in which we assume that the state vectors are time dependent and the operators which act on these states are time independent. In this language the time-dependent Schr\"odinger equation is given by
\begin{eqnarray}
i\hbar \frac{\partial}{\partial t} |\Psi_S(t)\rangle &=& H |\Psi_S(t)\rangle\label{equ:pictures:schrodinger},
\end{eqnarray}
where we have explicitly made clear the picture to which the states by the subscript $S$. The only part of this equation that dependends on time is the derivative in the left-hand side. To solve this equation, it is enough to have the state $|\Psi_S(t_0)\rangle$ at some time $t_0$, after which the behavior at any time is determined by
\begin{eqnarray}
|\Psi_s(t)\rangle &=& \exp\left(-\frac{i}{\hbar}H(t-t_0)\right)|\Psi_S(t_0\rangle.
\end{eqnarray}
The exponential function of the Hamiltonian is the time-evolution operator in the Schr\"odinger picture and is defined by its power series expansion in function of $H$.

\section{Heisenberg picture}
In the Heisenberg picture, the states are considered to be time-independent and all time dependence is assumed to be included in the operators. The relation with the Schr\"odinger states is given by
\begin{eqnarray}
|\Psi_H(t)\rangle &=& \exp\left(\frac{i}{\hbar}Ht\right)|\Psi_S(t)\rangle
\end{eqnarray}
Combined with the Schr\"odinger equation (\Equ{pictures:schrodinger}), this shows the time-independence of the Heisenberg state $|\Psi_H\rangle$
\begin{eqnarray}
i\hbar\frac{\partial}{\partial t}|\Psi_H(t)\rangle &=& -H \exp\left(\frac{i}{\hbar}Ht\right)|\Psi_S(t)\rangle + \exp\left(\frac{i}{\hbar}Ht\right)i\hbar\frac{\partial}{\partial t}|\Psi_S(t)\rangle\\
&=& 0.
\end{eqnarray}
The operators in the Heisenberg picture are time-dependent
\begin{eqnarray}
\langle \Psi'_S(t)|O_S|\Psi_S(t)\rangle &=& \langle \Psi_H'|\exp\left(\frac{i}{\hbar}Ht\right) O_S \exp\left(-\frac{i}{\hbar}Ht\right)|\Psi_H\rangle,
\end{eqnarray}
from which their relation to the Schr\"odinger picture operators follow as
\begin{eqnarray}
O_H(t) &=& \exp\left(\frac{i}{\hbar}Ht\right) O_S \exp\left(-\frac{i}{\hbar}Ht\right).
\end{eqnarray}
Now the operators also follow an equation of motion
\begin{eqnarray}
i\hbar\frac{\partial}{\partial t} O_H(t) &=& - H \exp\left(\frac{i}{\hbar}Ht\right) O_S \exp\left(-\frac{i}{\hbar}Ht\right)\nonumber\\
&&+ \exp\left(\frac{i}{\hbar}Ht\right) O_S H\exp\left(-\frac{i}{\hbar}Ht\right)\\
&=& \left[O_H(t),H\right].
\end{eqnarray}
                \clearnewpage

\renewcommand{\leftmark}{\bibname}
\renewcommand{\rightmark}{\bibname}

\bibliographystyle{alpha}
\addtocchapter{\bibname}
\bibliography{library}

\pagestyle{plain}
\addtocchapter{\papername}
\chapter*{\papername}
\begin{enumerate}
\item Matthias Degroote, Dimitri Van Neck and Carlo Barbieri. Faddeev Random Phase Approximation for Molecules. {\it Computer Physics Communications}, 182:1995–1998, 2011.
\item Matthias Degroote, Dimitri Van Neck and Carlo Barbieri. Faddeev Random Phase Approximation for Molecules. {\it Physical Review A}, 83:042517, 2011.
\item Carlo Barbieri, Dimitri Van Neck and Matthias Degroote. Accuracy of the Faddeev Random Phase Approximation for Light Atoms. {\it Physical Review A}, 85:012501, 2012.
\end{enumerate}



\section*{Supplementary material}

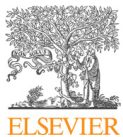
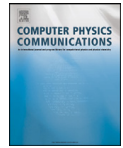
# Faddeev Random Phase Approximation for molecules

Matthias Degroote [a,*], Dimitri Van Neck [a], Carlo Barbieri [b]

[a] *Center for Molecular Modeling, Ghent University, Technologiepark 903, B-9052 Zwijnaarde, Belgium*
[b] *Department of Physics, Faculty of Engineering and Physical Sciences, University of Surrey, Guildford GU2 7XH, United Kingdom*

### abstract

The Faddeev Random Phase Approximation is a Green's function technique that makes use of Faddeev-equations to couple two-particle–one-hole and two-hole–one-particle excitations to the single-particle spectrum. Solving these equations implies an infinite partial summation of the perturbation expansion of the self energy. This method goes beyond the third-order Algebraic Diagrammatic Approximation by treating both the particle–hole and particle–particle interactions at the Random Phase Approximation-level. This paper presents the first results of our calculations for diatomic molecules.

© 2010 Elsevier B.V. All rights reserved.

## 1. Introduction

The study of electron correlation by means of first-principle calculations has taken a high rise thanks to modern computer technology. The Green's function formalism is one of these first-principles methods that has successfully been applied in quantum chemistry [1–4]. The correlations in a many-body system are described by an energy dependent self energy. A particular third-order approximation scheme to the self energy is the Algebraic Diagrammatic Construction (ADC) [5] developed by Schirmer and coworkers. This method has proven very successful in predicting electron properties in molecules [6]. This method is equivalent with resumming all two-particle–one-hole (2p1h) and two-hole–one-particle (2h1p) interactions up to Tamm–Dancoff (TDA) level. It has proven very difficult to go beyond the TDA-level [7], even though it is known that the Random Phase Approximation (RPA) should be better to describe long range correlations. The Faddeev Random Phase Approximation (FRPA) [8] solves this problem by using the Faddeev technique to include RPA-phonons in the self energy. This method has successfully been applied to nuclei [9] and atoms [10]. In the first section of this work we will give a short overview of the working equations for the FRPA method. In Section 3 we will present the numerical results for a set of diatomic molecules.

## 2. Theory

### 2.1. Single-particle Green's function

The behavior of an electronic many-body system is governed by the Hamiltonian

$$\hat{H} = \sum_{\alpha,\beta} T_{\alpha,\beta} a_\alpha^\dagger a_\beta + \frac{1}{4} \sum_{\alpha,\beta,\gamma,\delta} V_{\alpha\beta,\gamma\delta} a_\alpha^\dagger a_\beta^\dagger a_\delta a_\gamma, \qquad (1)$$

where $a_\alpha$ ($a_\alpha^\dagger$) is the annihilation (creation) operator for a state with quantum numbers $\alpha$ and where $T_{\alpha,\beta}$ and $V_{\alpha\beta,\gamma\delta}$ are the matrix elements of the one-body operator and the antisymmetrized two-body operator respectively. For the present study, the one-body operator $T$ represents the kinetic energy and the attraction to the nuclei and $V$ is the Coulomb repulsion between the electrons. The evolution of this $N$-body system can be described by the single-particle propagator [11]

$$G_{\alpha,\beta}(t,t') = -\frac{i}{\hbar} \langle \Psi_0^N | \mathcal{T}[a_\alpha(t) a_\beta^\dagger(t')] | \Psi_0^N \rangle, \qquad (2)$$

where $\mathcal{T}[\ldots]$ represents the time ordering operator and $a_\alpha(t)$ and $a_\alpha^\dagger(t)$ are the addition and removal operators in the Heisenberg picture. For practical calculations, the Lehmann representation of the Green's function

$$G_{\alpha,\beta}(E) = \sum_{m>F} \frac{\langle \Psi_0^N | a_\alpha | \Psi_m^{N+1} \rangle \langle \Psi_m^{N+1} | a_\beta^\dagger | \Psi_0^N \rangle}{E - (E_m^{N+1} - E_0^N) + i\eta} + \sum_{m<F} \frac{\langle \Psi_0^N | a_\alpha^\dagger | \Psi_m^{N-1} \rangle \langle \Psi_m^{N-1} | a_\beta | \Psi_0^N \rangle}{E - (E_0^N - E_m^{N-1}) - i\eta} \qquad (3)$$

\* Corresponding author.
*E-mail address:* matthias.degroote@ugent.be (M. Degroote).

# Faddeev random-phase approximation for molecules

Matthias Degroote,[1,*] Dimitri Van Neck,[1] and Carlo Barbieri[2]

[1]*Center for Molecular Modeling, Technologiepark 903, B-9052 Zwijnaarde, Belgium*
[2]*Department of Physics, Faculty of Engineering and Physical Sciences, University of Surrey, Guildford GU2 7XH, United Kingdom*

The Faddeev random-phase approximation is a Green's function technique that makes use of Faddeev equations to couple the motion of a single electron to the two-particle–one-hole and two-hole–one-particle excitations. This method goes beyond the frequently used third-order algebraic diagrammatic construction method: all diagrams involving the exchange of phonons in the particle-hole and particle-particle channel are retained, but the phonons are now described at the level of the random-phase approximation, which includes ground-state correlations, rather than at the Tamm-Dancoff approximation level, where ground-state correlations are excluded. Previously applied to atoms, this paper presents results for small molecules at equilibrium geometry.

## I. INTRODUCTION

The study of electronic systems by means of first-principles calculations has taken a high rise thanks to modern computer technology [1–5]. The Green's function formalism [6–8] is one of these *ab initio* methods that has been successfully applied in quantum chemistry [9–12]. The correlations in a many-body system are described in terms of an electron self-energy, which acts as an energy-dependent potential describing the motion of a single electron in the many-electron system. The accuracy of the (in principle, exact) Green's function formalism is now governed by the approximation chosen for the electron self-energy.

A particular third-order approximation scheme to the self-energy can be obtained using the algebraic diagrammatic construction [ADC(3)] method as developed by Schirmer and co-workers [13]. This method has proven to be very successful in predicting one-electron properties in molecules [14] as measured, e.g., in electron momentum spectroscopy. Although the ADC(3) equations were derived in a purely algebraic manner, they can be shown to be equivalent to resumming all particle-hole (ph) and particle-particle (pp) interactions between two-particle–one-hole (2p1h) and two-hole–one-particle (2h1p) states up to the Tamm-Dancoff approximation (TDA) [7] level. This is diagrammatically equivalent to considering phonons (defined here as excitations in the ph and pp channels) at the TDA level, and then allowing the exchange of these phonons in all possible ways between the tree propagators describing the 2p1h or 2h1p states.

The TDA allows no ground-state correlations in the construction of the phonons. An improvement in this respect is the random-phase approximation (RPA) [15]. At least for nuclear systems [15], it is known that the RPA performs better at describing collective behavior. Calculations for the electron gas also show that the RPA leads to a correct prediction of the plasmon pole, whereas the TDA completely fails to describe the plasmon spectrum. It is therefore of interest to formulate an analogous theory to ADC(3) that resums the ph and pp interactions up to the RPA level, i.e., replacing the exchange of TDA phonons by the exchange of their RPA counterparts, with the ultimate purpose of arriving at a self-energy approximation that is valid for both finite and extended systems.

Going beyond the TDA level has proven to be very difficult [16]. The Faddeev random-phase approximation (FRPA) [17] has solved this problem by using the Faddeev technique [18] to include RPA phonons in the self-energy. The FRPA method has been successfully applied to both nuclei [19,20] and atoms [21]. It is the aim of this paper to explore the application of this technique to simple molecular systems.

In the second section of this paper, we give an overview of the working equations for the FRPA method. In Sec. III, we present the numerical results for a set of small molecules. A summary is provided in Sec. IV.

## II. THEORY

### A. Single-particle Green's function

The single-particle motion in an $N$-body system is described by the single-particle propagator [7,8] (atomic units are used throughout the paper)

$$G_{\alpha,\beta}(t,t') = -i\langle\Psi_0^N|\mathcal{T}[a_\alpha(t)a_\beta^\dagger(t')]|\Psi_0^N\rangle, \quad (1)$$

where $\mathcal{T}[\ldots]$ represents the time-ordering operator, $\Psi_0^N$ is the exact ground state, and $a_\alpha(t)$ and $a_\alpha^\dagger(t)$ are the addition and removal operators in the Heisenberg representation for an electron in a single-particle state $\alpha$. For practical calculations, it is more convenient to use the Lehmann representation of the Green's function

$$\begin{aligned}G_{\alpha,\beta}(E) &= \sum_{m>F}\frac{\langle\Psi_0^N|a_\alpha|\Psi_m^{N+1}\rangle\langle\Psi_m^{N+1}|a_\beta^\dagger|\Psi_0^N\rangle}{E-\left(E_m^{N+1}-E_0^N\right)+i\eta} \\ &+ \sum_{m<F}\frac{\langle\Psi_0^N|a_\beta^\dagger|\Psi_m^{N-1}\rangle\langle\Psi_m^{N-1}|a_\beta|\Psi_0^N\rangle}{E-\left(E_0^N-E_m^{N-1}\right)-i\eta} \\ &= \sum_{m>F}\frac{f_{\alpha,m}f_{\beta,m}^*}{E-\omega_m+i\eta} + \sum_{m<F}\frac{f_{\alpha,m}f_{\beta,m}^*}{E-\omega_m-i\eta},\quad (2)\end{aligned}$$

where the $\Psi_m^{N\pm1}$ represent exact eigenstates of the Hamiltonian with energy $E_m^{N\pm1}$. This transition to the energy domain

*matthias.degroote@ugent.be

# Accuracy of the Faddeev random phase approximation for light atoms

C. Barbieri[*]

*Department of Physics, Faculty of Engineering and Physical Sciences, University of Surrey, Guildford, Surrey GU2 7XH, United Kingdom
and Theoretical Nuclear Physics Laboratory, RIKEN Nishina Center, 2-1 Hirosawa, Wako, Saitama 351-0198, Japan*

D. Van Neck and M. Degroote

*Center for Molecular Modeling, Ghent University, Technologiepark 903, B-9052 Gent, Belgium*

The accuracy of the Faddeev random phase approximation (FRPA) method is tested by evaluating total and ionization energies in the basis-set limit. A set of light atoms up to Ar is considered. Comparisons are made with the results of coupled-cluster singles and doubles (CCSD), with third-order algebraic diagrammatic construction [ADC(3)], and with the experiment. It is seen that even for two-electron systems, He and $Be^{2+}$, the inclusion of RPA effects leads to satisfactory results, and therefore it does not overcorrelate the ground state. The FRPA becomes progressively better for larger atomic numbers, where it gives $\approx$5 mH more correlation energy, and it shifts ionization potentials by 2–10 mH with respect to the similar ADC(3) method. The ionization potentials from FRPA tend to reduce the discrepancies with the experiment.

## I. INTRODUCTION

*Ab initio* studies of electronic systems aim at a direct solution of the Schrödinger equation in terms of the underlying Coulomb interaction, thus avoiding phenomenological input [1]. Due to its favorable accuracy and the gentle scaling of computational requirements with increasing number of particles the single-reference coupled-cluster (CC) method [2] has become the most frequently used theory in contemporary investigations of molecular and atomic systems. One can thus calculate molecules for which full configuration interaction (FCI) would not be feasible. Another approach with analogous characteristics is the Green's function (GF) theory (or, equivalently, propagator theory) [3–6]. An early scheme based on this approach is the outer-valence GF (OVGF) [7,8], which expands ionization energies up to third order in perturbation theory. The OVGF is very practical and computationally simple, and it has found several applications to studies of ionization spectra [8–12]. However, it becomes inaccurate whenever inner- and outer-valence ionization energies (IEs) are subject to shake-up contaminations [13–15]. In such cases one needs to resort at least to the third-order algebraic diagrammatic construction [ADC(3)] method [16,17]. The ADC($n$) approach is an intermediate-state representation [18,19] of the GF made to be consistent with perturbation theory up to order $n$. Thus, it is size consistent and can be systematically improved by going to higher orders. For the one-body propagator, ADC(3) implies performing explicit configuration mixing between valence electrons and shake-up configurations, such as two-hole–one-particle (2h1p) and/or two-particle–one-hole (2p1h) configurations [20]. These states are mixed together by ADC(3) theory in a Tamm-Dancoff approximation (TDA) fashion. The accuracy of the ADC(3) approach has been tested in several studies for both IEs [20,21] and excited states [22,23].

Green's function theory has also extensive applications to solid-state physics, where the most successful scheme is the GW approximation (GWA) [24–26]. The GWA describes the modification of electrons through repeated interactions with collective particle-hole (ph) excitations of the system, which are described in the random phase approximation (RPA). The RPA is essential for extended systems because it screens the Coulomb interaction at large distances [5,6] and it guarantees finite correlation energies in metals and the uniform electron gas [27–30]. In contrast, the TDA plasmon spectrum is incorrect and even diverges at small momenta. The GWA, however, is not always satisfactory. As an example, particle-particle (pp) and hole-hole (hh) configurations, which would be included by ADC(3) but not in GWA, are necessary to explain satellite structures above and below the Fermi surface [31]. How to include these effects efficiently in GWA is still being researched [32]. Conversely, the inclusion of RPA in atomic and molecular studies may be advantageous for describing long-range (van der Waals) forces and dissociation processes [33–35]. Thus, a practical method that combines ADC(3) with RPA might become beneficial to the fields above.

In a recent publication we have considered the *ab initio* calculation of the Ne atom using Green's function theory in the so-called Faddeev random phase approximation (FRPA) [36], which was originally proposed for studies of nuclear structure [37–41]. This approach completely includes 2p1h and 2h1p states in the self-energy but expands these configurations in terms of couplings between valence particles and (simpler) ph and pp (hh) excitations that are calculated using RPA. By calculating these excitations in TDA one would be led back to the ADC(3) scheme. Stated otherwise, the FRPA can be seen as an extension of ADC(3) that employs RPA for pairs of electron and hole excitations (see also Ref. [42] for a discussion). The inclusion of RPA opens a possible way for treating long-range correlations in both finite and extended electron systems on a equal footing. An application of this method to extended systems, such as the electron gas, could be pursued following the approach of Ref. [43]. Before this is attempted, however, one wishes to better assess the quality of the method for

---

[*]Present address: University of Surrey, Guildford GU2 7XH, United Kingdom. Email: c.barbieri@surrey.ac.uk

									\clearnewpage

\end{document}